\RequirePackage{fix-cm}
\documentclass[smallextended]{svjour3}       
\smartqed  
\usepackage{graphicx}
\usepackage{amsmath}	
\usepackage{amssymb}	
\usepackage{wasysym}
\usepackage[usenames]{xcolor}

\begin{document}

\title{Towards a sustainable exploitation of the geosynchronous orbital region
}
%


\titlerunning{Towards a sustainable GEO}        

\author{Ioannis Gkolias         \and
        Camilla Colombo 
}


\institute{I. Gkolias \at
				Department of Aerospace Science and Technology,
				Politecnico di Milano, Via la Masa 34, 20156 Milan, Italy
              \email{ioannis.gkolias@polimi.it}           
           \and
           C. Colombo \at
           		Department of Aerospace Science and Technology,
				Politecnico di Milano, Via la Masa 34, 20156 Milan, Italy
               \email{camilla.colombo@polimi.it}  
}

\date{Received: date / Accepted: date}

\maketitle

\begin{abstract}
In this work the orbital dynamics of Earth satellites about the geosynchronous altitude are explored, with primary goal to assess current mitigation guidelines as well as to discuss the future exploitation of the region. A thorough dynamical mapping was conducted in a high-definition grid of orbital elements, enabled by a fast and accurate semi-analytical propagator, which considers all the relevant perturbations. The results are presented in appropriately selected stability maps to highlight the underlying mechanisms and their interplay, that can lead to stable graveyard orbits or fast re-entry pathways. The natural separation of the long-term evolution between equatorial and inclined satellites is discussed in terms of post-mission disposal strategies. Moreover, we confirm the existence of an effective cleansing mechanism for inclined geosynchronous satellites and discuss its implications in terms of current guidelines as well as alternative mission designs that could lead to a sustainable use of the geosynchronous orbital region.  
\end{abstract}

\section{Introduction}
\label{intro}
Artificial Earth satellites provide a wide variety of services to the humankind. Weather monitoring and prediction, marine and air traffic management, telecommunications, television broadcasting, geolocalization, just to name a few, heavily reside on satellite information. Especially important to these activities is the contribution of satellites orbiting at geosynchronous altitude (GEO\footnote{ The acronym GEO in this work is used as Geosynchronous Earth Orbits, therefore, representing all the orbits with a period of about one sidereal day, and not only the equatorial ones (also known as Geostationary Equatorial Orbits).}). GEO satellites have a semi-major axis of about $R_{\textrm{GEO}} = 42165$ $\textrm{km}$ and orbit about the Earth at the same rate that the Earth rotates around its axis. Due to this commensurability, the geosynchronous region provides us unique ground-tracks that have been heavily exploited since the beginning of the Space Era. However, this came with a toll, the region about Earth's Clarke belt has been populated with artificial objects, most of them debris. GEO is nowadays the second most populated orbital region, after the low Earth orbits (LEO), with a couple of thousand catalogued objects and the list is continuously growing\footnote{ https://www.sdo.esoc.esa.int/environment\_report/Space\_Environment\_Report\_latest.pdf (retrieved at 28/3/2019)}.

The importance of the GEO region dictated to satellite operators to take measures at their missions' operational end-of-life. Indeed, even the very first GEO equatorial satellites applied some kind of re-orbiting manoeuvres to clear the longitude slots for future missions \cite{Frem2013}. In the early 90s, when the first disposal studies were performed for equatorial GEO satellites \cite{cho1990}, the use of super-synchronous \emph{graveyard} orbits was proposed as an economical and effective solution to reduce the collision probability in the region. For the following years this was the norm followed by the operators, and is used up to now, modified to take also into account the perigee height variations due to solar radiation pressure perturbations \cite{DelFre2005,Chao2005}. The aforementioned ideas have sculpted the Inter Agency Debris Coordination Committee (IADC) set of mitigation guidelines for decommissioning GEO spacecraft \cite{IADC2017} and the European Space Agency (ESA) instructions set for the ESA-operated GEO missions \cite{ESA2015}. The guidelines suggest that a decommissioned GEO space system should: 1) increase the altitude of perigee by 235 $\textrm{km}$ plus a factor accounting for the solar radiation pressure perturbations and 2) to circularise the graveyard orbit such that the eccentricity is of the order of $10^{-3}$.

However, from a sustainability point of a view, the use of graveyard orbits as suggested by the current mitigation guidelines, will keep increasing the collision probability in GEO \cite{Jenkin2018}. Moreover, some further considerations support the need to investigate alternative ways of GEO exploitation. For instance, defunct satellites can act also as sources of secondary debris, even if there are no collisions. Satellites in graveyard orbits are subject to solar radiation torques that can constantly speed up their rotation \cite{Albuja2018}. The structural integrity of a space system rotating at several times per minute is not guaranteed. In fact, a population of high-area-over-mass (HAMR) object observed in GEO \cite{Schi2004} was identified as multi-layer insulation (MLI) foils from satellites in graveyard orbits. The dynamics of these objects are quite different from the parent satellites \cite{Valk2009} due to their higher Area-to-mass ($A/m$) ratio. Their long-term orbital evolution suggests that they cannot be contained in the nominal graveyard orbit and they could potentially cross the GEO protected region\footnote{ The GEO protected region is defined at the semi-major axis $a_{\textrm{prot}}=R_{\textrm{GEO}} \pm 200 \textrm{ km}$ and a latitude sector from $15^\circ$ South to $15^\circ$ North \cite{IADC2017,ESA2015}.}.

Alternative solutions would require to either try to manoeuvre the satellite to a heliocentric orbit, but it is a cost-inefficient solution, or try to achieve atmospheric re-entry via natural perturbations. Unfortunately, the most effective perturbation that leads to the re-entry of close Earth satellites, namely the atmospheric drag, is not present at GEO altitude. Nevertheless, re-entry within reasonable time-scales could be achieved also by exploiting other type of perturbations. In their study of LEO satellites \cite{Ale2018a,Ale2018b} suggested the possibility of re-entry due to solar radiation pressure resonances, but unfortunately this mechanism is also not effective for GEO. A more promising strategy would be to exploit the lunisolar gravitational perturbations \cite{Ros2015,Daq2016}, an idea that has been already been suggested for satellites in the Medium Earth Orbits (MEO) region  \cite{Ale2016,Ros2017,Skoulidou2017,Arm2018} and Highly Elliptical Orbits (HEO) \cite{Jenkin2008,Colombo2015}. A living example of a mission that exploited natural lunisolar effects for its re-entry strategy is the INTEGRAL spacecraft in HEO. ESA operators manoeuvred INTEGRAL in 2015 such that it will perform a safe re-entry in 2029. The optimal manoeuvre design to enhance the effect of lunisolar perturbations \cite{Colombo2014,Armellin2015} was used as starting point for the definition of the re-entry trajectory sequence \cite{Merz2015}.

In order to assess the possibility of re-entry due to lunisolar perturbations, the natural evolution of orbits at geosynchronous altitude has to be well understood. The dynamics of the geosynchronous equatorial orbits have been studied in the literature, both analytically and numerically \cite{Lara2002,Breit2005,Cel2014,Gach2017}. Some aspects of the dynamics of the highly inclined GEO orbits have also been presented \cite{DelMor1993,Breit2007,Zhang2017}. In this work, we will revisit the dynamics of geosynchronous region, in terms of searching for feasible re-entering highways or stable graveyard orbits when re-entry is not feasible. Having in mind a propagation span of at least 120 years, we employ an efficient semi-analytical propagator \cite{Colombo2016} that takes into account all the relevant forces at geosynchronous orbits in a single-averaged formulation (see Appendix A). The effect of Earth's precession, which is usually omitted, but turns out to be relevant for long-term propagations at high altitudes \cite{Gur2007}, has also been efficiently formulated and included (Appendix B).

Three different mapping methods are selected to highlight the different aspects of the dynamics. First, we study the contribution of the tesseral effects on the eccentricity growth for geosynchronous orbits. Then we employ dynamical maps over the angle-like variables, namely the argument of the perigee and the right ascension of the ascending node of the satellite, to explore the contribution of the different perturbations and their interactions. Finally, a global picture of the geosynchronous region in action-like variables (eccentricity and inclination) is presented to identify the orbital regimes where re-entry solutions are viable. 

A natural separation is observed in terms of the natural evolution of initially low- and high-inclined orbits. Long-term stability and low eccentricity variations is the norm for low inclinations while an abundance of re-entering orbits exist in high inclinations. The eccentricity growth maps allow us to identify particular re-entry orbits that are interesting for future mission planning. We study the lifetimes of those orbits and try to identify the conditions that could lead to fast re-entry. Another interesting interaction that is revealed is the interplay of solar radiation and lunisolar perturbations for low-eccentric orbits. We present a case where, despite the usual behaviour where opening a solar-sail at the end-of-life enhances the de-orbiting process, this is not happening for GEO orbits. Finally, a general assessment of the current guidelines in GEO is made based on the current population and the underlying dynamics. The case of the Sirius constellation provides a strong case why a single equation guideline is not adequate to fully regulate the dynamically rich geosynchronous region. 

The paper is organised in the following way, in Sec.~\ref{sec:semianalytical} the single-averaged model is presented, in Sec.~\ref{sec:dynamicalmapping} we present the results of the dynamical mapping of the geosynchronous region, in Sec.~\ref{sec:disposal} the findings in terms of post-mission disposal planning are discussed and in Sec.~\ref{sec:conclusions} we present our conclusions.

\section{Orbit propagation with semi-analytical techniques}\label{sec:semianalytical}

The methods used to analyse the long-term orbital evolution of geosynchronous orbits are briefly discussed here. A detailed description of the force model is given in Appendix A. Aiming for a 120-years integration time span, a single-averaged semi-analytical propagation was opted for, which is a typical practise for Earth satellite orbits. The PlanODyn \cite{Colombo2016} orbital analysis suite was adopted and further developed to include the relevant perturbation effects. For the main force model, the contributions of Earth's geopotential, third body perturbation from the Sun and the Moon and the effect of solar radiation pressure were considered. Additionally, a revised version of Earth's general precession contribution to the secular evolution was developed and it is presented in the Appendix B.

For the geopotential force a fourth order and degree truncation is chosen, on the grounds that adding higher order harmonics did not produce any noticeable effects. For the zonal harmonics the first order averaged contributions were considered \cite{Kau1966} and the second order contribution due to $J_2$ ($J_2^2$) \cite{Brou1959} was also included. On the other hand, for the tesseral effects, only the resonant contributions relevant for geosynchronous orbits were considered \cite{Kau1966}.
 
The third body potential is expanded up to fourth order in the parallactic factor and is averaged in closed form over the mean anomaly of the satellite \cite{Kauf1972,Colombo2015}. This is more efficient computationally \cite{Lane1989} instead of using a series expansion representation (see \cite{CelRos2017} and references therein). The positions of the perturbing bodies, i.e. the Sun and Moon, are computed from analytical time-series truncated to a sufficient order for our work \cite{Meeus1998}. The solar radiation pressure is also included under a cannonball approximation \cite{Kau1962,Krivov1997}. The Earth's shadow is not taken into account since the constant Sun exposure is valid at geosynchronous altitude. Finally, Earth's general precession is also considered (Appendix B).

\begin{figure}
\centering
\includegraphics[width=\textwidth]{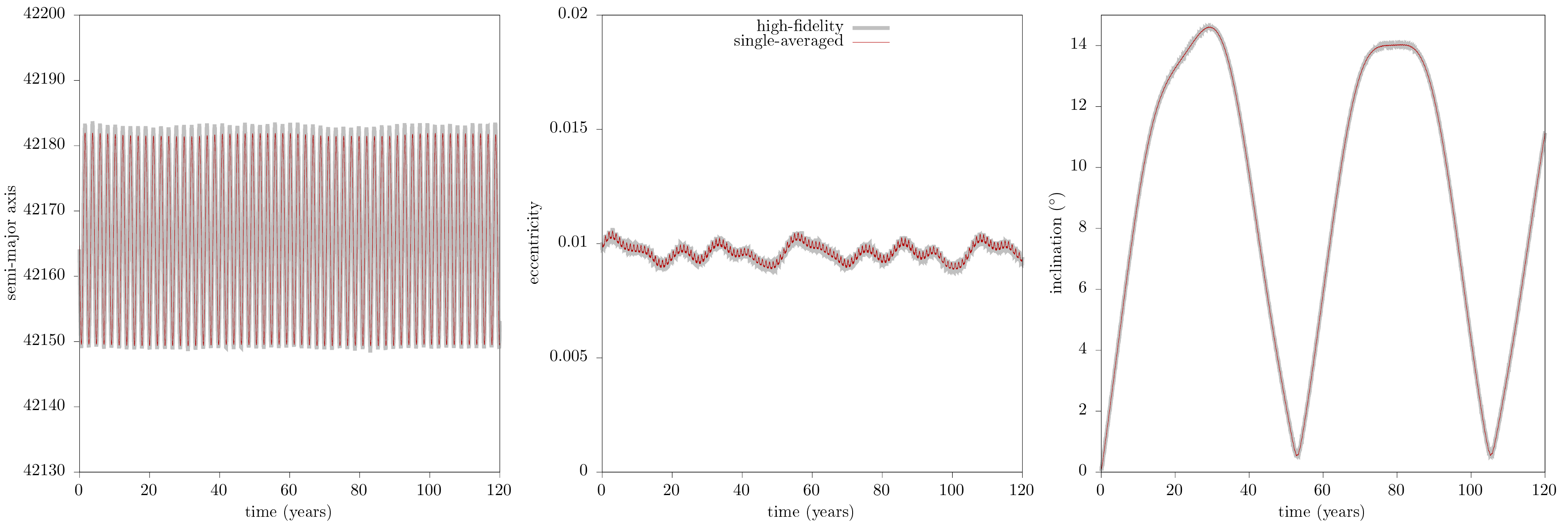}
\caption{The 120 years orbital evolution of an equatorial GEO satellite ($a=R_{\textrm{GEO}}$, $e=0.01$, $i=0.1^\circ$, $\Omega = 10^\circ$, $\omega = 50^\circ$, $M = 0^\circ$ and initial epoch 21/06/2020 at 06:43:12.0). The semi-analytical propagation (red line) is in excellent agreement with a high-fidelity model (grey line) for the same initial conditions.} 
\label{fig:orbitcomp}
\end{figure}

In Fig.~\ref{fig:orbitcomp} a validation of the full force model for the GEO region is presented, including all the above mentioned contributions. An initial condition for a GEO equatorial satellite ($a=R_{\textrm{GEO}}$, $e=0.01$, $i=0.1^\circ$, $\Omega = 10^\circ$, $\omega = 50^\circ$, $M = 0^\circ$ and initial epoch 21/06/2020 at 06:43:12.0) is propagated for 120 years with the semi-analytical method (red line) and is compared with a high-fidelity (grey line) integration. The high-fidelity model is in Cartesian coordinates, using Cunningham's algorithm \cite{Montenbruck2012} for the geopotential calculations and NASA's SPICE tool-kit\footnote{ https://naif.jpl.nasa.gov/naif/index.html} for the ephemerides of the Sun and the Moon and to retrieve the transformation matrices related to the motion of Earth's rotation axis. The semi-analytical propagation used in this work is in excellent agreement with the high-fidelity one, validating the force model selection and the use of a single-averaged formulation. Moreover, the single-averaged formulation is a few orders of magnitudes faster, which allows us to proceed with a massive and accurate characterisation of the phase space.

\section{Dynamical Mapping}\label{sec:dynamicalmapping}     
															

In this Section, a dynamical study of the GEO region is performed, with the goal of presenting a detailed and complete study of post-mission disposal opportunities. Our work focuses on two main elements: the study of the GEO area in its entirety with a complete force model and the search not only for good graveyard solutions, but also for potentially exploitable re-entry trajectories.

In this work we considered only dynamical indicators associated with the eccentricity evolution of the orbit, which is mainly driving the perigee variations and is the key in studying and planning post-mission disposal strategies. Moreover, other type of dynamical indicators based on the exponential divergence of nearby orbits, could fail to provide a clear picture for our purposes. Namely, in the region of the separatrix of the geosynchronous resonance, it is known that there exist chaotic behaviour \cite{Breit2005,Valk2009,Cel2014}. However, in disposal design applications, we are not interested in the chaotic evolution with respect to any of the action variables, but rather in the chaotic behaviour manifesting as eccentricity growth. The fact that the use of chaotic indicators in disposal design can be missleading was also mentioned in a recent work \cite{Daq2018}. 

Therefore, we focus our studies on the eccentricity variations and more specifically, two types of indicators are used. First the typical diameter of the eccentricity 
\begin{equation}
Diam(e) = |e_{max} - e_{min}|
\end{equation}
defined as the absolute difference between the minimum and the maximum values of the eccentricity obtained along the propagation time span.

The second indicator, mainly used in Sec.~\ref{subsec:actionspace}, is a \emph{normalised eccentricity diameter} and is designed to study the eccentricity variations along a wide range of initial eccentricities and/or semi-major axis. It is defined as \cite{Gko2016}:
\begin{equation}
\Delta e = \frac{|e_0 - e_{max}|}{|e_0 - e_{re-entry}|}
\end{equation} 
where $e_0$ is the initial eccentricity, $e_{max}$ its maximum value along a propagation and $e_{re-entry}$ is defined as the value of eccentricity for which the perigee value becomes equal to a re-entry condition. Assuming the re-entry condition of $120$ $\textrm{km}$ above the surface of the Earth ($a_{re-entry}=R_{Earth} + 120$ $\textrm{km}$) for a satellite at the GEO region the re-entry value for the eccentricity is $e_{re-entry} \approx 0.846$. The behaviour of $\Delta e$ then is the following: when the eccentricity stays bounded about its initial value $\Delta e \rightarrow 0$, while if the eccentricity grows enough to reach the re-entry value $\Delta e \rightarrow 1$. In other words, $\Delta e$ gives the fraction of the physically available phase-space that the eccentricity evolution has covered.

For the numerical exploration the single-averaged formulation was propagated numerically, using a Bulirsch–Stoer numerical integrator with automatic time-step control, imposing a local relative tolerance of $10^{-13}$.  A propagation is terminated if the re-entry condition is met. Studying the full 6-dimensional space of orbital elements for all possible configurations could be a really involved task, so instead we have selected to study appropriate slices of it. Various 2-dimensional grids of initial conditions were chosen with a resolution of $201\times 201$ orbits, yielding 40400 orbits per plot presented here. For each orbit, the computational time is a few seconds for a 120-years time span, giving less than a day computational time on average for each map\footnote{ The simulations were carried out in the Milkyway server, which is equipped with 4 Intel Xeon CPU E5-4620 v4 at 2.10 GHz with a total amount of 40 physical cores.}. A total amount of more than 50 million orbits were propagated in the GEO region, for the needs of the ReDSHIFT project \cite{Rossi2018}.   


For the selected grids of initial conditions, two values of the effective Area-to-mass ratio was used, a low one $A/m = 0.012$ $\textrm{m}^2/\textrm{kg}$, which is the typical value for decommissioned satellites, and a high one $A/m = 1.0$ $\textrm{m}^2/\textrm{kg}$, which represents a satellite that carries an area-augmenting device (i.e. a solar sail kit) that is deployed at its end-of-life. The reflectivity coefficient was constant and equal to  $c_R=1$. Finally, the initial epoch was selected on the 21/06/2020 at 06:43:12.0. 


\subsection{Tesseral maps}


The first effect to be explored is the interaction between the GEO tesseral resonance and the other perturbations. For this reason, a series of eccentricity diameter maps on a two dimensional grid of initial semi-major $a$ and the 24-hour resonant angle $\lambda = \omega + \Omega + M - \theta_g$ were produced, with $\omega,\Omega ,M$ the argument of perigee, the right ascension of the ascending node and the mean anomaly of the satellite respectively, and $\theta_g$ is the Greenwich hour angle. There exist several ways to vary the resonant angle, but it was decided to do so only by varying the mean anomaly at the initial epoch. This selection allows to focus on the different dynamics induced purely from the tesseral contribution. In fact, varying the initial $\omega$ or $\Omega$, would not complement the study as changing those two angles would also affect the initial configuration with respect to the lunisolar and solar radiation pressure perturbations and would contaminate the results. 

\begin{figure}
\centering
\includegraphics[width=0.49\textwidth]{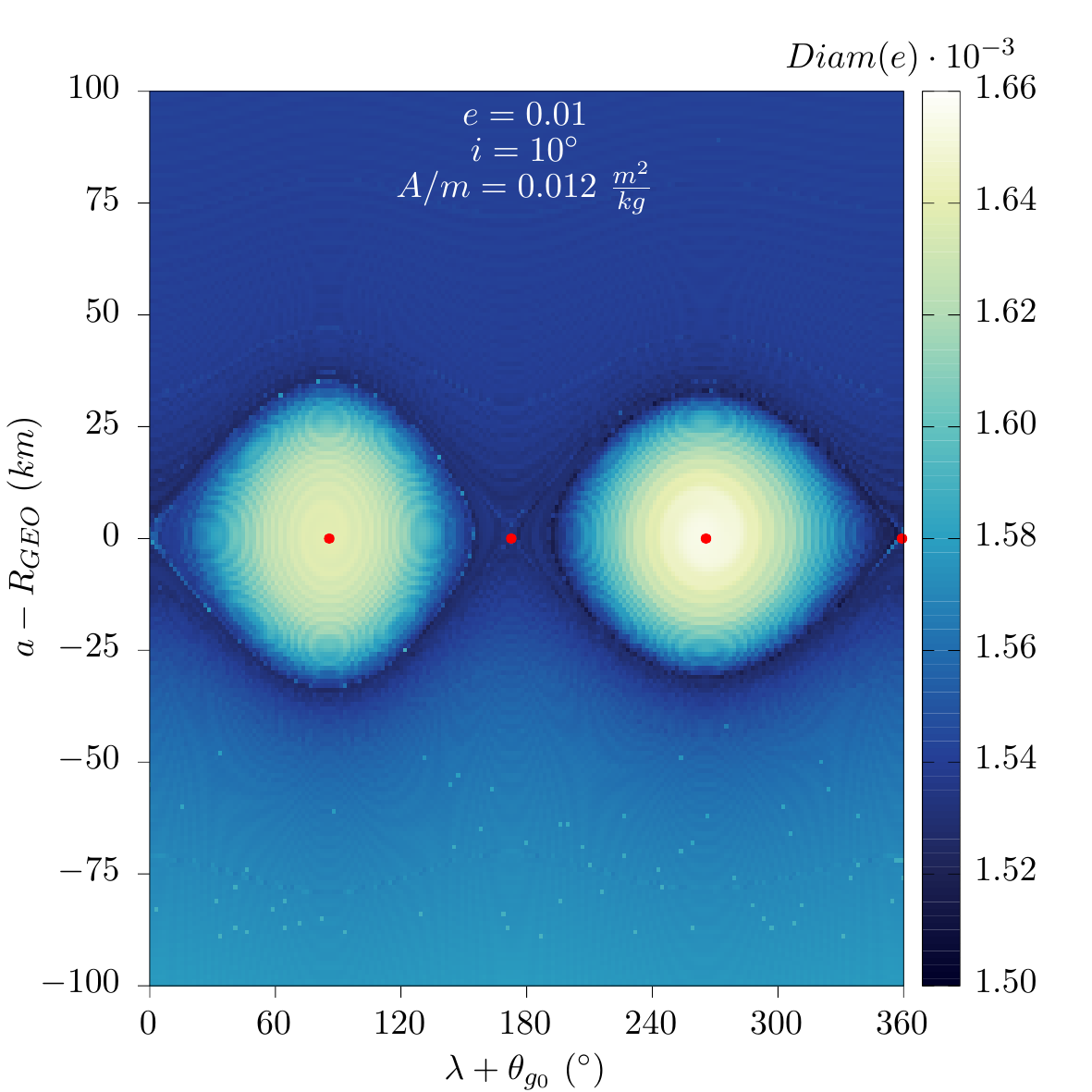}
\includegraphics[width=0.49\textwidth]{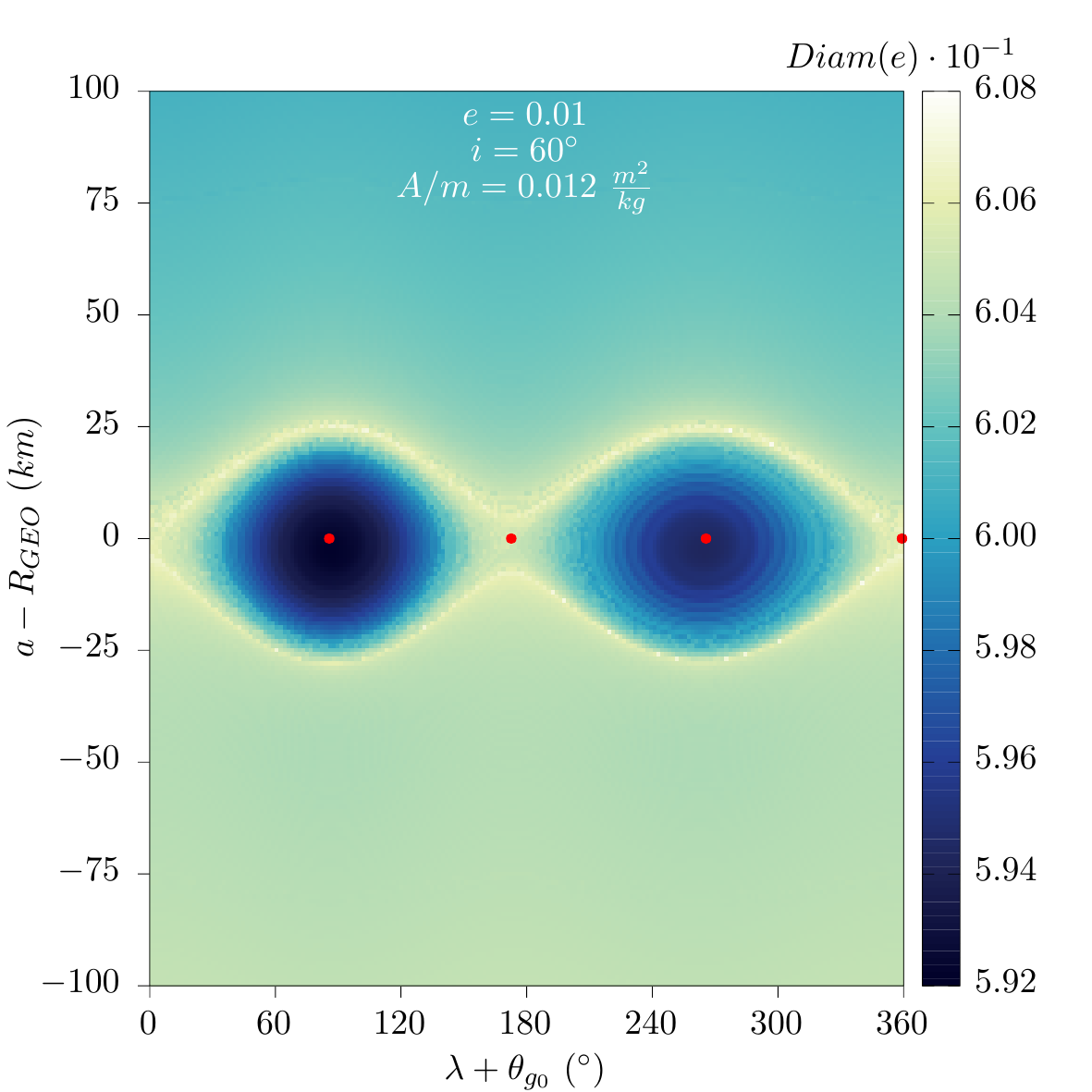}
\includegraphics[width=0.49\textwidth]{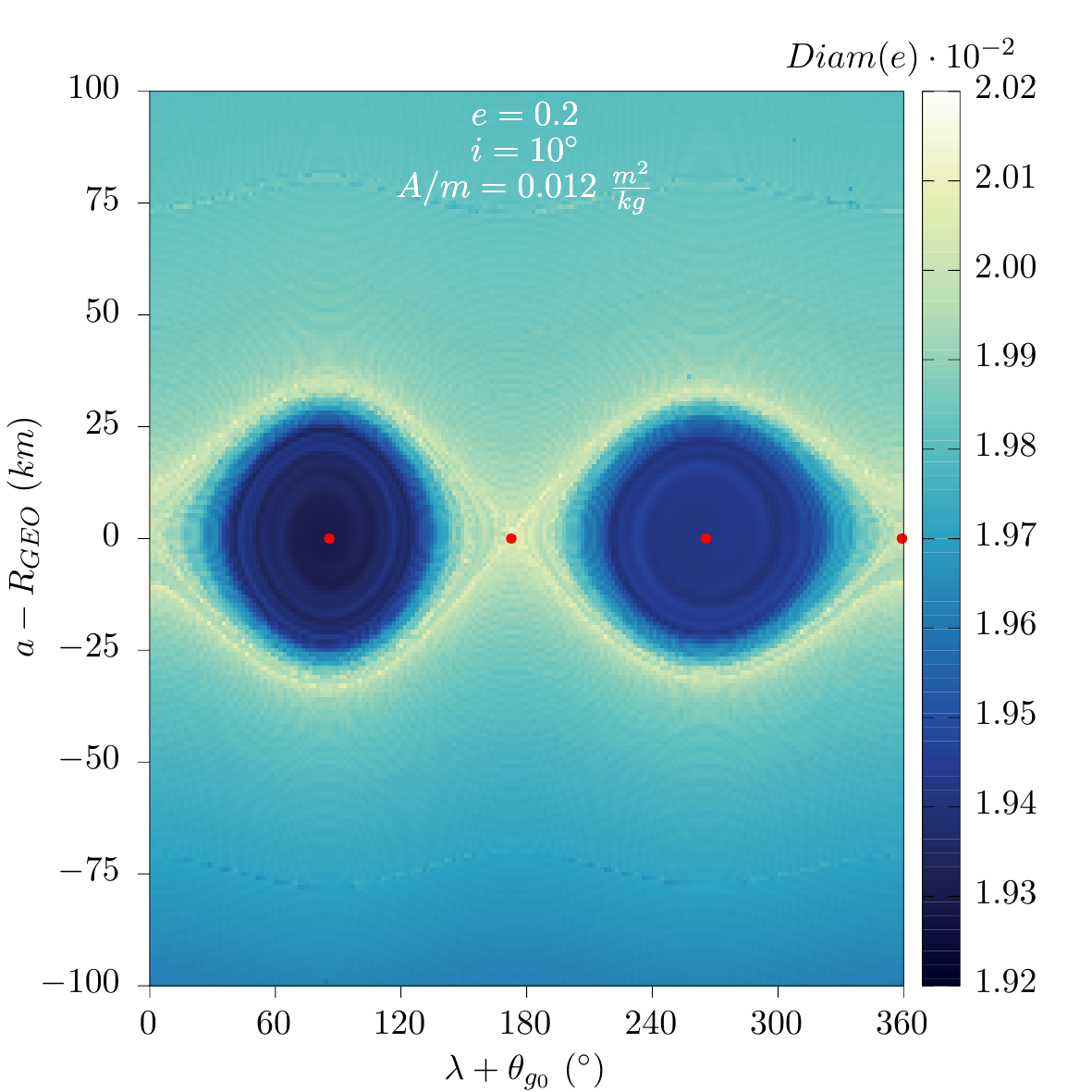}
\includegraphics[width=0.49\textwidth]{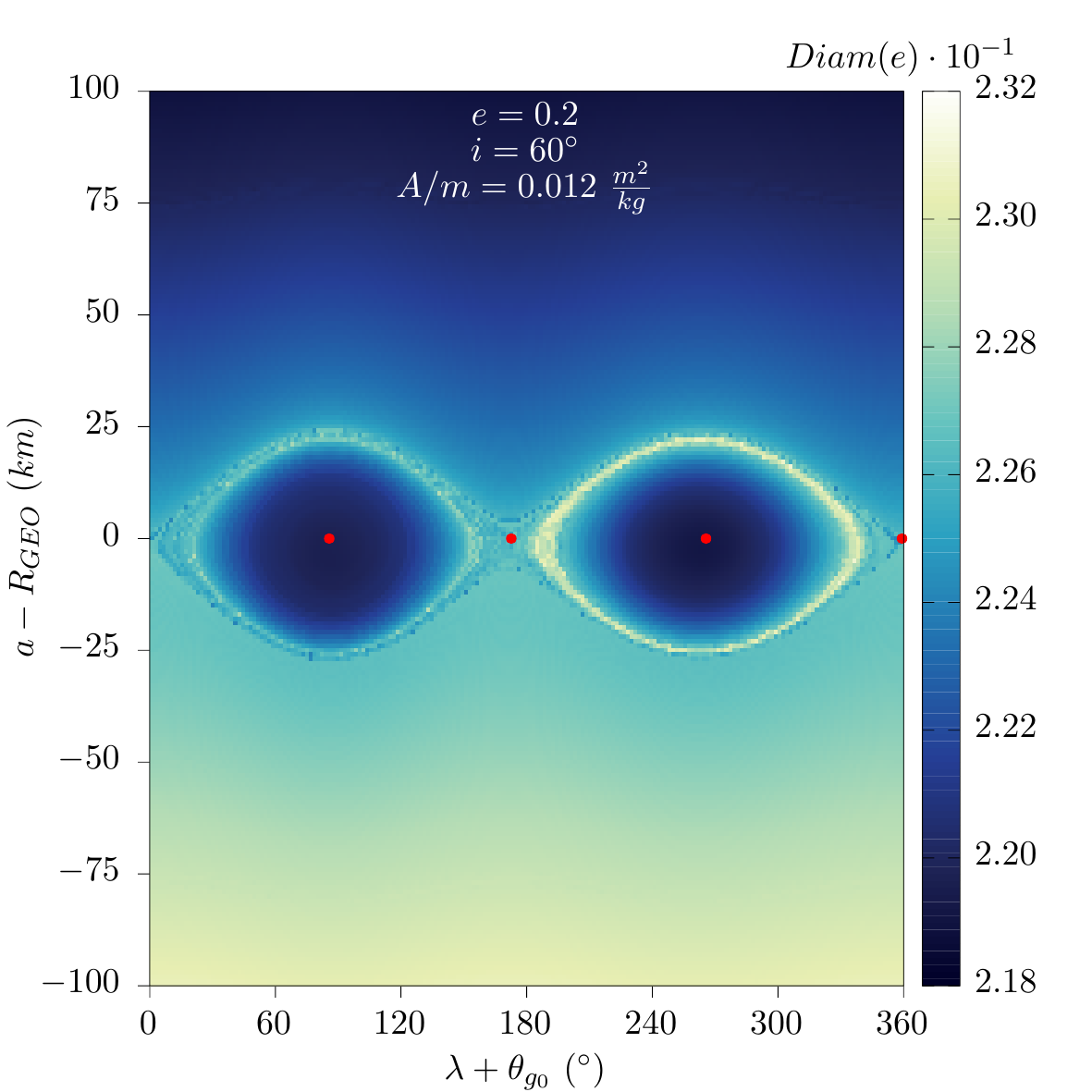}
\caption{Dynamical maps on the semi-major axis - geosynchronous resonant angle ($a - \lambda$) plane for the low area-to-mass case $A/m  = 0.012$ $\textrm{m}^2/\textrm{kg}$. The left column corresponds to low inclination orbits ($i = 10^\circ$) and the right column to the high-inclination ones ($i = 60^\circ$). The top row shows the evolution of low-eccentricity  ($e = 0.01$), while the bottom row that of high-eccentricity orbits ($e = 0.2$). The colormap corresponds to the value of the eccentricity diameter over 120 years. Notice the narrow range between the minimum and maximum eccentricity diameter among all the maps. }
\label{fig:eyeatm12}
\end{figure}

\begin{figure}
\centering
\includegraphics[width=0.49\textwidth]{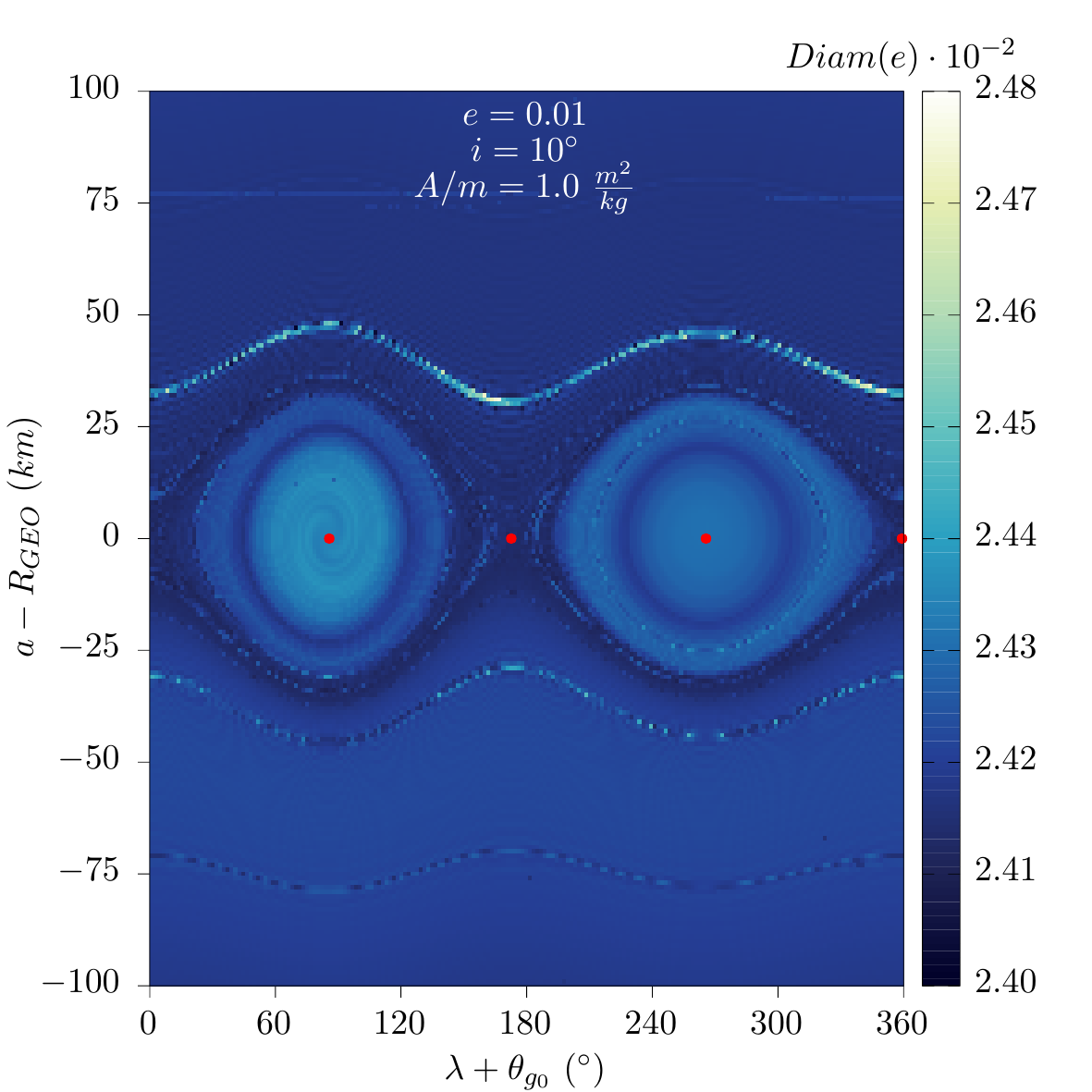}
\includegraphics[width=0.49\textwidth]{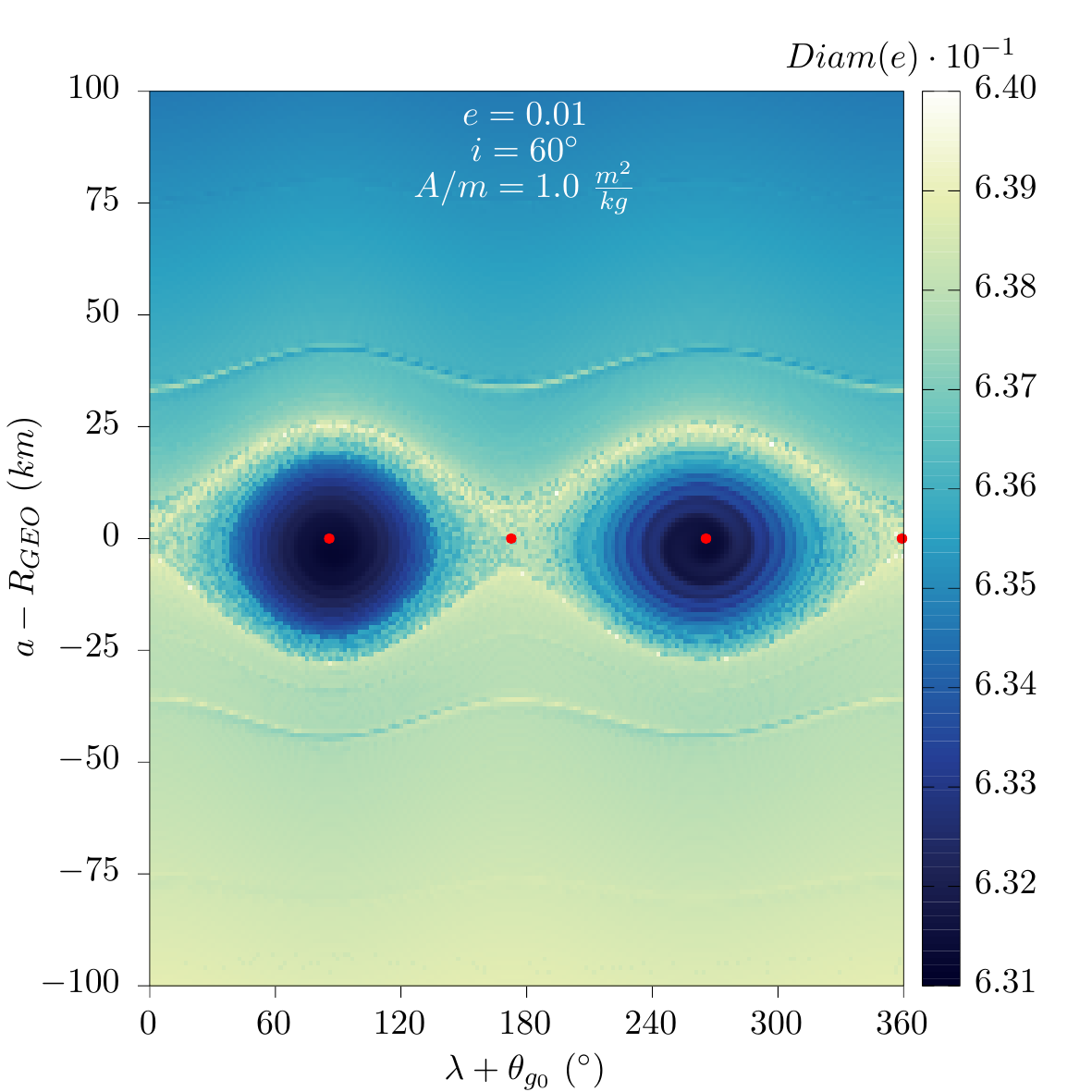}
\includegraphics[width=0.49\textwidth]{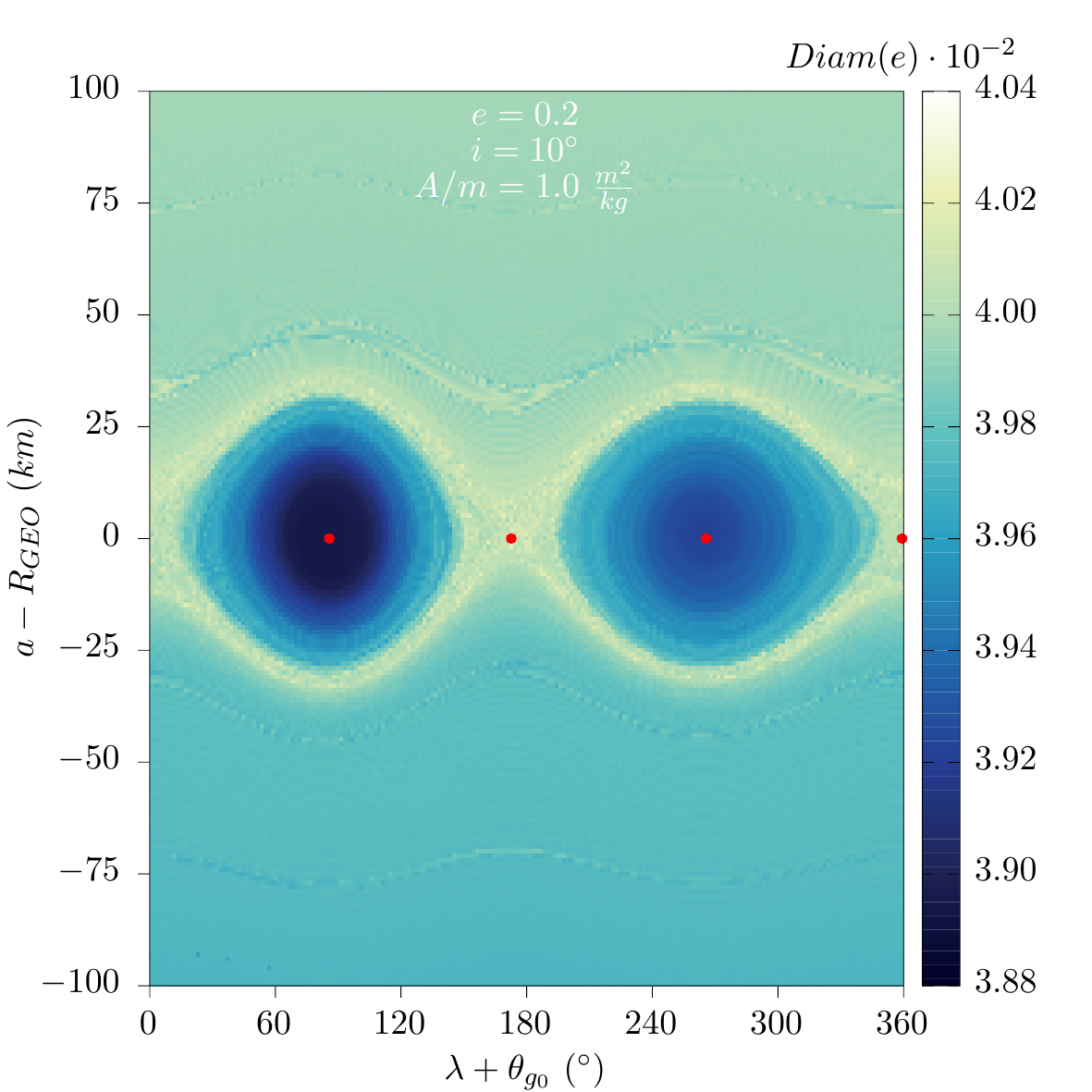}
\includegraphics[width=0.49\textwidth]{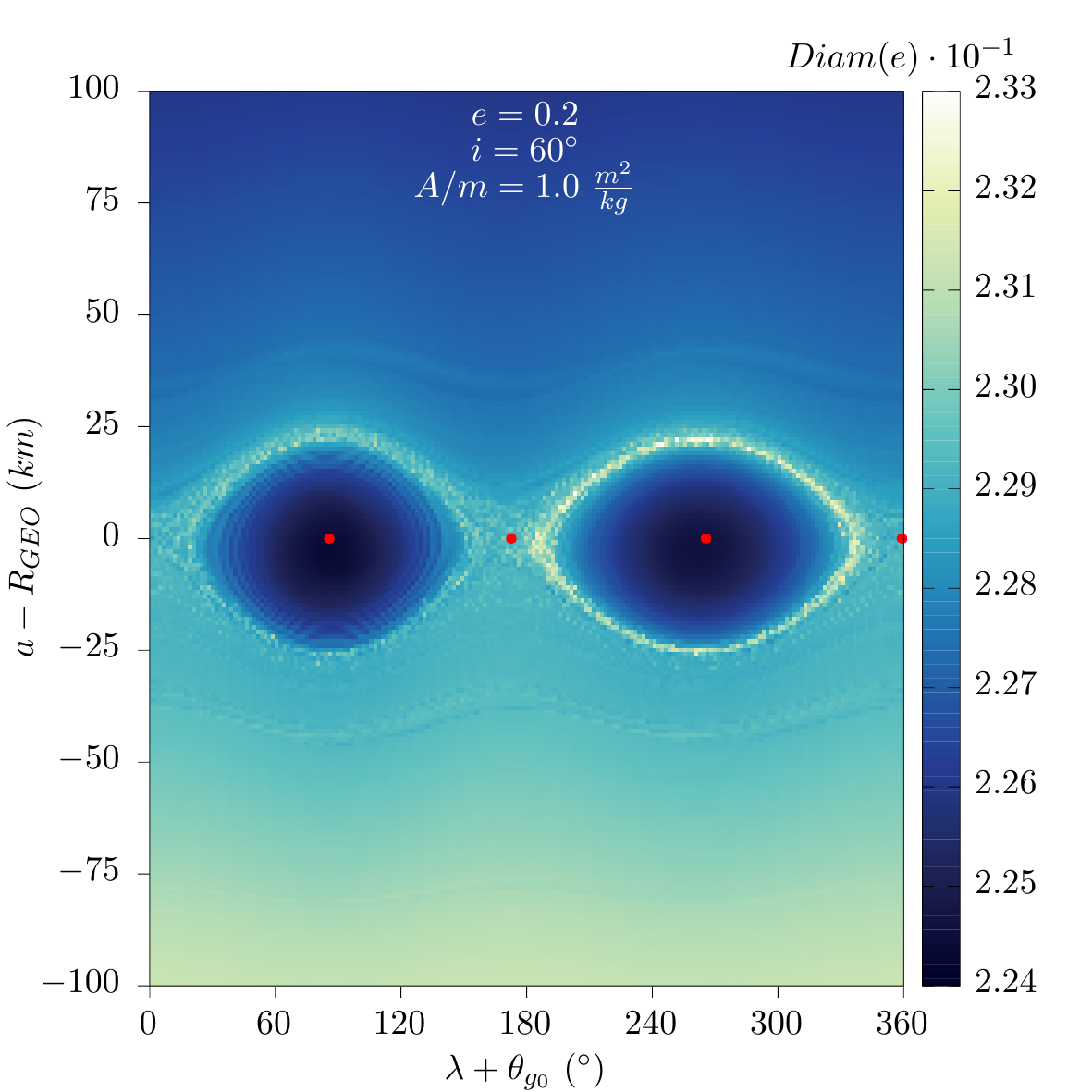}
\caption{Dynamical maps on the semi-major axis - geosynchronous resonant angle ($a - \lambda$) plane for the high area-to-mass case $A/m  = 1.0$ $\textrm{m}^2/\textrm{kg}$. The left column corresponds to low-inclination orbits ($i = 10^\circ$) and the right column to high-inclination ones ($i = 60^\circ$). The top row shows the evolution of low-eccentricity  ($e = 0.01$), while the bottom row that of high-eccentricity orbits ($e = 0.2$). The colormap corresponds to the value of the eccentricity diameter over 120 years. Notice the narrow range between the minimum and maximum eccentricity diameter among all the maps.}
\label{fig:eyeatm1000}
\end{figure}

In Fig.~\ref{fig:eyeatm12} the mapping for a typical satellite with $A/m  = 0.012$ $\textrm{m}^2/\textrm{kg}$ is presented. Both a low-inclined (left column) and high-inclined (right column) configurations are presented. The first thing to mention is the difference in the order of magnitude of the eccentricity diameters with respect to the inclination. The low-inclined orbits exhibit a variation of the order of $10^{-3}$, while for the highly inclined ones it is two orders of magnitude larger. A low-eccentricity (top row) and a high-eccentricity (bottom row) initial condition is also presented for each case. In all cases we observe a clear distinction between the dynamics within the geosynchronous resonance and outside of it. The low-eccentricity and low-inclination is the only case where the eccentricity variation is larger than the surrounding altitudes. The separatrix of the tesseral resonance is also obvious in all the cases, with the two stable equilibria\footnote{ Notice that the x-axes of Fig~\ref{fig:eyeatm12} and \ref{fig:eyeatm1000} report the value of $\lambda + \theta_{g_0}$, where  $\theta_{g_0} = 10.73^\circ$ is the Greenwich hour angle at the initial epoch.} at $\lambda=74.94^{\circ}$ and $\lambda=254.91^{\circ}$ and the unstable at $\lambda=161.91^{\circ}$ and $\lambda=348.48^{\circ}$ (red points in Figs.~\ref{fig:eyeatm12} and \ref{fig:eyeatm1000}). Moreover, the width of the resonance is reduced from $80$ $\textrm{km}$ in the low-inclination case to about $50$ $\textrm{km}$ in the high-inclination case. The splitting of the separatrix is also apparent, particularly in the low-inclination high-eccentricity case, due to the inclusion of odd tesseral-harmonics in the equations of motion. Finally, for the high-inclination case, we observe smaller eccentricity variations in the low initial eccentricity case than in the high initial eccentricity case, which is an effect due to the lunisolar perturbations.

Fig.~\ref{fig:eyeatm1000} shows a similar mapping for the case of a GEO satellite with an area-augmenting device that allows an $A/m  = 1.0$ $\textrm{m}^2/\textrm{kg}$ at the end-of-life.
In this case, the eccentricity variations in low-inclination case are an order of magnitude higher than in previous case, while in the high-inclination case the variations are similar. The other features are similar, again with the separatrix apparently dividing the phase-space, the width of the resonance decreasing with the inclination and the low-eccentricity high-inclination case exhibiting higher eccentricity variations than the high-eccentricity high-inclination case. Finally, the additional structures that appear as curvy lines above and below the separatrix are associated with secondary resonances related to the solar-radiation pressure. 

The phase-space exploration of the geosynchronous resonance reveals some interesting features of the phase-space, the most prevalent being the separatrix that links the two unstable equilibria. However, one should also notice that the overall differences in the eccentricity variations along the same map are very small. From our investigation it is apparent that the chaos detected in previous works \cite{Breit2005,Cel2014,Valk2009} both for low and high area-to-mass objects is not resulting in any exploitable eccentricity growth with respect to nearby orbits with the same phasing with respect to lunisolar and solar radiation pressure perturbations. Therefore, the conclusion of the resonant angle investigation is that, placing a satellite on the unstable equilibria of the tesseral resonance does not present any significant re-entry opportunities.


\subsection{Disposal maps}


After having excluded the position of the satellite within the geosynchronous resonance as a source of interesting re-entry possibilities, here we investigate the orbital configuration with respect to the Sun and the Moon. The study is performed over a grid of initial argument of the perigee and right ascension of the ascending node. A similar approach has been used for the disposal design in the MEO region \cite{Ale2016,Arm2018}. The advantage of this kind of approach is that it allows, given a particular point in the action-like variables space ($a,e,i$), to study its re-entry properties based on the initial orientation of the orbit with respect to the perturbing bodies. In addition, it is a good tool to study the interactions between all the angle related effects caused by lunisolar perturbations and solar radiation pressure.

\begin{figure}
\centering
\includegraphics[width=0.325\textwidth]{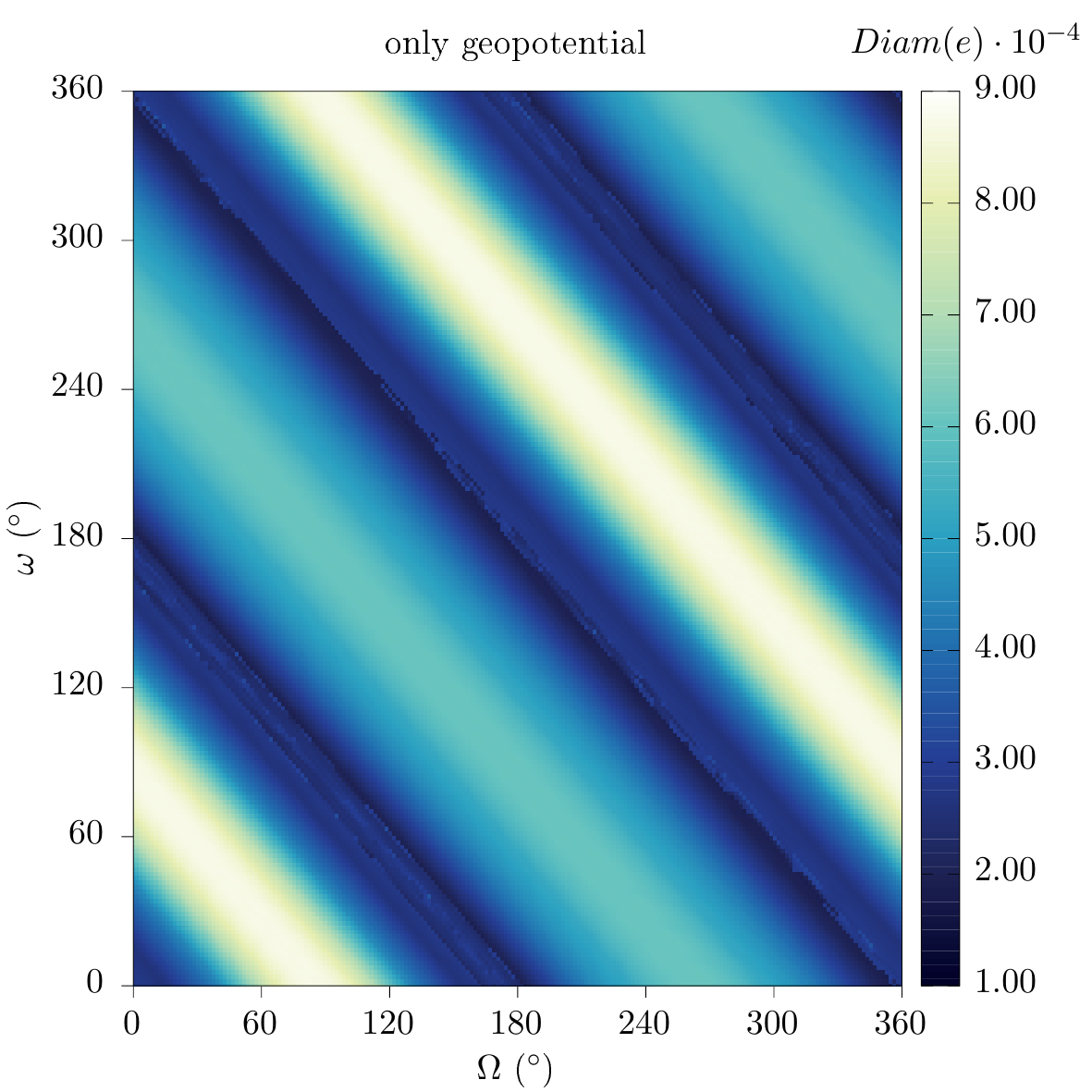}
\includegraphics[width=0.325\textwidth]{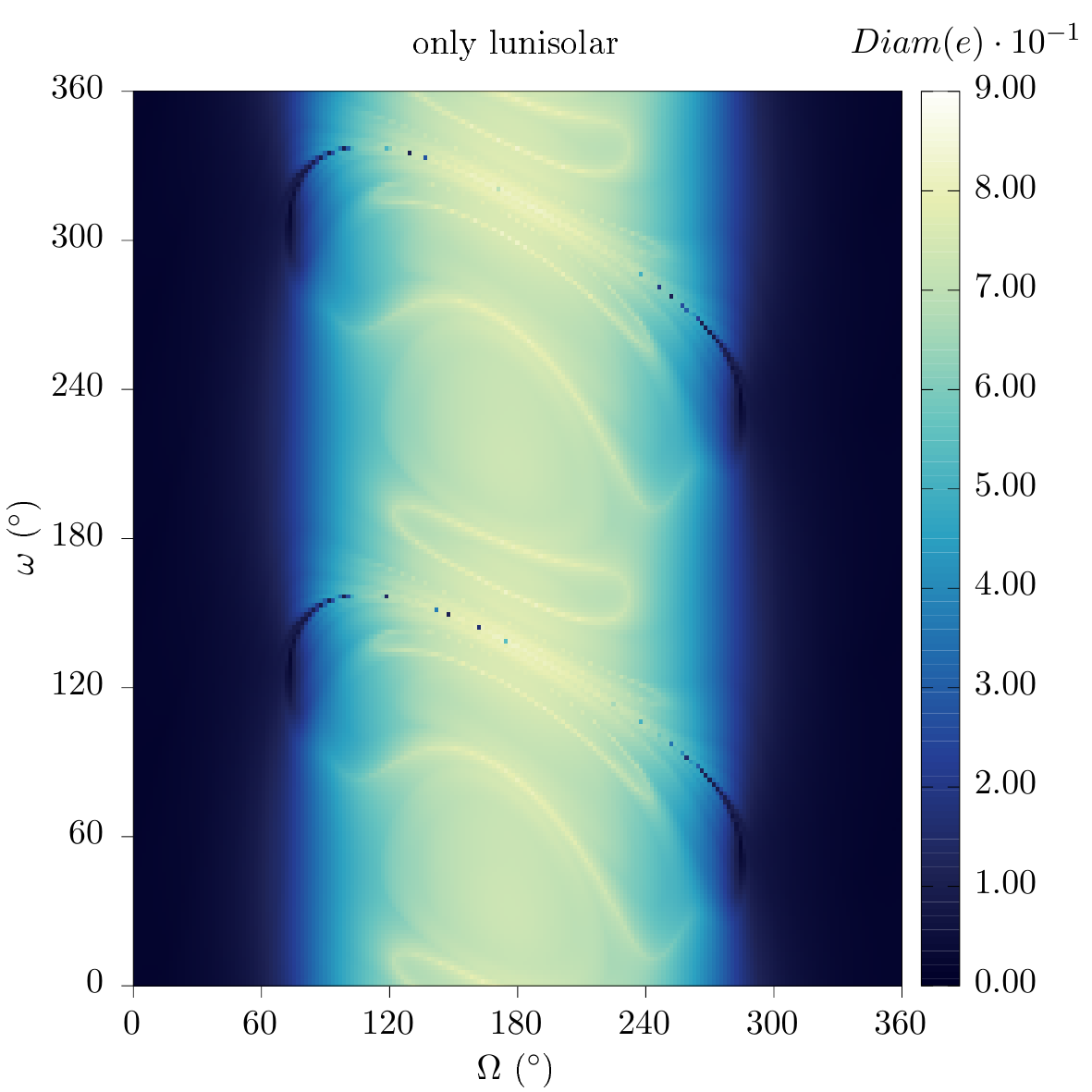}
\includegraphics[width=0.325\textwidth]{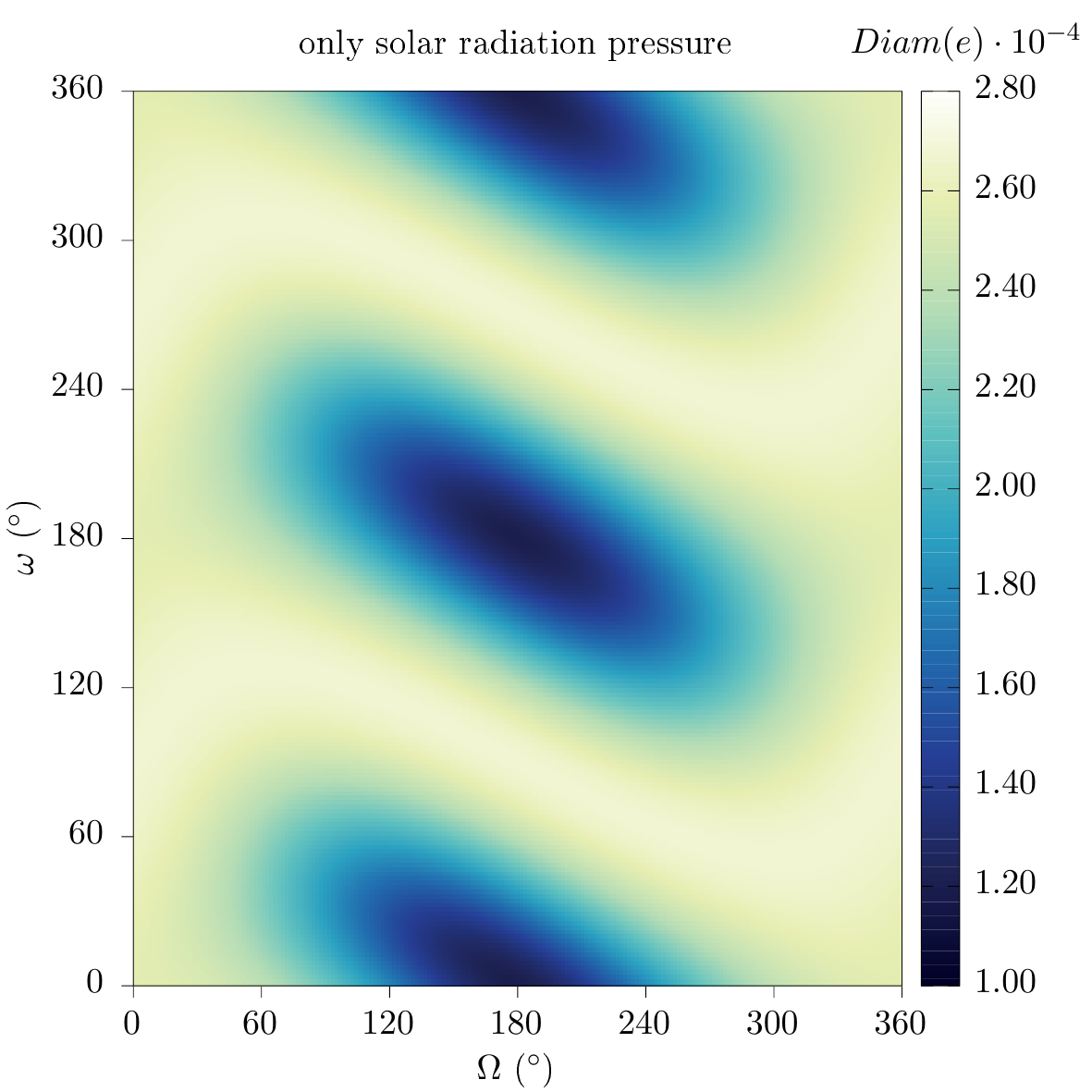}

\includegraphics[width=0.75\textwidth]{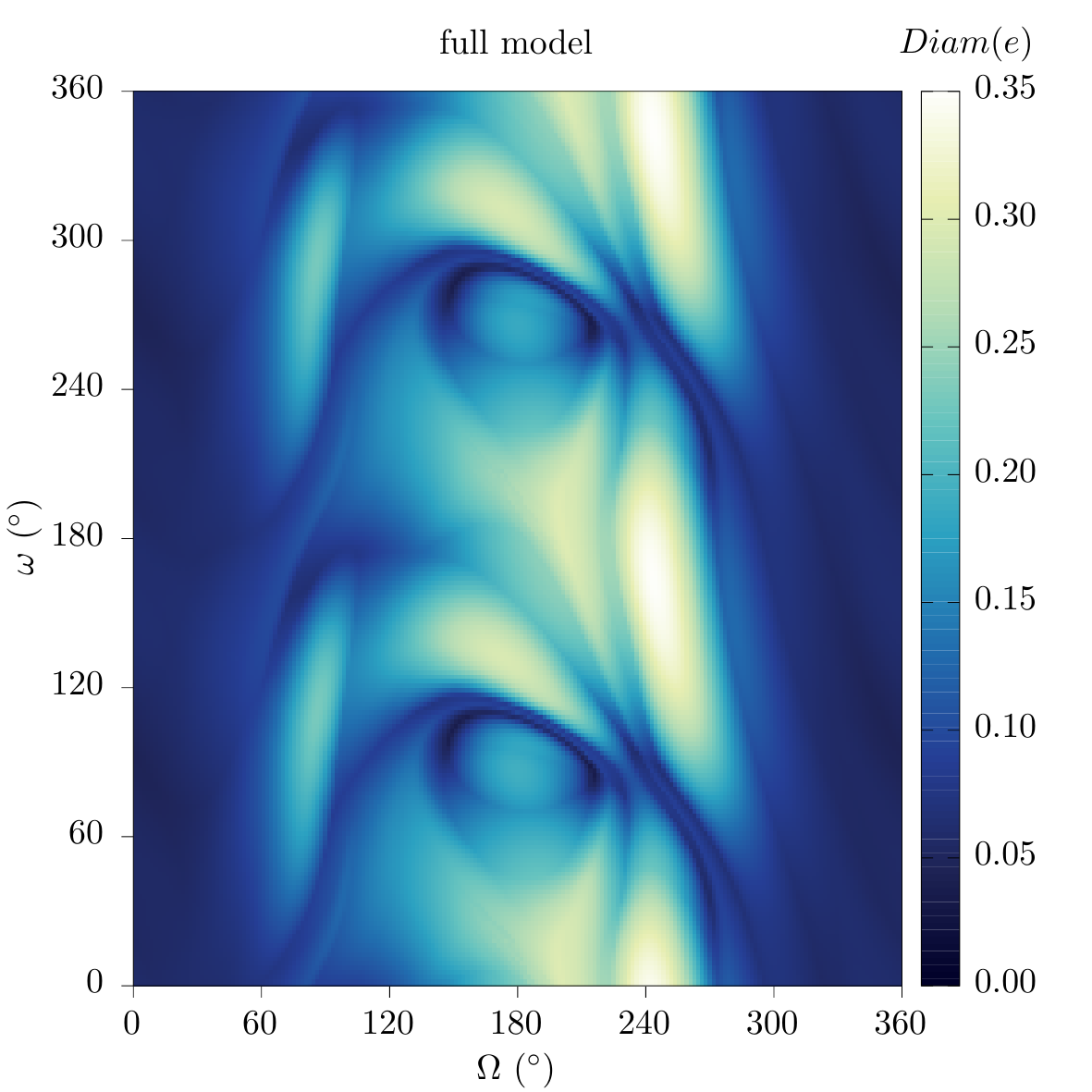}
\caption{The contribution of the different forces acting alone in the $\omega,\Omega$ plane for semi-major axis $a=R_{\textrm{GEO}}$. The initial eccentricity is $0.1$ and the initial inclination $40^\circ$. The colormap corresponds to the value of the eccentricity diameter over 120 years.}
\label{fig:omomcontrib}
\end{figure}

In Fig.~\ref{fig:omomcontrib} we present how the contributions of the different effects mesh up to create a final disposal map for the GEO region ($a = R_{\textrm{GEO}}$). The initial orbital eccentricity is $e=0.1$ and the initial inclination is $40^\circ$ with respect to the equator. The driving force that shapes the eccentricity growth at this altitude is the gravitational lunisolar interaction. Indeed, the eccentricity variation due to the geopotential and the solar radiation pressure are 3 orders of magnitude smaller than the those from the third body perturbations. Basically, under the isolated third body dynamics, all orbits with initial node of 180 degrees are reaching eccentricity values close to $e_{re-entry}$. The way, however, that the geopotential and solar radiation pressure affect the combined effect evolution, is by fixing the frequency of the perigee oscillations. The tuned frequency of perigee, could suppress the Lidov-Kozai type dynamics \cite{Lidov1962,Kozai1962} induced by the combined solar and lunar attractions. Therefore, the result of the evolution under the full dynamical model is quite complex and produces interesting dynamical structures.     

Another interesting feature of those maps is that the position of the instabilities is mainly associated with the orientation of the node of the satellite with respect to the node of the Moon at the starting epoch. Therefore, changing the starting epoch could horizontally shift the appearing structures \cite{Ale2016} and this feature repeats itself with a period of about 18.6 years, which is the nodal precession period of the Moon (also known as the Saros cycle). This adds a third dimension in the post-disposal design scheme and opens up for interesting design opportunities \cite{Ros2017}. Namely, at the end-of-life one could wait for the value of the node to take correct value to maximise the effect of the lunisolar contributions.

\begin{figure}
\centering
	\includegraphics[width=0.325\textwidth]{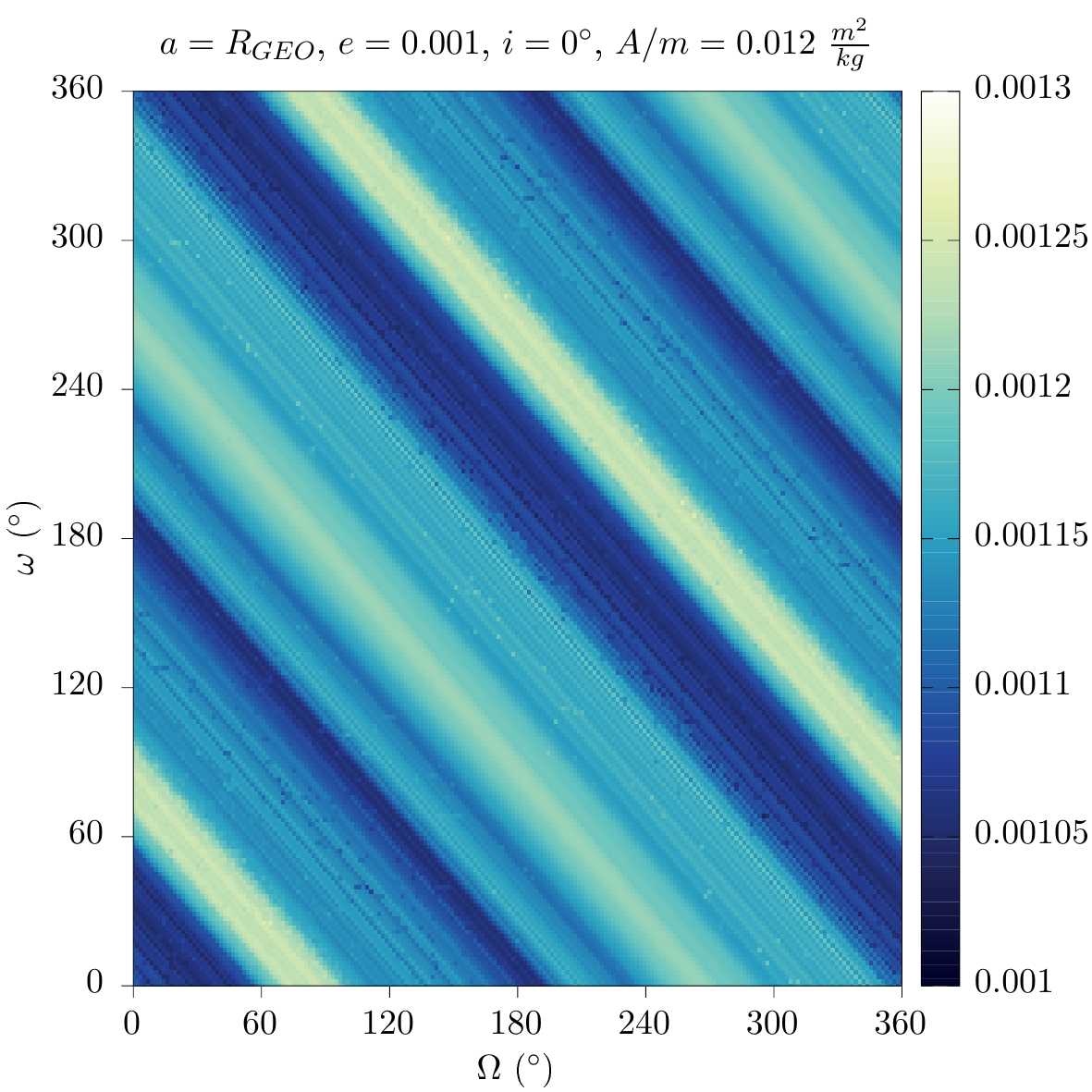}
	\includegraphics[width=0.325\textwidth]{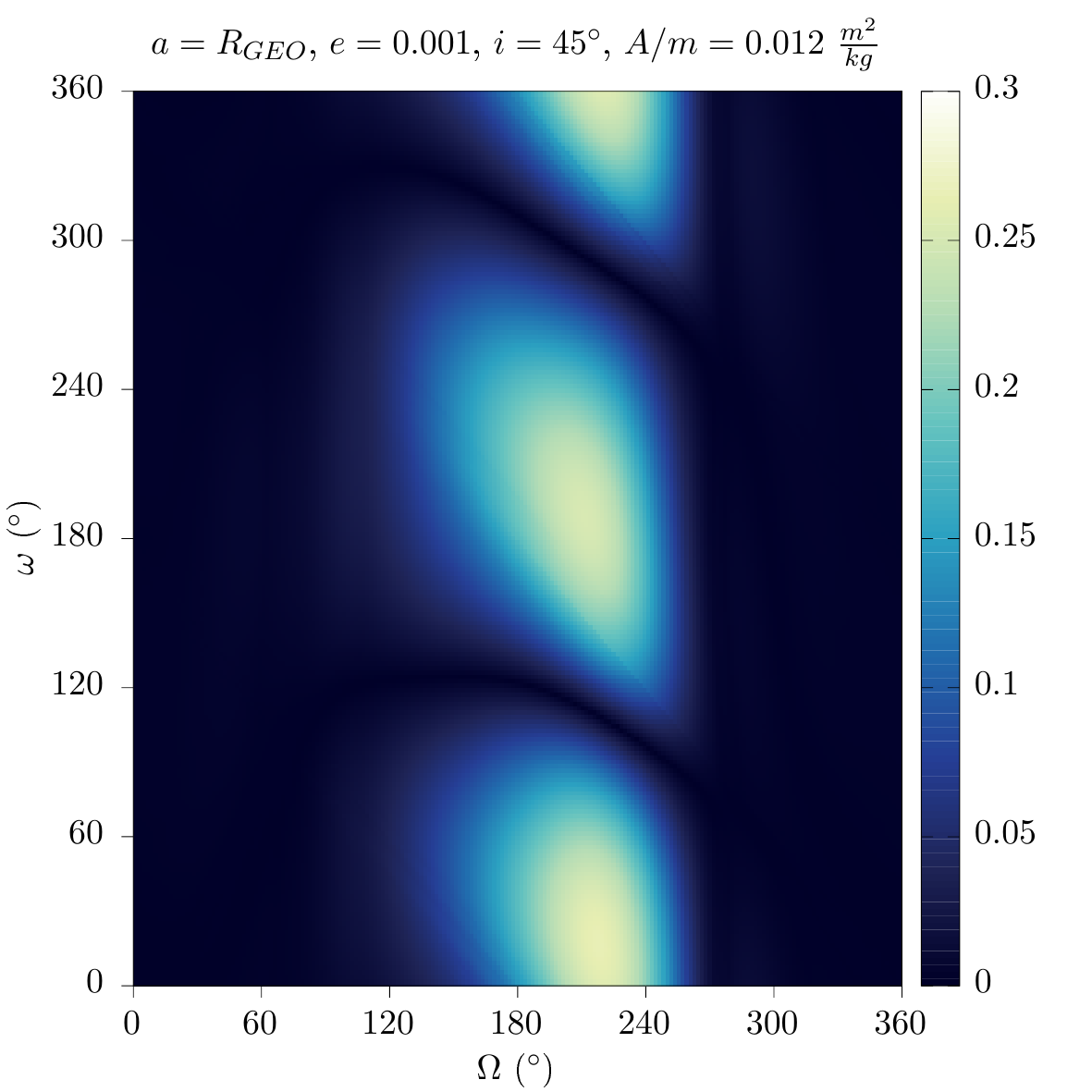}	
	\includegraphics[width=0.325\textwidth]{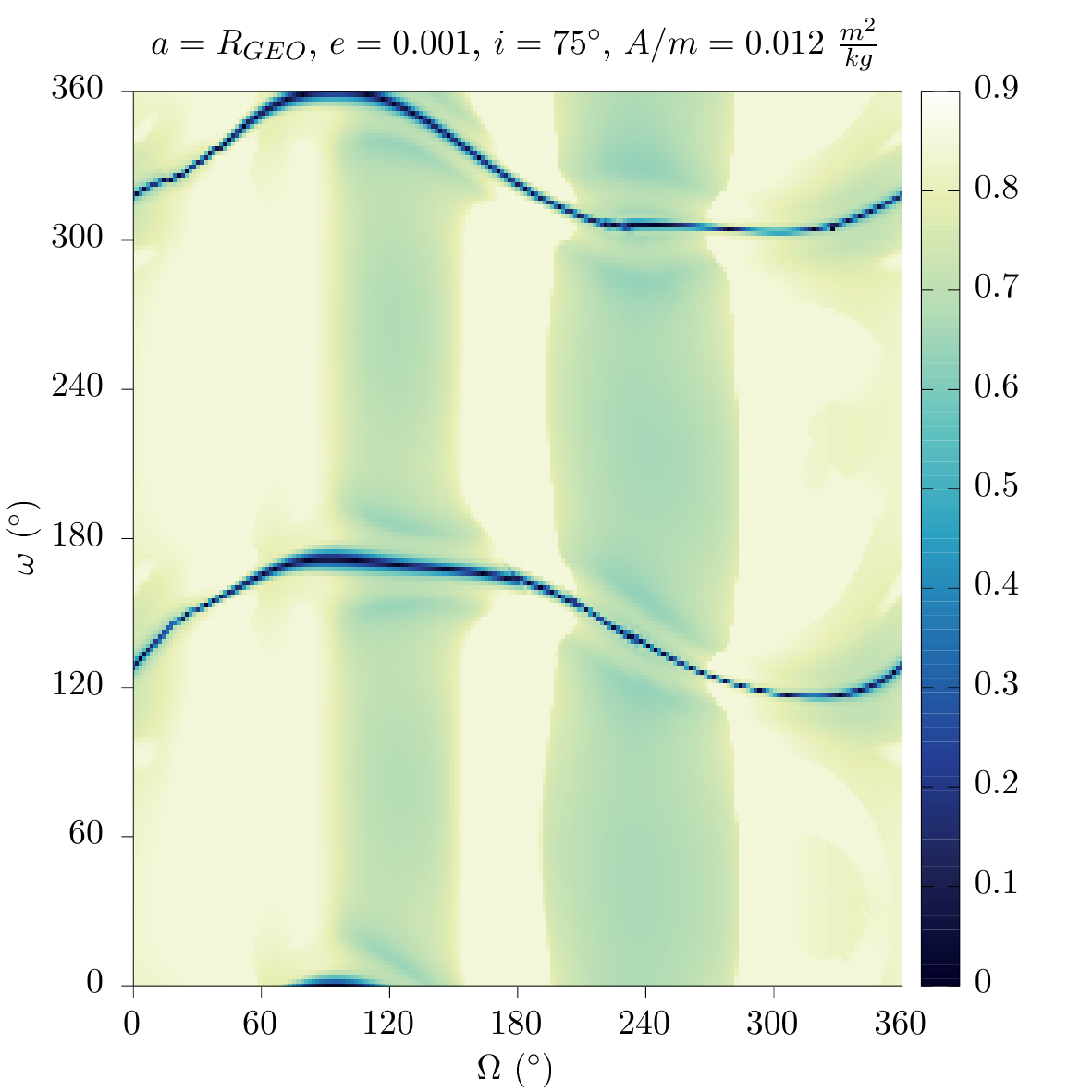}

	\includegraphics[width=0.325\textwidth]{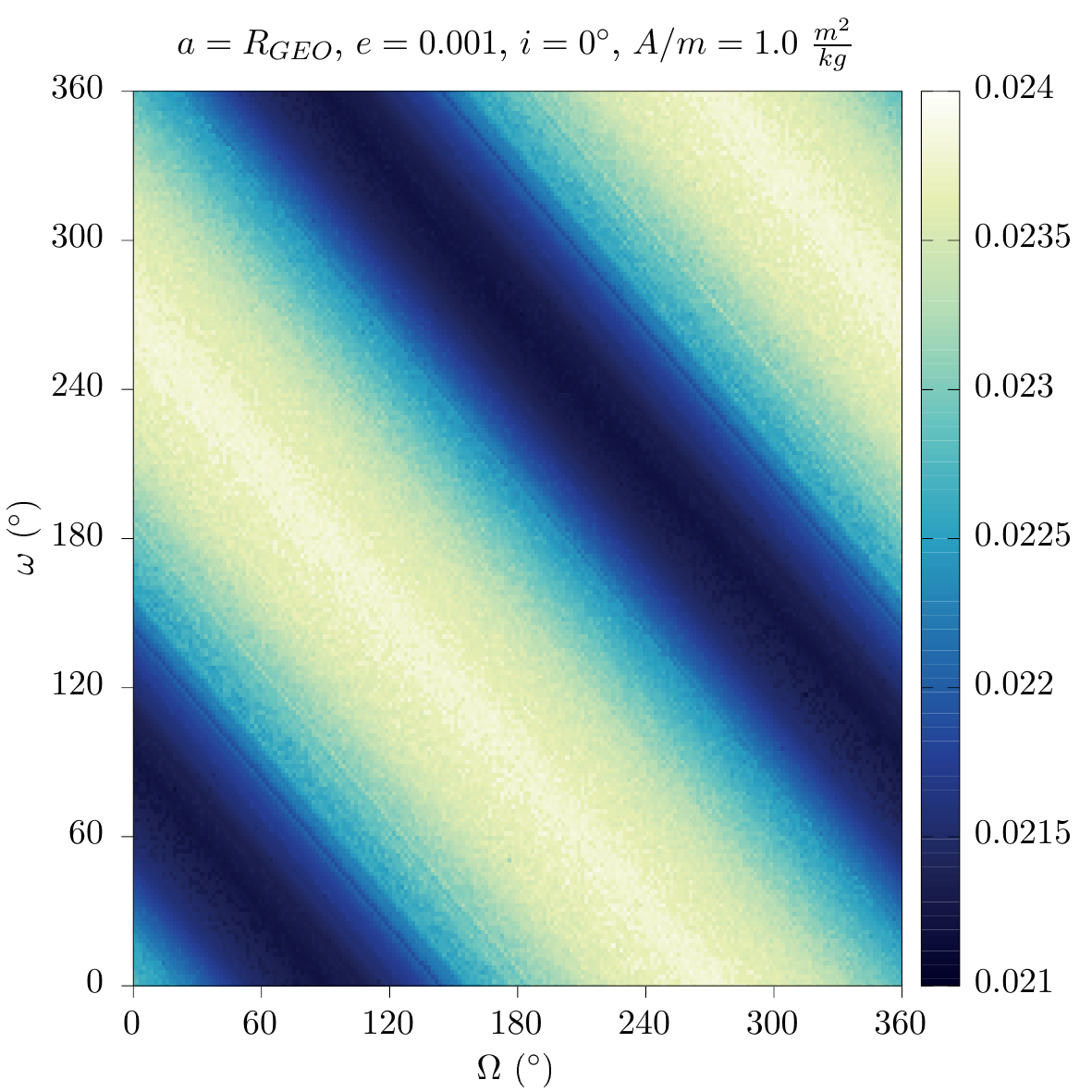}
	\includegraphics[width=0.325\textwidth]{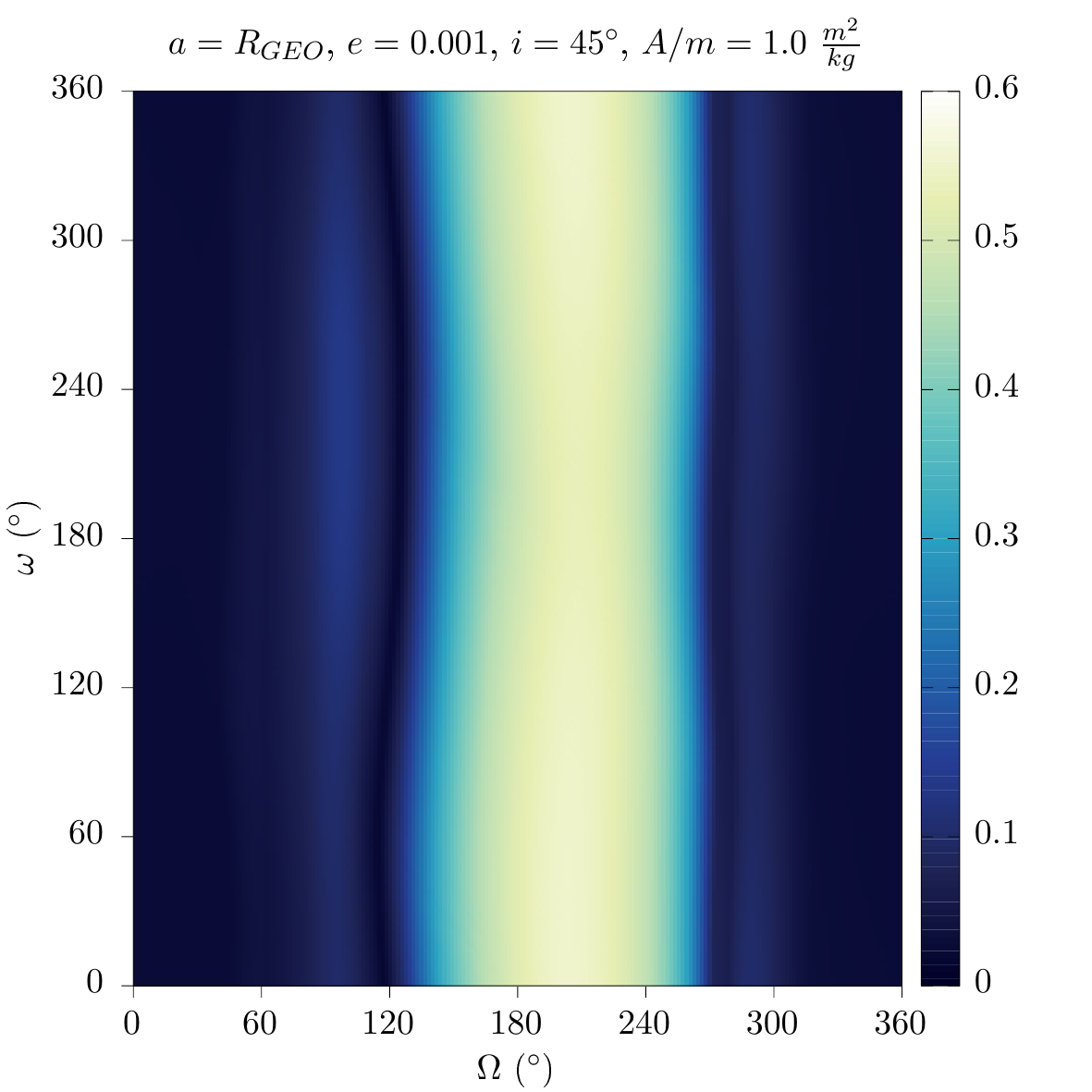}	
	\includegraphics[width=0.325\textwidth]{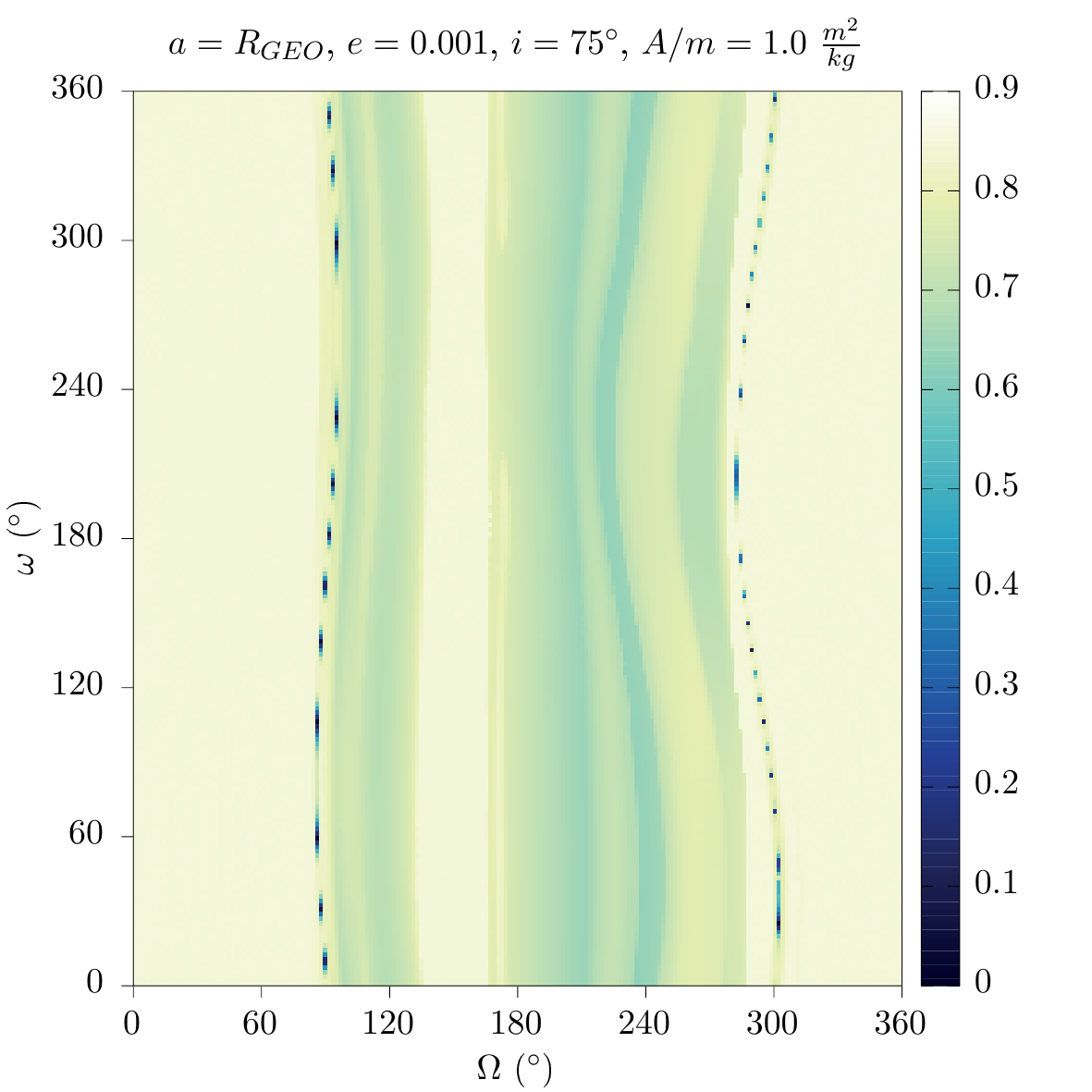}

\caption{Dynamical maps on the ($\Omega , \omega$) plane at geosynchronous semi-major axis $a=R_{\textrm{GEO}}$ and for a low initial eccentricity $e=0.001$. The inclination is $0.1^\circ$ in the left column, $45^\circ$ in the middle column and $75^\circ$ in the right column. The Area-to-mass ratio is $0.012$ $\textrm{m}^2/\textrm{kg}$ in the top row and $1.0$ $\textrm{m}^2/\textrm{kg}$ in the bottom row. The colormap corresponds to the value of the eccentricity diameter over 120 years.}
\label{fig:omomecc001}
\end{figure}

In Fig.~\ref{fig:omomecc001} a set of disposal maps for a satellite with low initial eccentricity $e = 0.001$ is presented. Three different initial inclinations are presented $i=0^{\circ}$ (left column), $i=45^{\circ}$ (centre column) and $i=75^{\circ}$ (right column). For the typical satellites with  $A/m = 0.012$ $\textrm{m}^2/\textrm{kg}$ (top row), the eccentricity variations are of the order $10^{-3}$ for orbits with initial inclinations up to about $40^{\circ}$ and then they abruptly increase to reach up to re-entry values. At an initial inclination of about $75^{\circ}$ almost all the orbits are re-entering except for two symmetric values of the argument of the perigee for each node, which represent frozen orbits configurations. A similar behaviour with respect to the inclination increase is obtained also for the high $A/m$ satellites (bottom row). Namely, the abrupt increase is observed passing from $J_2$ and solar radiation pressure dominated regime at low inclinations to a third body dynamics dominated regime past the $40^{\circ}$ of inclination. 

Notice that there exist significant differences between the low and high $A/m$ cases, showcasing the importance of its contribution for low-eccentricity orbits. More specifically, at low-inclination the increased $A/m$ forces only two stable configurations, those for $\omega+\Omega-\lambda_{sun}=0 \hspace{0.1cm} or \hspace{0.1cm} \pi$ at a geosynchronous altitude, where $\lambda_{sun}$ is the ecliptic longitude of the Sun. For moderate inclinations, we observe that the solar radiation pressure seems to enhance the instability domain induced by the lunisolar perturbations, but this is not always the case (see also Sec.~\ref{sec:srpimpl}). Finally, for the $75^{\circ}$ inclination, due to the solar-radiation pressure, the stable perigee configurations do not exist any more and have been replaced by two stable nodal configuration for each perigee. 

\begin{figure}
\centering
	\includegraphics[width=0.325\textwidth]{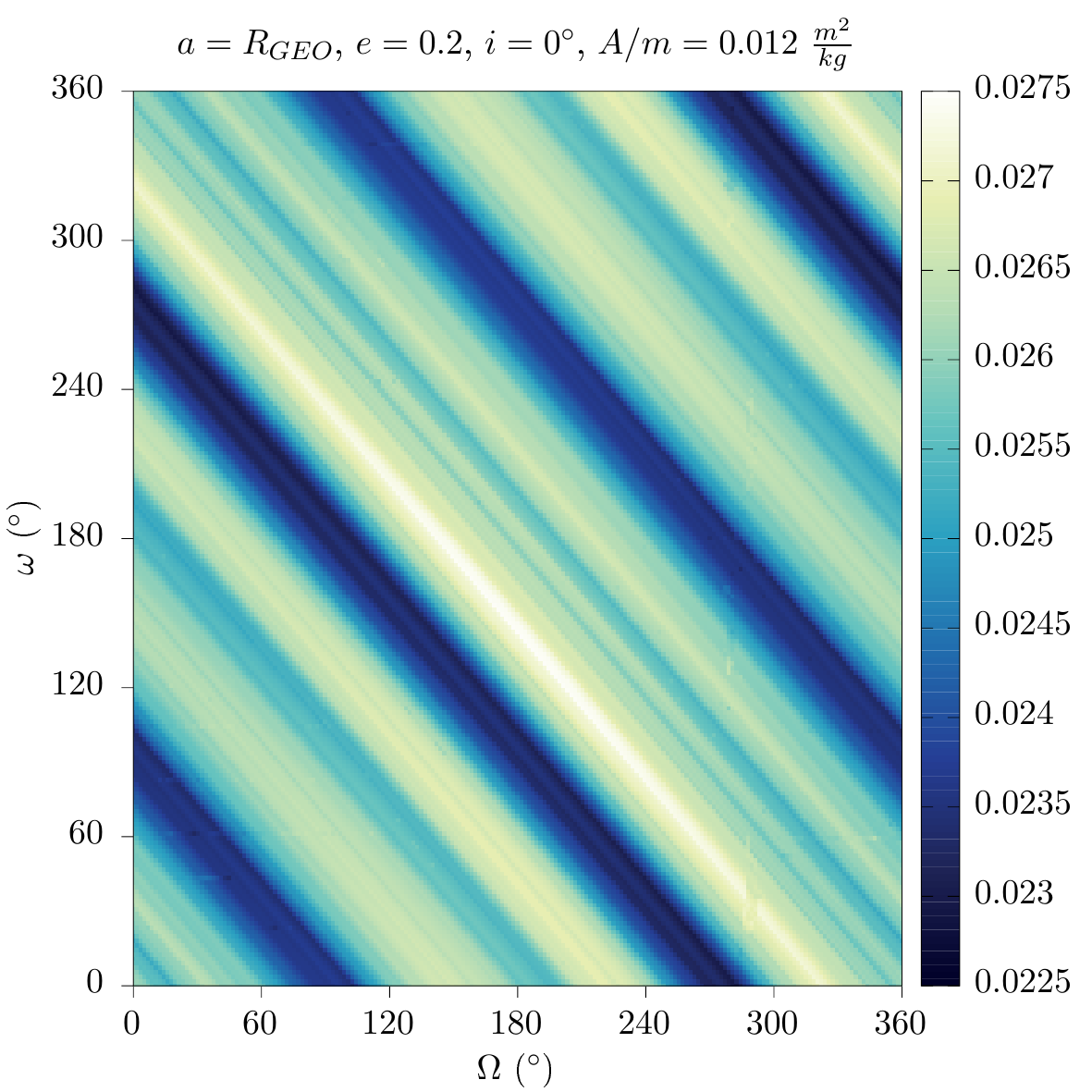}
	\includegraphics[width=0.325\textwidth]{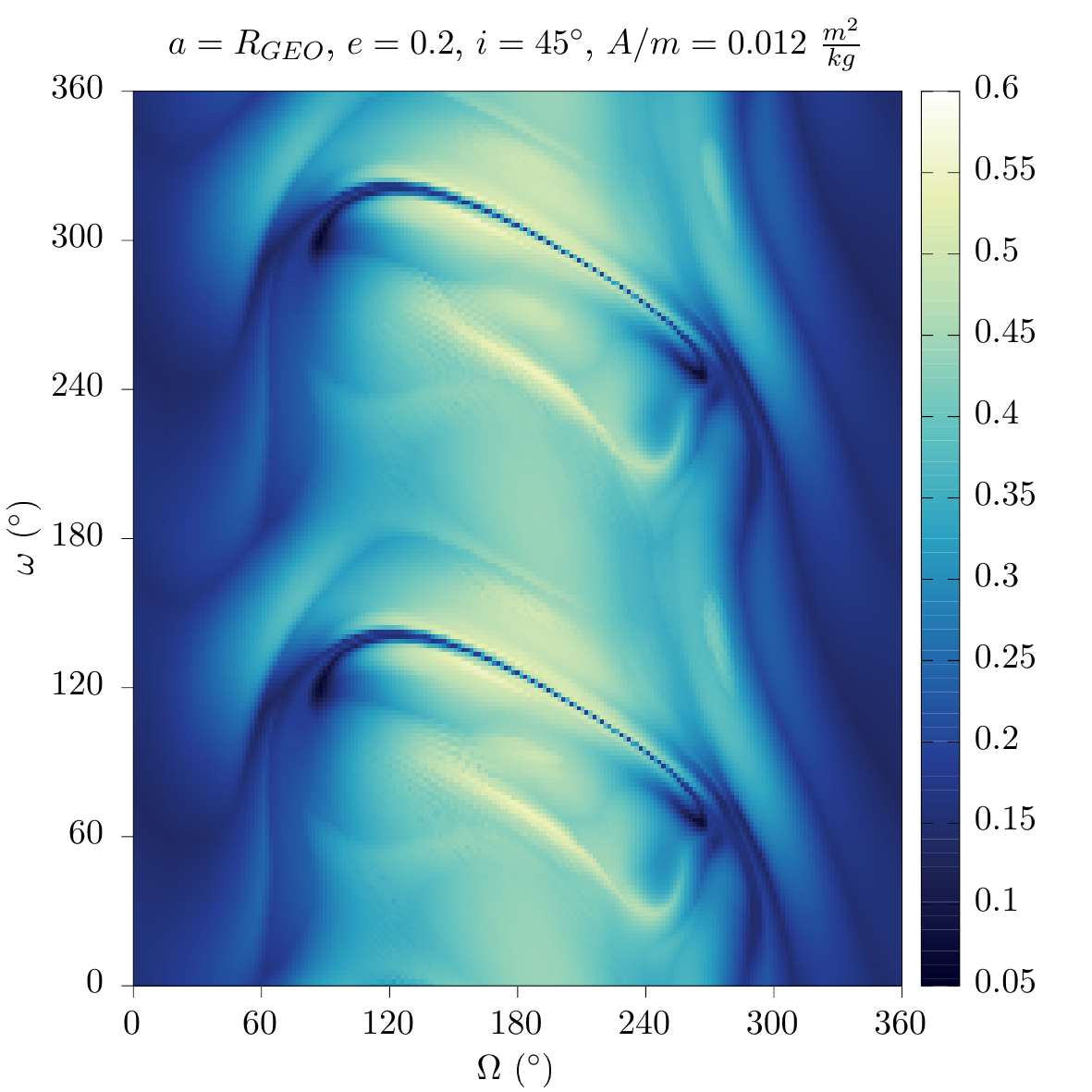}	
	\includegraphics[width=0.325\textwidth]{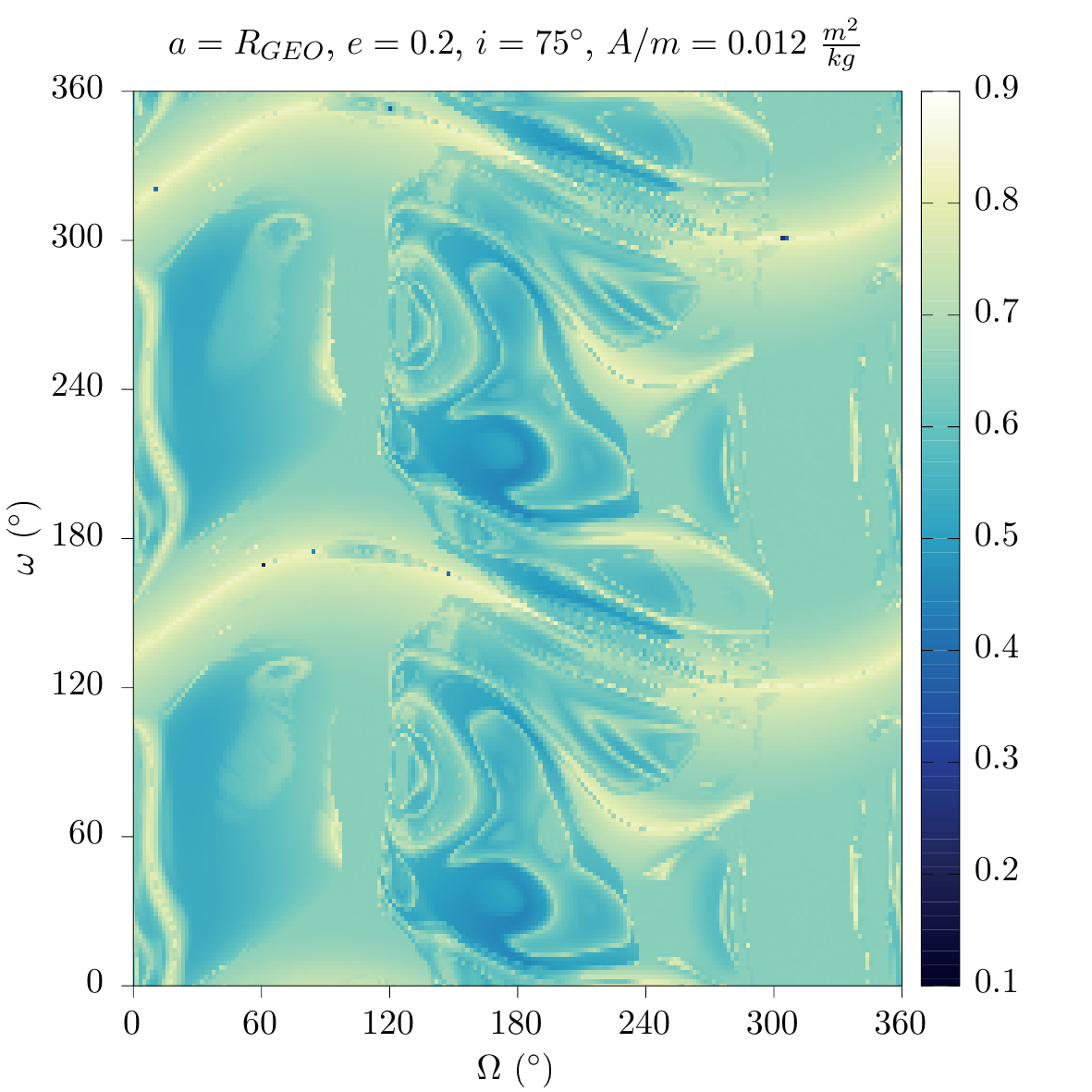}

	\includegraphics[width=0.325\textwidth]{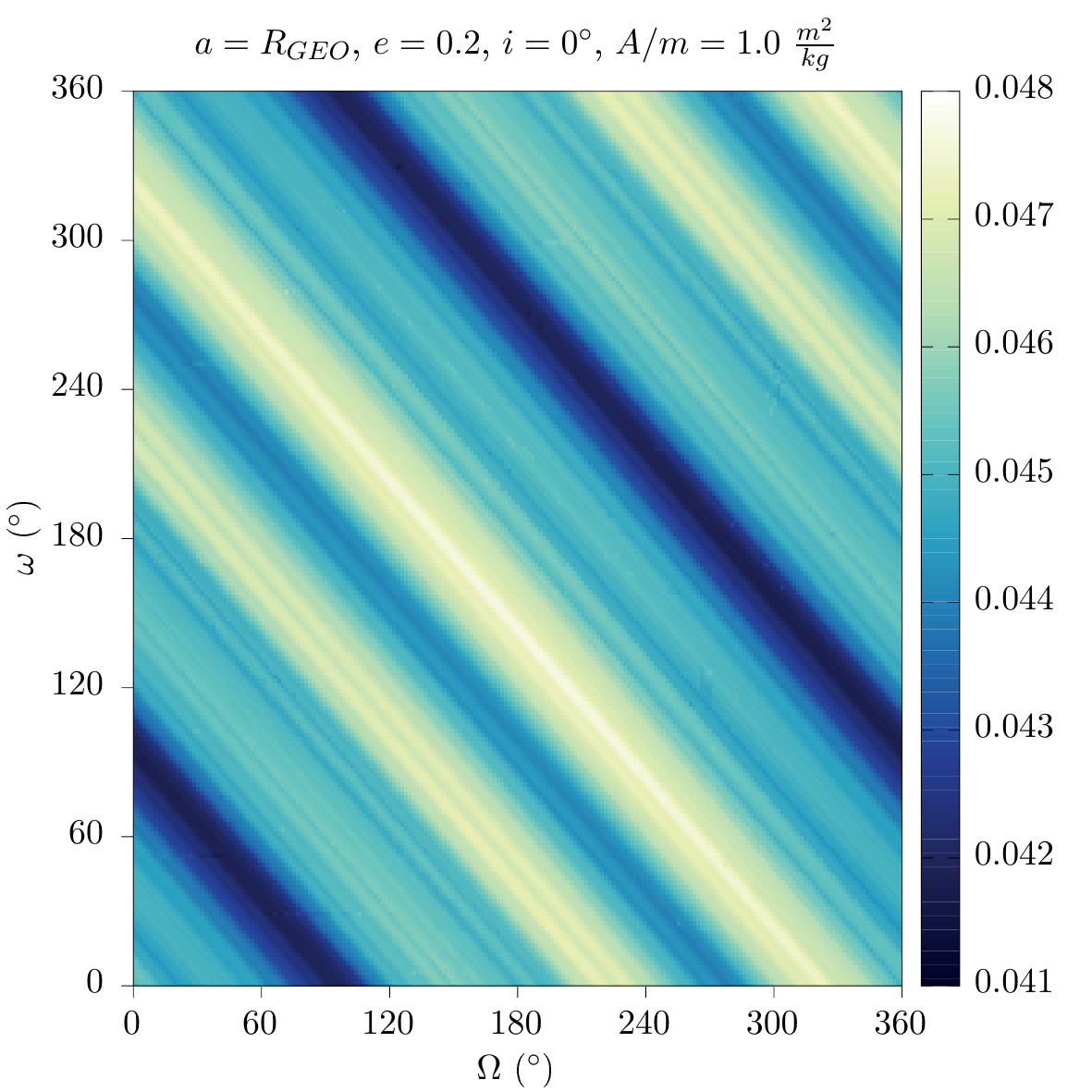}
	\includegraphics[width=0.325\textwidth]{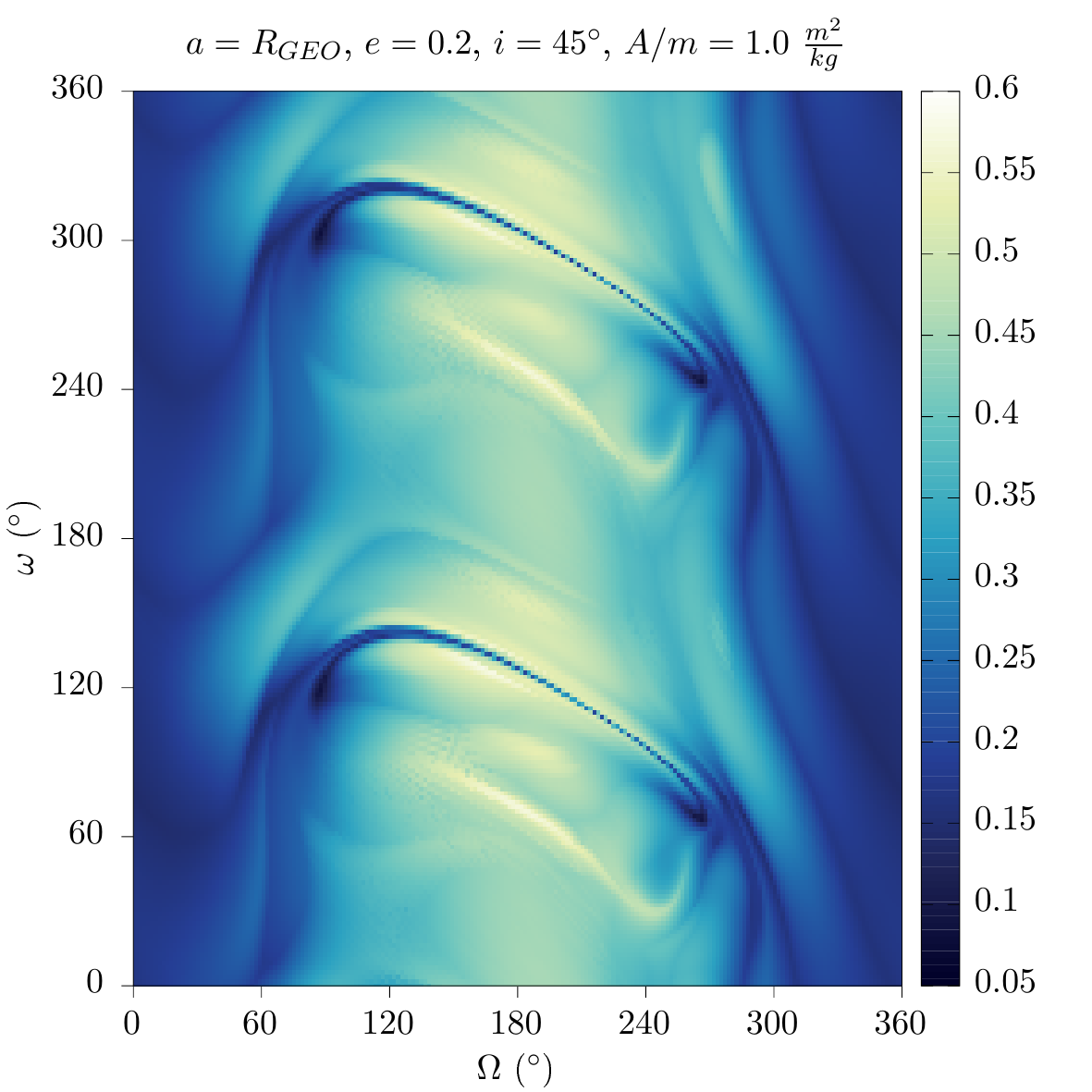}	
	\includegraphics[width=0.325\textwidth]{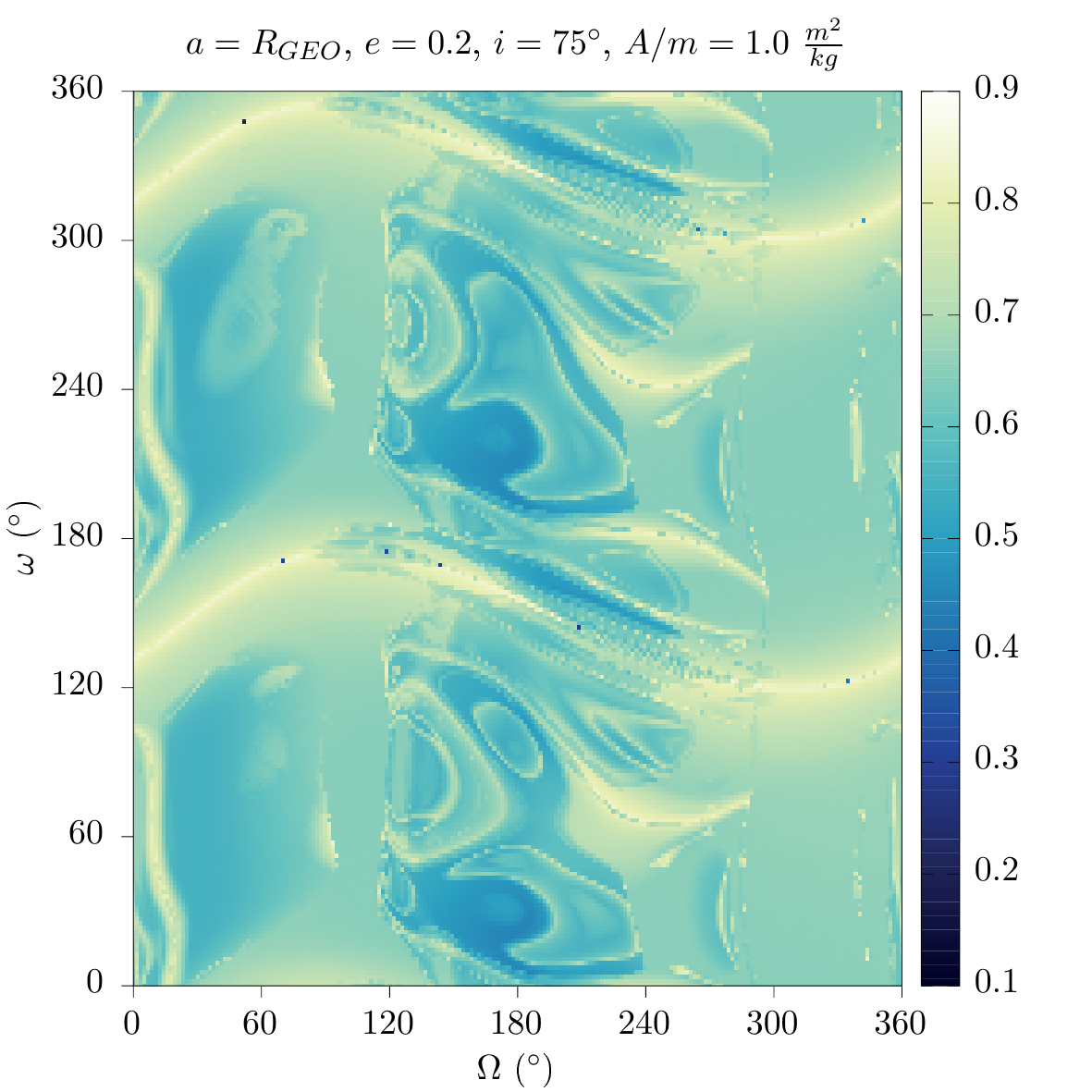}
\caption{Dynamical maps on the ($\Omega , \omega$) plane at geosynchronous semi-major axis $a=R_{\textrm{GEO}}$ and  for a low initial eccentricity $e=0.2$. The inclination is $0.1^\circ$ in the left column, $45^\circ$ in the middle column and $75^\circ$ in the right column. The Area-to-mass ratio is $0.012$ $\textrm{m}^2/\textrm{kg}$ in the top row and $1.0$ $\textrm{m}^2/\textrm{kg}$ in the bottom row. The colormap corresponds to the value of the eccentricity diameter over 120 years.}
\label{fig:omomecc2}
\end{figure}

It is interesting now to examine the case of a higher initial eccentricity $e = 0.2$. In the panels presented in Fig.~\ref{fig:omomecc2}, in the same fashion as in Fig.~~\ref{fig:omomecc001}, the different columns correspond to increasing values of  the initial inclination (left to right) and the different rows to increasing values of the $A/m$ (top to bottom). The behaviour with respect to the inclination increase is similar to the low-eccentricity case, i.e. as soon as the inclination exceeds the $40^{\circ}$ the orbits exhibits large eccentricity growth due to third body perturbations. On the other hand, in this case the effect of the solar-radiation pressure seems to be less profound. Indeed, except for the slightly enhanced eccentricity variations in the low-inclination case, there do not seem to appear any other significant differences between the lower and the upper row maps. 

From the dynamical maps presented here, one can deduce that for equatorial GEO satellites graveyard disposal is the only option. In our disposal mapping it is easy to identify the lowest perigee variation corridors (dark blue lines in left panels in Figs~\ref{fig:omomecc001} and \ref{fig:omomecc2}). This set of orbits should be targeted with post mission disposal manoeuvres, although this is not enough for a long-term safe graveyard. An adequate spacing between the disposed satellites should be ensured, such that the collision probability becomes minimal. On the other hand, for inclinations higher than about $40^{\circ}$ there exist an abundance of re-entering solutions. The angle dependence of the position of unstable structures is not trivial at all, and poses interesting problems in re-entry disposal design which we will discuss in Sec.~\ref{sec:disposal}.  

Another interesting aspect that we would like to highlight, is the effect of the higher $A/m$ ratio in the low-eccentricity region. This is connected with the existence of a stable equilibrium of the solar-radiation pressure force at low eccentricities. On the other hand, for high initial eccentricities and inclinations, the evolution in the two $A/m$ cases is almost identical.


\subsection{Eccentricity-inclination maps}\label{subsec:actionspace}


Although the complex dependence on the initial angles has already been discussed, here we attempt a global characterisation of the geosynchronous orbital region. We study the action-like variable space ($a,e,i$) and we address the angles dependence in a statistical manner, like in \cite{Gko2016}. The semi-major axis is considered fixed and equal to the geosynchronous value for this mapping, since the same investigation for even up to $1000$ $\textrm{km}$ above or below $R_{\textrm{GEO}}$ yields very similar results. First, we present a set of maps for a fixed set of angles and then proceed with an angles-averaged dynamical mapping, i.e. for each point in the action-like variable space ($a,e,i$) we randomly select a sufficient sample of angular configurations ($\Omega,\omega$) and average the normalised eccentricity diameters over all the angles dataset. 
 
\begin{figure}
\centering
	\includegraphics[width=0.5\textwidth]{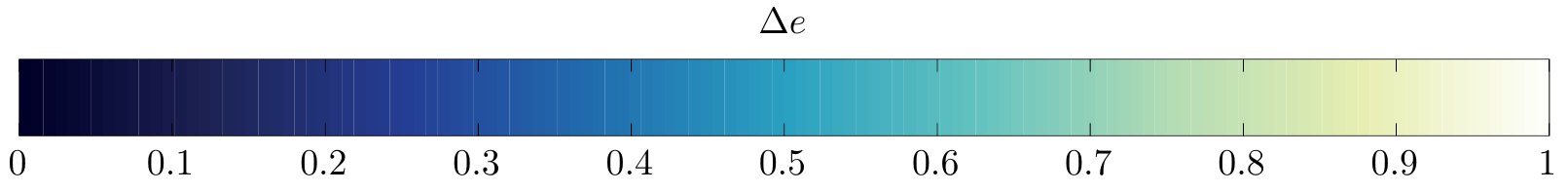}
	\includegraphics[width=\textwidth]{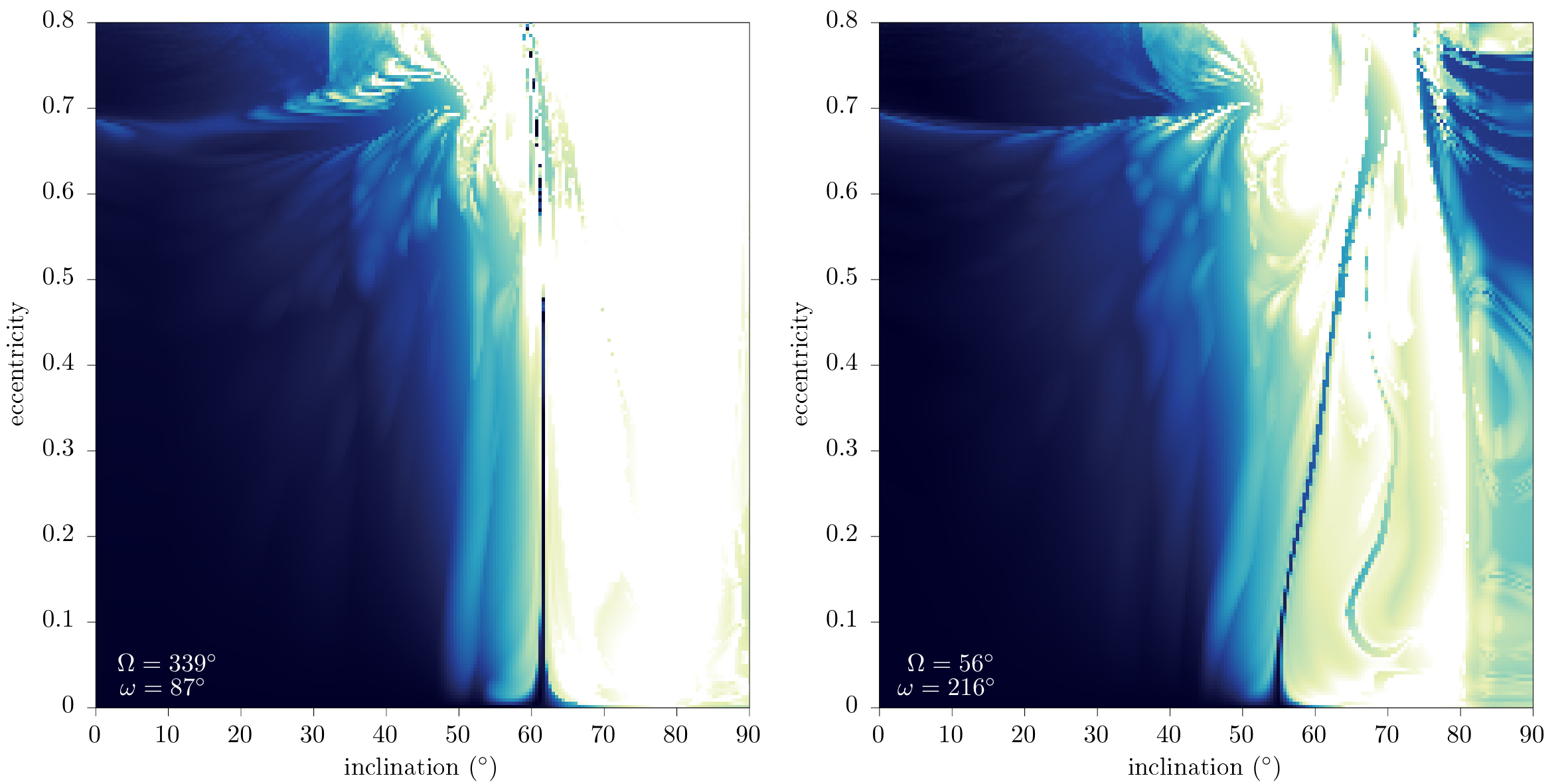}	
	\caption{Dynamical maps on the eccentricity - inclination plane at geosynchronous semi-major axis. Two sets of initial angular configurations are presented: in the left panel ($\Omega = 339^\circ$, $\omega = 87^\circ$) and in the right panel ($\Omega = 56^\circ$, $\omega = 216^\circ$). The colormap corresponds to the value of the normalised eccentricity diameter $\Delta e$. The Area-to-mass ration is $0.012$ $\textrm{m}^2/\textrm{kg}$ in both panels.}
	\label{fig:anglesavgcases}
\end{figure}

In Fig.~\ref{fig:anglesavgcases} the eccentricity-inclination study for $a=R_{\textrm{GEO}}$ and two different angular configurations is presented. From disposal design point of view, a general feature of those maps that we should pay attention to is the generalised instability appearing at higher inclinations. And not only that, embedded in the unstable domain there exist particular configurations for which the eccentricity variations are small. Those regions of the phase space present some intriguing scenarios from future GEO exploitation, since they provide a stable operational regime next to an unstable region which could be used for end-of-life disposal (see for example the blue curves at $50-60^{\circ}$ inclination in Fig.~\ref{fig:anglesavgcases}).

\begin{figure}
\centering
	\includegraphics[width=0.5\textwidth]{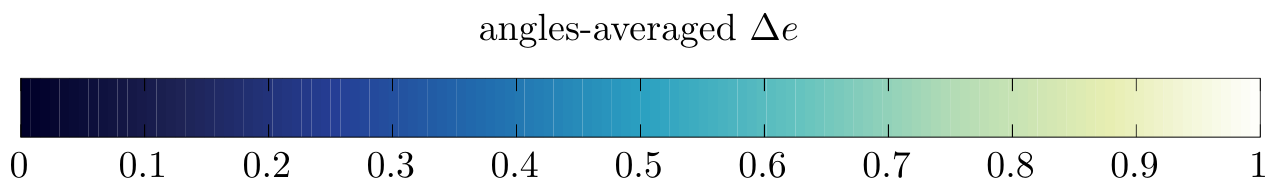}
	\includegraphics[width=\textwidth]{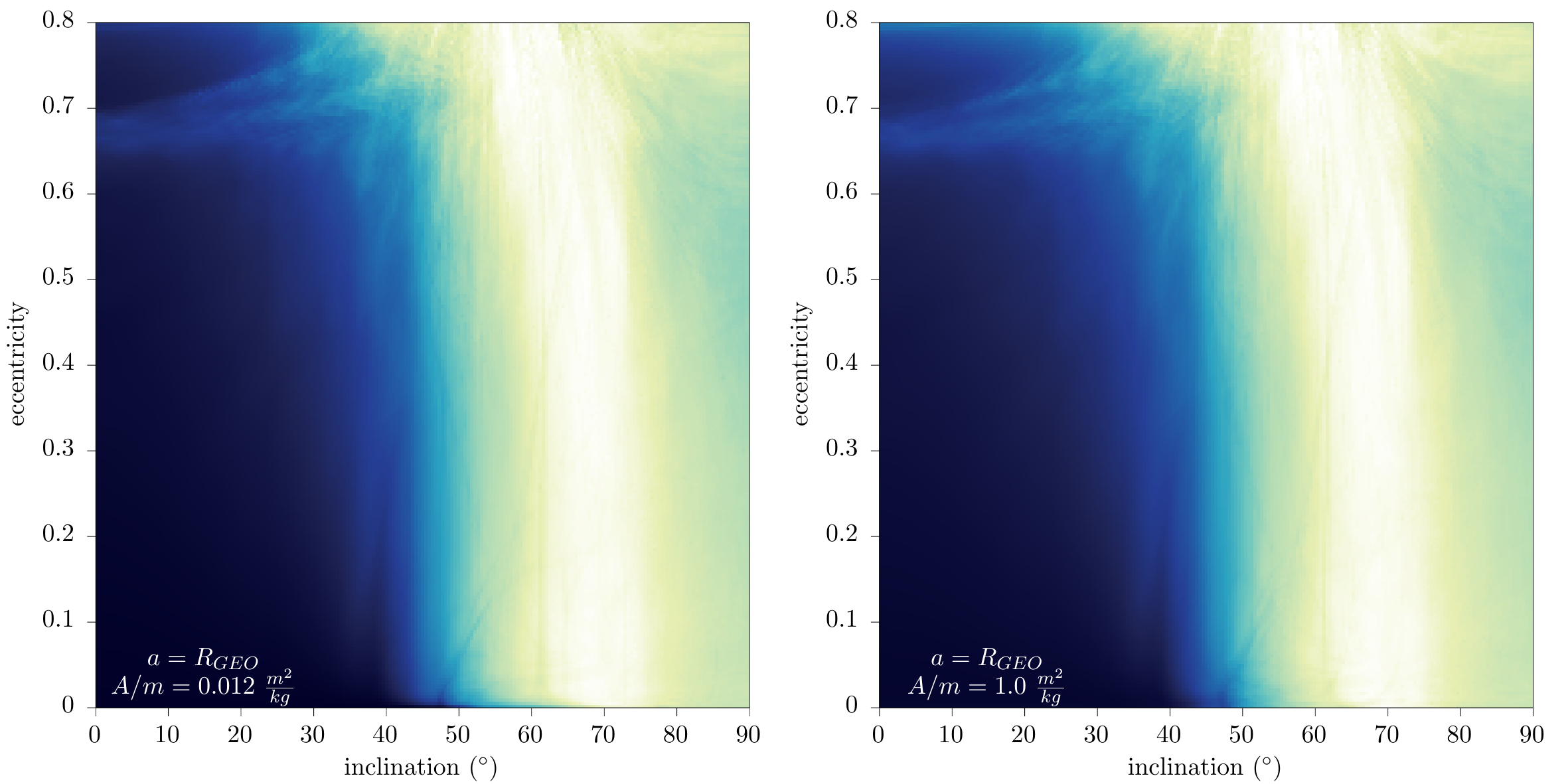}	
	\caption{Dynamical maps on the eccentricity - inclination plane at GEO semi-major axis. The colormap corresponds to the value of the angles-averaged $\Delta e$ over 50 randomly selected $\omega-\Omega$ configurations. In the left panel the Area-to-mass ratio is $0.012$ $\textrm{m}^2/\textrm{kg}$ and in the right panel $1.0$ $\textrm{m}^2/\textrm{kg}$.}
	\label{fig:anglesavg}
\end{figure}

As a last step, we present in Fig.~\ref{fig:anglesavg} the angles-averaged dynamics over the eccentricity-inclination maps. In those maps, each point of the action-like space is sampled with 50 randomly selected angular configurations and the value of the dynamical indicator $\Delta e$ is averaged over all the angles. The result is a global dynamical map of the geosynchronous region, where the stable and unstable regions are clearly separated, albeit the information for the initial angle dependencies is lost. The region of above $40- 45^{\circ}$ inclination presents a richness of re-entry solutions. Of particular interest is also the region from $65-75^{\circ}$ inclination where almost every orbit is re-entering within 120 years. The effect of the higher $A/m$ (right panel) is almost negligible at higher eccentricities in the angles-averaged map. However, it creates some angle-related differences at low eccentricities.

The results of the angles-averaged study, convincingly confirm the natural separation of the phase space between low-inclined (below $\sim 40^\circ$) and high-inclined (above $\sim 40^\circ$) orbits. Namely, for low-inclined orbits, there is a natural deficiency of eccentricity growth orbits, and the search for stable graveyard solution is the only possible post mission design plan. However, at higher inclination another opportunity presents itself, an abundance of re-entering orbits exists. The global map indicates that for Earth satellites at geosynchronous altitude, the third body perturbations are prevailing over the other perturbations leading to large eccentricity variations for inclined orbits. The characteristics of those orbital highways and possible exploitation scenarios are discussed in the following Section. 


\section{Post-mission disposal issues}\label{sec:disposal}


The findings of Sec.~\ref{sec:dynamicalmapping} are further discussed here, considering also issues related to post-mission disposal planning. We have already mentioned the abundance of highly-inclined orbits that have a considerable eccentricity growth within the 120-year time-span of our propagation. However, there is another crucial dynamical information that is hidden in the discussion of the previous paragraph, that being the orbital lifetime of each orbit. Of special interest are short-lived solutions that naturally re-enter the atmosphere. Examples of short-lived orbits are presented and their properties are discussed. Moreover, we present an interesting case where, even though an increased $A/m$ usually enhances the de-orbiting process, not only this does not happen, but in fact the solar-radiation pressure effect cancels the real eccentricity growth mechanism. Finally, in the light of the dynamical mapping results, the current guidelines are discussed, and alternative ways of GEO exploitation are proposed.

\subsection{An effective cleansing mechanism}

\begin{figure}
\centering
	\includegraphics[width=\textwidth]{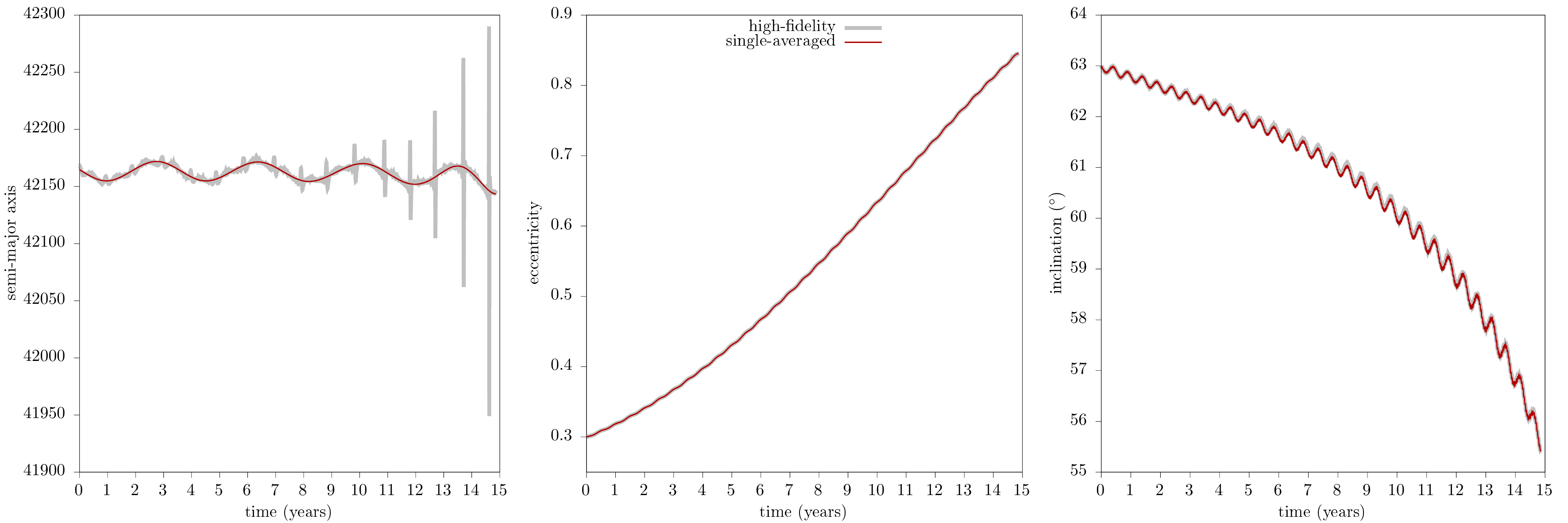}
	\caption{Orbital evolution of an inclined GEO orbit ($a=R_{\textrm{GEO}}$, $e=0.3$, $i=63^{\circ}$, $\Omega = 240.0^{ \circ} $, $\omega = 0.0^{\circ}$, $M=0.0^{\circ}$ $A/m=0.012$ $\textrm{m}^2/\textrm{kg}$ and initial epoch 21/06/2020 at at 06:43:12.0) re-entering within 15 years of simulation. The existence of this type of trajectory is also confirmed through a high-fidelity simulation of the same initial condition.}   
	\label{fig:reenorbit}
\end{figure} 

In Sec.~\ref{sec:dynamicalmapping} we have encountered some orbits with exceptional short life-time. A typical example of this type of orbits is shown in Fig.~\ref{fig:reenorbit}, where the evolution of a trajectory with initial condition $a=R_{\textrm{GEO}}$, $e=0.3$, $i=63^{\circ}$, $\Omega = 240.0^{ \circ} $, $\omega = 0.0^{\circ}$, $M=0.0^{\circ}$ and $A/m=0.012$ $\textrm{m}^2/\textrm{kg}$ is presented. The interesting feature is its orbital lifetime, which is less than 15 years. To exclude the chance that this is an outcome of the truncated force model or the single-averaged formulation, the same initial condition was also propagated under the high-fidelity dynamics. The orbital evolution of two orbits coincides, suggesting that there is a quite effective cleansing mechanism at GEO altitude, that can make satellites re-enter even within the 25-year rule that is imposed for LEO orbits \cite{IADC2017}. Moreover, the collision probability of an orbit like this is really minimal; for the solution shown in Fig.~\ref{fig:reenorbit} the total time spend in the LEO protected region\footnote{ The LEO protected region is defined as the spherical shell between Earth's surface and up to $2000$ $\textrm{km}$ altitude.} as well as the dwell time in the GEO protected region is just a few days.

\begin{figure}
\centering
	\includegraphics[width=0.5\textwidth]{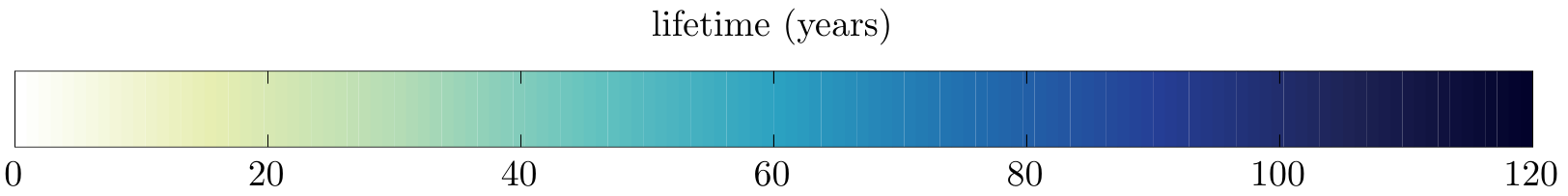}
	\includegraphics[width=\textwidth]{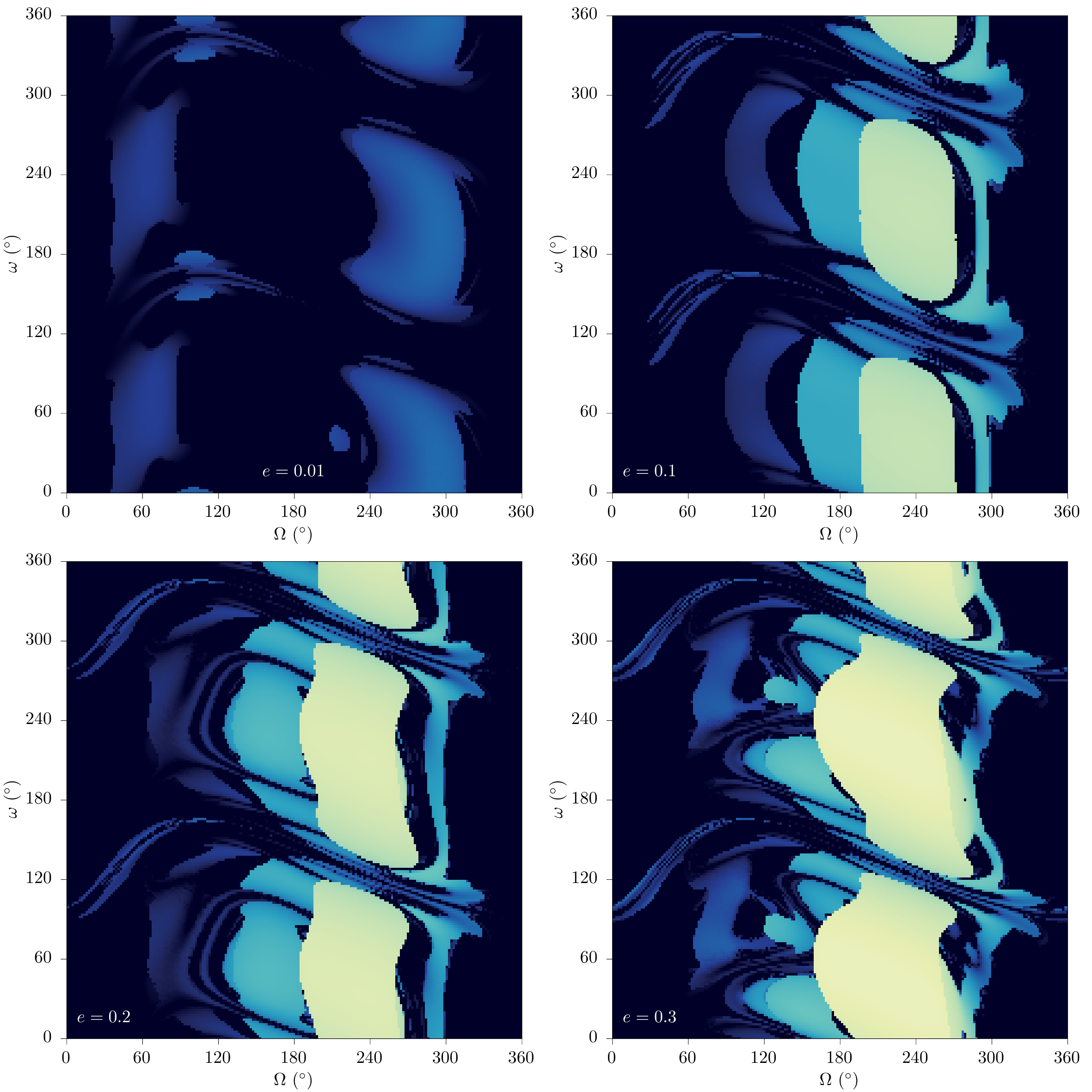}
	\caption{The ($\Omega,\omega$) orbital lifetime maps for orbits at $a=R_{\textrm{GEO}}$ , $63^\circ$ inclination and eccentricities: a) $e=0.01$ (top left), b) $e=0.1$ (top right) c) $e=0.2$ (bottom left) and d) $e=0.3$ (bottom right).}
	\label{fig:lifetime63}
\end{figure}

Fast re-entering orbits were also reported in the literature \cite{Breit2007,Jen2017} for high initial eccentricities and inclinations at geosynchronous altitude. Therefore, we would like to further explore under which conditions those orbits appear and study their properties. In Fig.~\ref{fig:lifetime63} a set of disposal maps for $63^{\circ}$ inclination is presented, however, the colormap here does not correspond to the eccentricity diameter but rather to the orbital lifetime. The results are shown for four different values of the initial eccentricity $e=0.01$, $e=0.1$, $e=0.2$, $e=0.3$. For values of the eccentricity higher than $0.1$ there exists, around an initial right ascension of the ascending node of $\Omega=230^{\circ}$, a set of orbits with very promising lifetimes of $20-30$ years.   
 
\begin{figure}
\centering
	\includegraphics[width=0.5\textwidth]{cbarlifetimehor}
	\includegraphics[width=\textwidth]{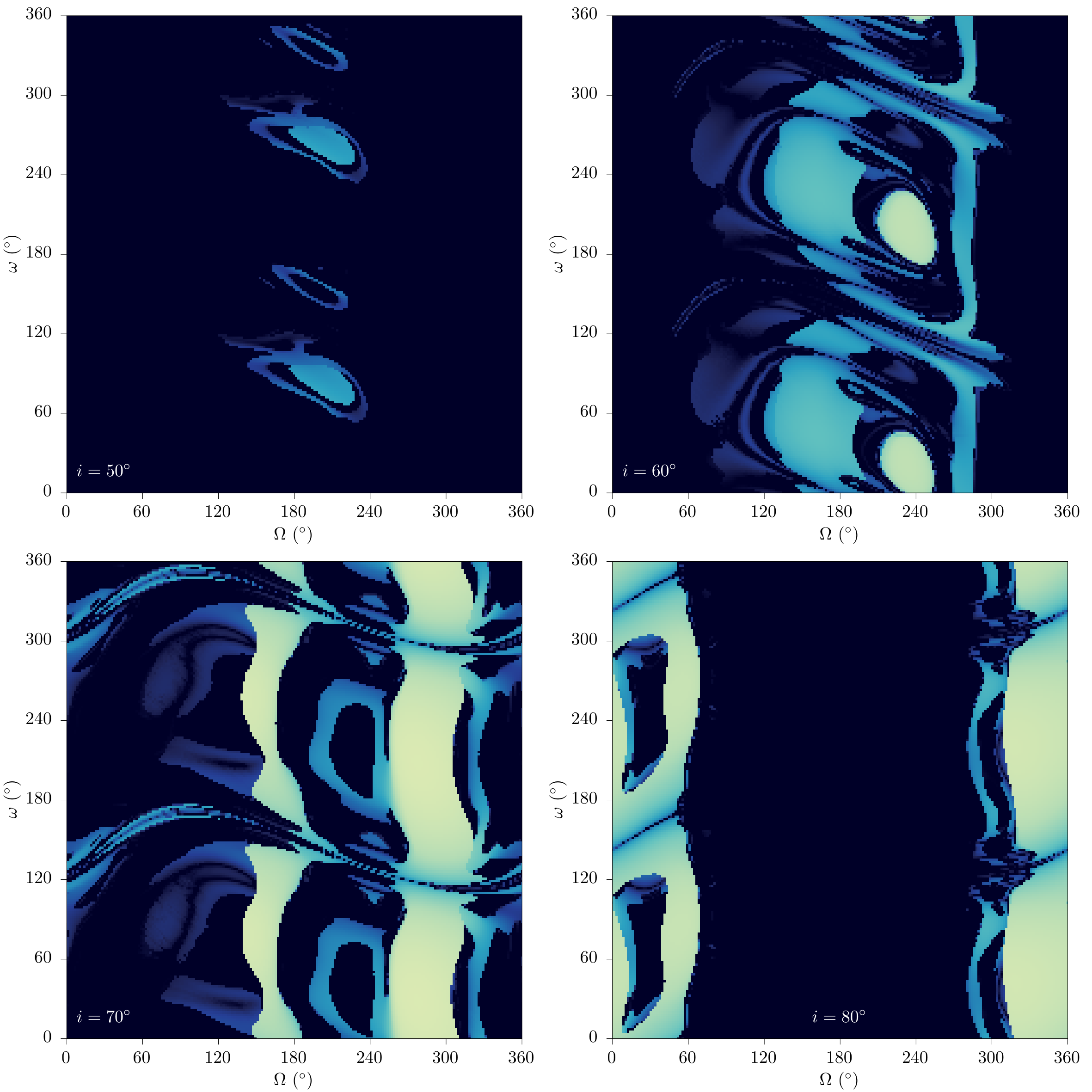}
	\caption{The $\Omega,\omega$ orbital lifetime maps for orbits at $a=R_{\textrm{GEO}}$ , eccentricity $e=0.2$ and inclinations: a) $i=50^\circ$ (top left), b) $i=60^\circ$ (top right) c) $i=70^\circ$ (bottom left) and d) $i=80^\circ$ (bottom right).}
	\label{fig:lifetimeecc200}
\end{figure}

Moreover, this set of orbits is not a local characteristic that happens only for the $63^{\circ}$ of inclination. Fig.~\ref{fig:lifetimeecc200} shows the lifetime disposal maps for initial eccentricity $e=0.2$ and different values of the initial inclination. The fast re-entry patches exist in a range of initial inclinations from $50-90^{\circ}$. However, their structure and their position with respect the initial value of the satellites node changes with varying inclination, due to the varying orientation of the perigee and node with respect to the perturbers' planes \cite{Kozai1962}.

It is interesting now to understand the mechanism that leads those orbits to re-enter within 20-30 years. In order to further study this effect we select a set of orbits for $i=63^{\circ}$, $e=0.2$, $\omega = 60^{\circ}$ and values of the node $\Omega$ equally spaced every $10^{\circ}$. The eccentricity evolution of this set of orbits is presented in the left panel of the Fig.~{\ref{fig:ecchkevolution}. The fast re-entering orbits are those with nodes between $\Omega = 190^{\circ}$ and $260^{\circ}$ (red bold lines). 

This effect is even more clear if we look at the evolution of the eccentricity vector in the orbital plane through the set of variables $e\cos{\omega}, e\sin{\omega}$. It is now clear, that the evolution is following a Lidov-Kozai type of evolution, induced by the combined contribution of the Sun and the Moon. The interesting orbits, with fast re-entry times, are just those for which the eccentricity evolution allows them to reach the re-entry value within the first quarter of the dynamical evolution cycle. A suitable analytical method to check for the existence and a-priori locate their position is currently under development \cite{Gkolias2018}. The insight developed from the study of the triple averaged Hamiltonian model suggests that, the in-plane dynamics for a range of inclinations with respect to the ecliptic become such, that the maximum eccentricity acquired during the Lidov-Kozai type of dynamics is equal to or larger the atmospheric re-entry value. Those initial conditions correspond to various sets of equatorial inclinations and node combinations, and could produce the complicated patterns that we see in the disposal maps in Fig~\ref{fig:lifetime63} and \ref{fig:lifetimeecc200}. 

\begin{figure}
\centering
	\includegraphics[height=0.24\textheight]{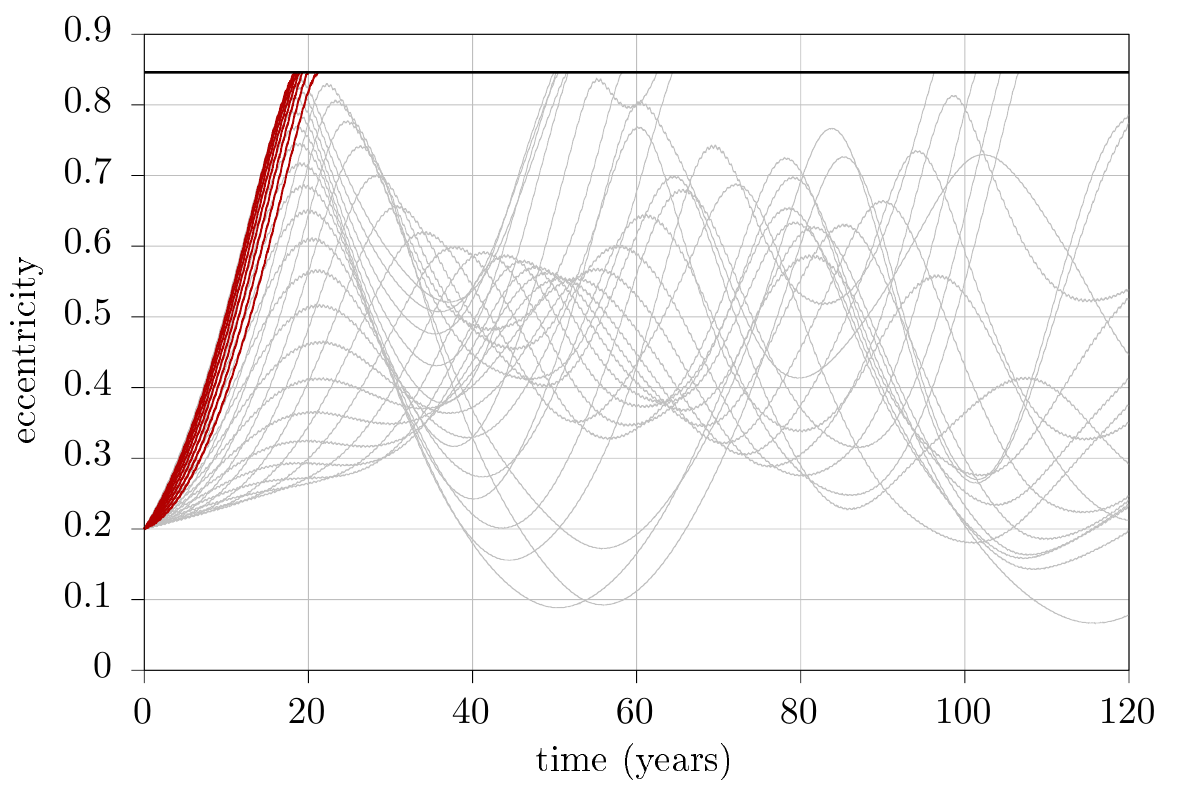}
	\includegraphics[height=0.24\textheight]{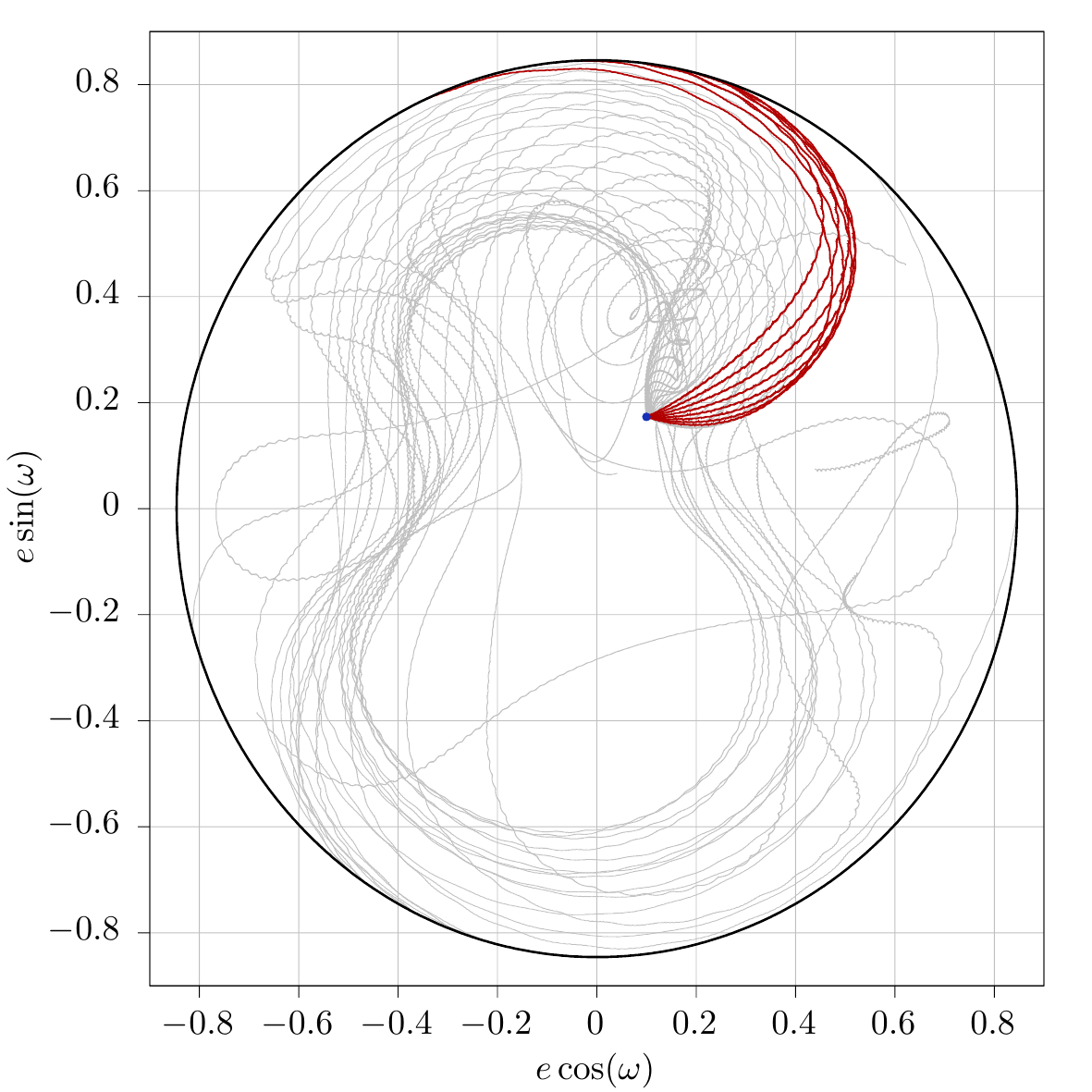}
	\caption{In the left panel: the eccentricity evolution of a set of 36 orbits with same initial conditions except for the right ascension of the ascending node, which is sampled every $10^\circ$. In the right panel: the evolution of the same orbits in the ($e\cos{\omega},e\sin{\omega}$) plane. Those with node between $\Omega = 190^{\circ}$ and $260^{\circ}$ have re-entry times of about 20 years (red lines).}
	\label{fig:ecchkevolution}	
\end{figure}

\subsection{Solar-radiation pressure implications}\label{sec:srpimpl}

In Sec.~\ref{sec:dynamicalmapping} we concluded that the effect of the solar radiation pressure can be important for low-eccentricity orbits. Moreover, usually during a lunisolar driven re-entry, an enhanced solar radiation pressure would promote the de-orbit process, as it has been observed for the transition region between LEO and MEO \cite{Rosengren2018}. However, this is not always the case, especially for the inclined geosynchronous orbits where the the Lidov-Kozai type dynamics are driving the re-entry.  

More specifically, by further inspecting both the disposal maps for low-eccentricity and high-inclination and the eccentricity-inclination action maps, we encountered cases where opening a solar sail would considerably delay the de-orbit process. An example of this type of interaction is given in Fig.\ref{fig:srpecchk}. The low $A/m$ orbit (blue curve) is re-entering within about 60 years of evolution, while the high $A/m$ orbit (red curve) has almost double the lifetime. In an attempt to understand the delayed re-entry, we plot again the evolution in the orbital plane using the $e\cos{\omega},e\sin{\omega}$ variable for the two orbits. Immediately we recognise that the low $A/m$ orbit directly follows a Lidov-Kozai type evolution. On the other hand, the high $A/m$ orbit is trapped about the origin for a long time span, until it finally escapes and follows again a third-body dynamics dominated trajectory.

\begin{figure}
\centering
	\includegraphics[height=0.24\textheight]{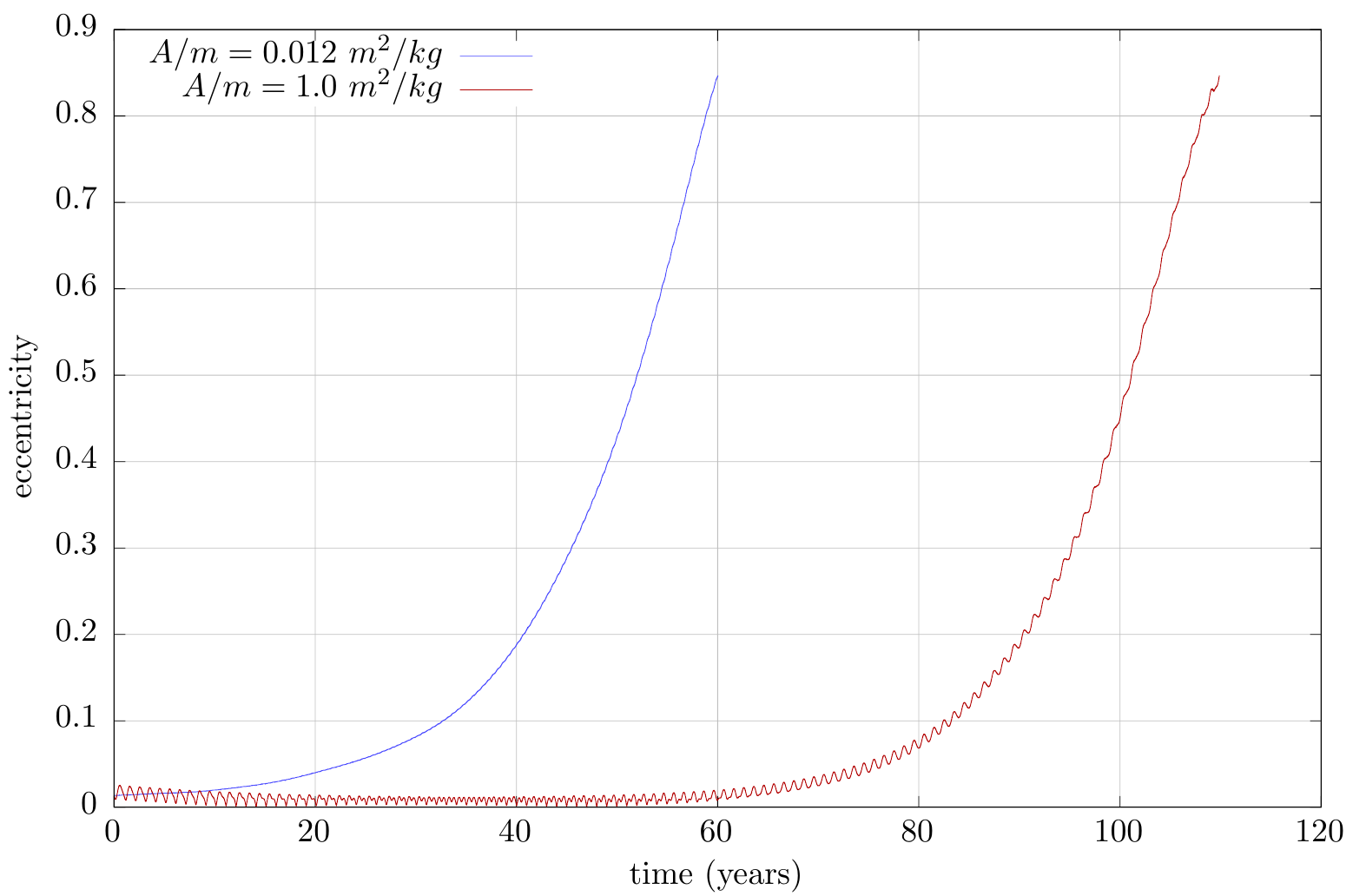}
	\includegraphics[height=0.24\textheight]{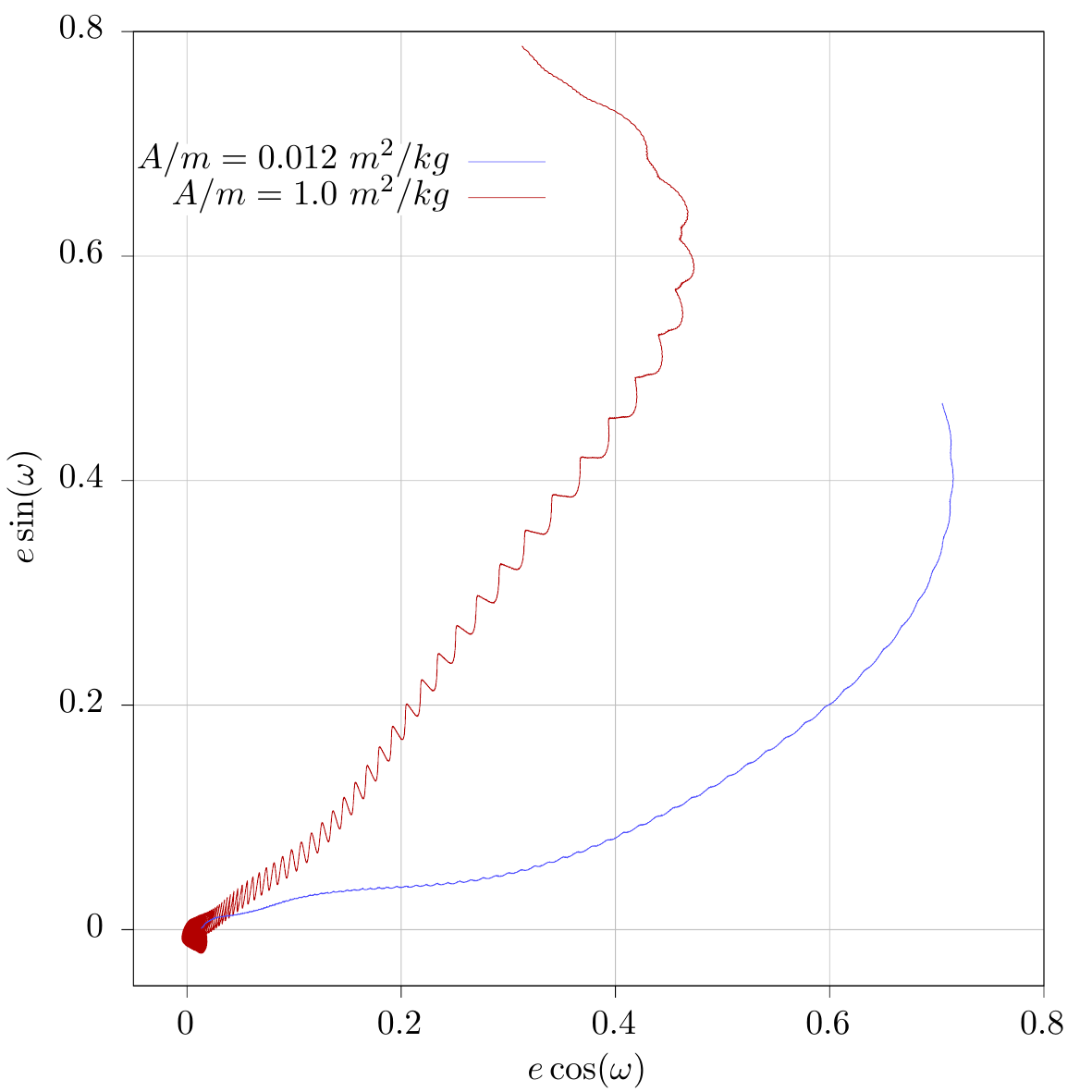}
	\caption{The eccentricity evolution of two inclined geosynchronous orbits with same initial conditions but different values of the $A/m$ ratio. In the left panel, the orbit evolution for a standard spacecraft $A/m =$ $0.012\textrm{ m}^2/\textrm{kg}$ is presented, whereas in the right panel, the orbit evolution for a spacecraft equipped with an area-augmenting device $A/m = 1.0\textrm{ m}^2/\textrm{kg}$ is shown.}  
	\label{fig:srpecchk}
\end{figure}

\begin{figure}
\centering
	\includegraphics[width=\textwidth]{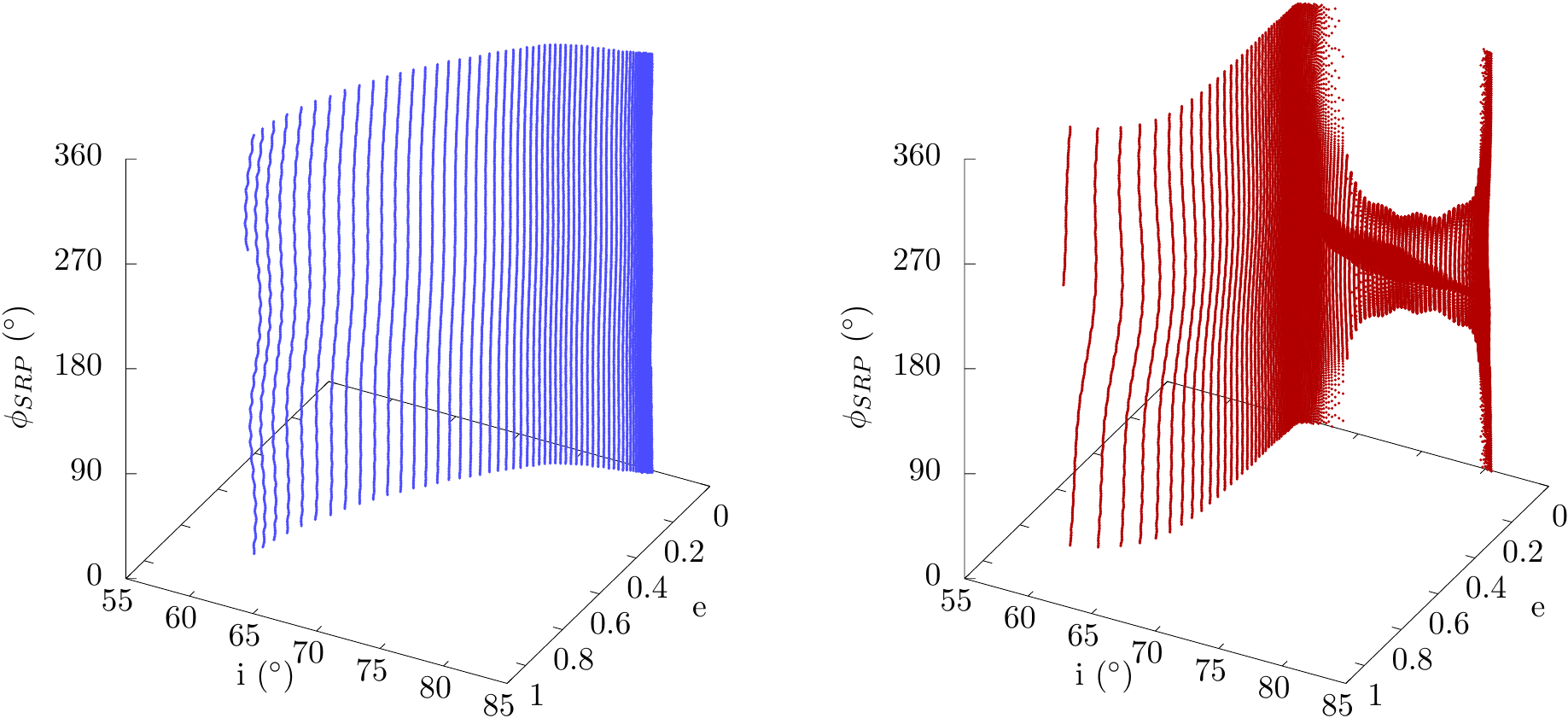}
	\caption{The 3-dimensional evolution of the orbits in Fig.~\ref{fig:srpecchk} in the $e$, $i$ , $\phi_{SRP}$ space. In the left panel, the orbit evolution for a standard spacecraft $A/m =$ $0.012\textrm{ m}^2/\textrm{kg}$ is presented, whereas in the right panel, the orbit evolution for a spacecraft equipped with an area-augmenting device $A/m = 1.0\textrm{ m}^2/\textrm{kg}$ is shown.}
	\label{fig:srpsurf}
\end{figure}

This dynamical interaction is further explained in Fig.\ref{fig:srpsurf} where we identify the main cause of the low-eccentricity trapping to be nothing else than the stable equilibrium of the solar radiation pressure resonance for the high $A/m$ case. Namely, by defining the resonant angle for the solar radiation pressure as $\phi_{SRP} = \Omega+\omega-\lambda_{sun}$, the evolution of the two orbits with respect to their eccentricity, inclination and $\phi_{SRP}$ is presented. For the low $A/m$ case, $\phi_{SRP}$ is always rotating and the dynamics are following the eccentricity-inclination evolution dictated mainly by the third-body perturbations. On the contrary, in the evolution for the high $A/m$ case, $\phi_{SRP}$ is initially rotating but with the decrease in the inclination it is trapped into the resonance and is forced to librate about the stable equilibrium. The induced frequency in the argument of the perigee evolution, temporarily suppresses the Lidov-Kozai effect and the eccentricity stays bounded to low values. The further decrease of the inclination, finally drives the orbit out of the solar-radiation pressure resonance and it follows once again the Lidov-Kozai type of dynamics as we have also seen in Fig.~\ref{fig:srpecchk}.   

\subsection{Population, dynamics and guidelines}

\begin{figure}
\centering
	\includegraphics[height = 0.32 \textheight]{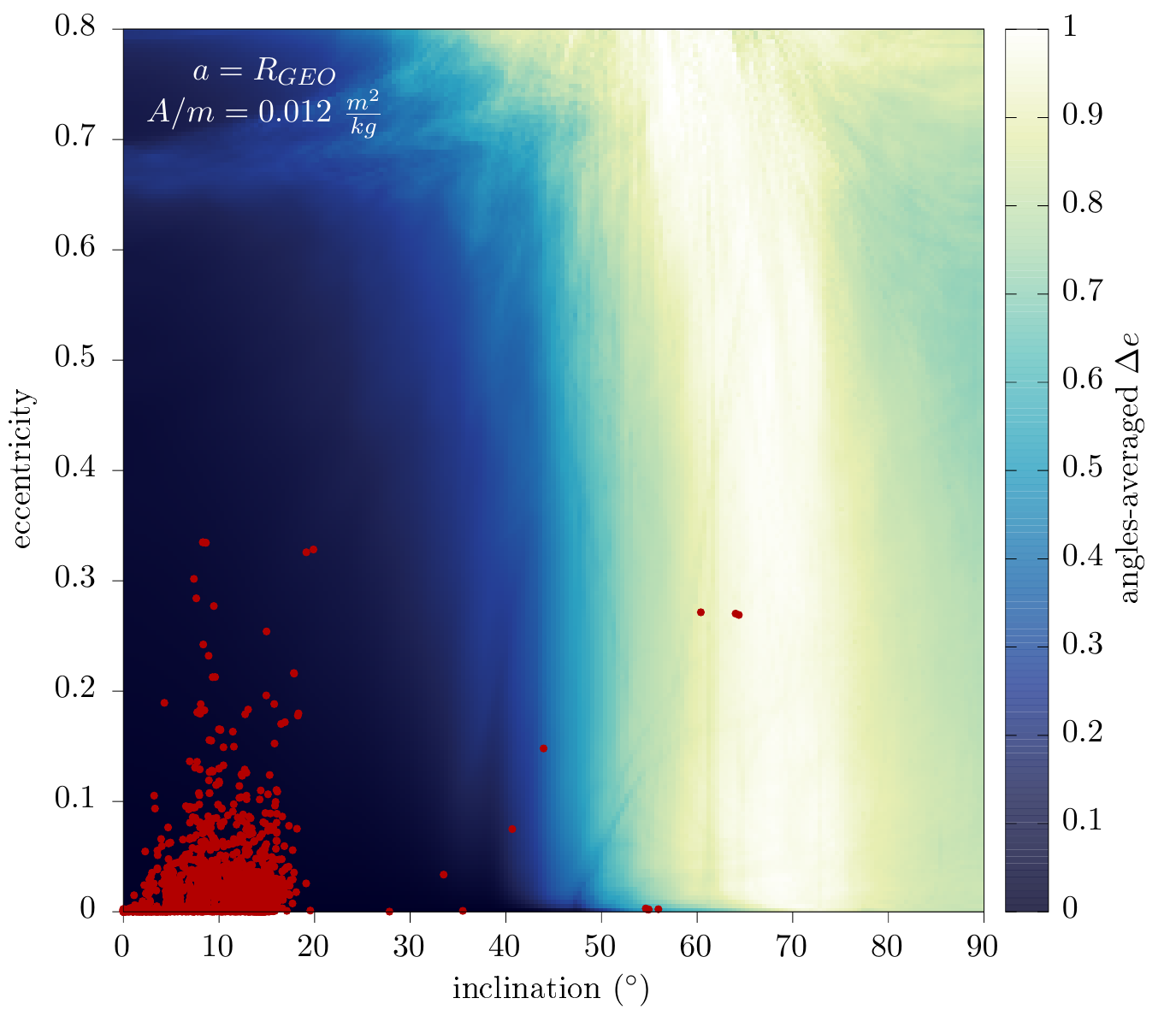}
	\includegraphics[height = 0.32 \textheight]{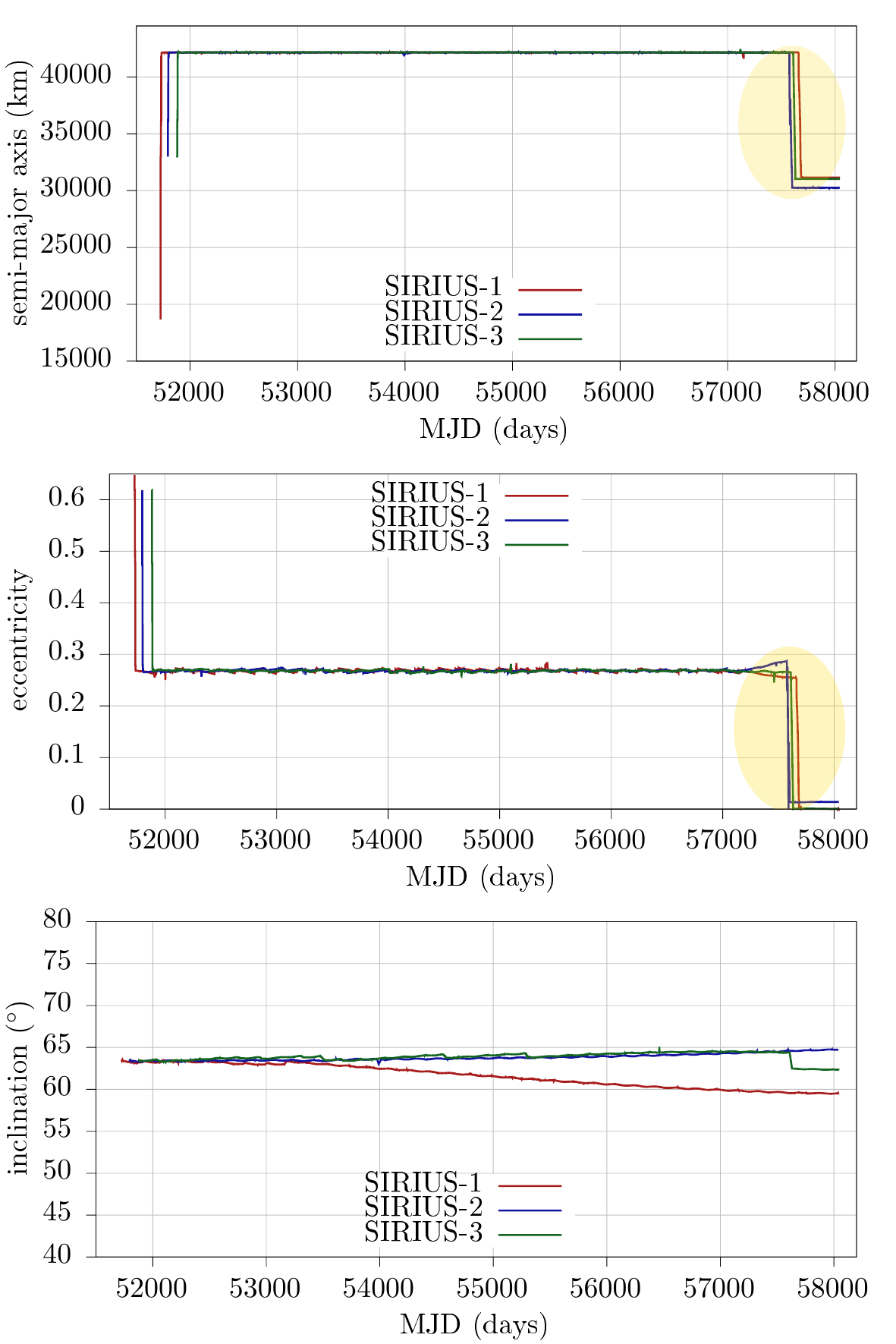}
	\caption{The angles-averaged stability map of the geosynchronous area with the current population superposed. All operations up to now have been heavily concentrated in the region of the phase space where good re-entry opportunities do not exist.}
	\label{fig:population}
\end{figure}

Currently, space operations in GEO are heavily concentrated in the low-eccentricity, low-inclination region. This is evident in the right panel of Fig.~\ref{fig:population} where the real population is superposed to the angles-averaged eccentricity-inclination map. In this regime, mission planning and operations are well established, however there is not an efficient re-entry mechanism for inclinations lower than $40^{\circ}$ inclination. Hence, it is inevitable that the population of the space debris in the region will keep increasing. 

On the other hand, carefully selected highly-inclined geosynchronous orbits can re-enter in time-scales comparable to the 25-years, which is the current IADC guideline for the LEO region. One could argue that inclined geosynchronous orbits are not as useful as the equatorial ones, but it has been shown that small constellations of a few eccentric and inclined geosynchronous satellites could reproduce the same coverage as a geostationary satellite \cite{Bru2005}. This kind of exploitation has been already implemented successfully with the Sirius constellation. Sirius-1, 2 and 3 were operating in eccentric and inclined geosynchronous orbits for several years providing satellite radio services in North America. Unfortunately, at the end of the operational lifetime of the constellation, the operators following the current guidelines removed the three satellites from the GEO region, by reducing the semi-major axis by almost $10000$ $\textrm{km}$ and circularising the orbits. In the right panel of Fig.~\ref{fig:population} the time evolution of the publicly available two-line elements (TLE) of three constellation components are shown and the end-of-life manoeuvres are highlighted in yellow. Nonetheless, the constellation was operating in an orbital region were fast re-entering orbits existed. We believe that considering re-entry as an alternative disposal solution, should not only be included as option in the guidelines for inclined satellites but also should be promoted. In this sense, the Sirius constellation was a missed opportunity to showcase a long-term sustainable exploitation of the geosynchronous region.

Another interesting idea to be explored in future geosynchronous mission design concepts is certainly the interplay between the solar radiation pressure and lunisolar perturbations. As we have seen in Sec.~\ref{sec:srpimpl}, using an area augmenting device can suppress the Lidov-Kozai effect and provide some low eccentricity variation operational orbits. A mission that uses the solar radiation trapping to stabilise and proceed to retract the area augmenting device at the end of the operations, could also lead to an atmospheric re-entry of the defunct satellite in a reasonable time frame.          

\section{Conclusions}\label{sec:conclusions}
 
The GEO orbital region was historically and is foreseen to be, one of the most precious assets in space exploitation. As it should be the norm with all natural resources available to humankind, the geosynchronous orbital region should also be treated in a sustainable way. Unfortunately, current practises in the region do not definitely ensure this.

In this work, a detailed dynamical cartography of the geosynchronous orbital region was performed to identify interesting possibilities in post-mission disposal strategies. Some of the key findings from the eccentricity variation maps in GEO are: 1) the positioning of the satellite only with respect to the geosynchronous resonance does not create interesting re-entry scenarios, 2) solar radiation pressure is important in the evolution of low-eccentricity low-inclination orbits, 3) for highly eccentric and inclined orbits the third-body perturbations dominate the dynamics, 4) there is a clear separation in the long-term evolution of low- and high-inclined orbits, implying that at geosynchronous altitude the Lidov-Kozai type of dynamics are prevailing.

Moving towards a sustainable GEO environment, mission design and planning should focus on exploiting fast re-entering orbits. Those are associated with particular geometries with respect to the Sun and Moon and they could be analytically located and introduced in the trajectory design process. Of course, this would require a whole re-assessment of the operations in the region. However, satellites in eccentric and inclined orbits could provide similar services to equatorial ones, but with the benefit that could achieve atmospheric re-entry at the end-of-life. Designing autonomous-navigation and propulsion systems to reach and follow those pathways is also a technological challenge but is well within the capabilities of future astrodynamical applications.  
 
One of the first steps that we need to take in this direction, is to redesign the guidelines for decommissioning GEO satellites. In fact, the opportunity to apply this kind of re-entering strategies was presented in the past with the Sirius constellation. Unfortunately, the operators decided to follow the current regulations and re-orbited the satellites. What could have been a pioneer example of a clean exploitation of GEO is not enforced by current guidelines. 

Moreover, given the dynamical complexities in the region, even the graveyard selection for low inclinations cannot be efficiently reduced to a single equation rule. Specialised tools that exploit the dynamical mapping of the region and sophisticated optimization algorithms should be used to provide the best-case disposal for each individual of post-mission disposal scenario. Considering the surrounding population in order to minimize collision probabilities of a given graveyard orbit should be also part of the disposal design process.

\begin{acknowledgements}
The research leading to these results has received funding from the Horizon 2020 Program of the European Union's Framework Programme for Research and Innovation (H2020-PROTEC-2015) under REA grant agreement number 687500 -- ReDSHIFT. The authors would like to thank  Martin Lara, Alessandro Rossi, Kleomenis Tsiganis, Elisa Maria Alessi, Despoina Skoulidou, and Aaron Rosengren for useful discussions and suggestions. The authors acknowledge the use of the Milkyway High Performance Computing Facility, and associated support services at the Department of Aerospace Science and Technology in Politecnico di Milano. The datasets generated for this study can be found in the repository at the link \emph{http://redshift-h2020.eu/}. We thank the anonymous reviewers for the useful comments.
\end{acknowledgements}

\section*{Appendix A: Semi-analytical modeling of geosynchronous orbits}\label{sec:appndA}

The single-averaged theory for Earth's satellites has been extensively studied and presented in the literature. In Particular the dynamical system adopted in PlanODyn is presented and validated in \cite{Colombo2015,Colombo2016}. Here we report the equations necessary to reproduce the calculations discussed in the present manuscript. In the following formulas $\mu_\oplus$ is the gravitational parameter of the Earth, $R_\oplus$ 
is the equatorial mean radius of the Earth and ($a,e,i,\Omega,\omega,M$) the classical orbital elements of the satellite.

\subsection*{Geopotential}

The perturbing function can be obtained in orbital elements through the classical Kaula expansion \cite{Kau1966}:
\begin{equation*}
R_{J_{lm}} = \frac{\mu_{\oplus} R^l_{\oplus}}{ a^{l+1}} \sum_{p=0}^l F_{lmp}(i)  \sum_{q=-\infty}^{\infty} G_{lpq}(e) S_{lmpq}(\omega,M,\Omega,\theta_g)
\end{equation*}
where
\begin{align*}
S_{lmpq} =& \left[ \begin{array}{c}
C_{lm} \\
- S_{lm}
\end{array} \right]^{l-m \textrm{ even}}_{l-m \textrm{ odd}} \cos \left((l-2p)\omega + (l-2p+q)M+ (\Omega - \theta_g)\right) \\
+& \left[ \begin{array}{c}
S_{lm} \\
C_{lm}
\end{array} \right]^{l-m \textrm{ even}}_{l-m \textrm{ odd}} \sin \left((l-2p)\omega + (l-2p+q)M+ (\Omega - \theta_g)\right).
\end{align*}

In the above expressions ($l,m,p,q$) are integers, $F_{lmp}(i)$, $G_{lpq})(e)$ are the Kaula $F$ and $G$ functions \cite{Kau1966}, $C_{lm}$,$S_{lm}$ the non-normalised spherical harmonics coefficients of Earth's gravitational field and $\theta_g$ the Greenwich hour angle. In the cases where the function $G_{lmn}$ is not a closed form function of $e$, it is expressed as a series up to $\mathcal{O}(e^{20})$ for the needs of this work.

\subsubsection*{Zonal harmonics}
For the secular effect of the zonal harmonics $J_{l0}=J_{l}$ we take into account the first-order averaged perturbations with respect to $J_2$, $J_3$ and $J_4$ as well as the second-order with respect to $J_2$ ($J_2^2$) \cite{Liou1980}:   
$$
\bar{\mathcal{R}}_\textrm{zonal} = \bar{\mathcal{R}}_{J_2} + \bar{\mathcal{R}}_{J_2^2} + \bar{\mathcal{R}}_{J_{3}} +\bar{\mathcal{R}}_{J_4}. 
$$
To isolate the secular terms in Kaula's expansion it suffices to select the terms in $R_{J_{l}}$ for which $M$ is not present i.e. $l-2p+q=0$ : 
\begin{equation*}
\bar{\mathcal{R}}_{J_{2}}  =  \frac{R_{\oplus}^2 J_2 \mu_{\oplus}  (3 \cos (2 i)+1)}{8 a^3 \left(1-e^2\right)^{3/2}}
\end{equation*}

\begin{equation*}
\bar{\mathcal{R}}_{J_3} = \frac{3 R_{\oplus} ^3 e J_3 \mu_{\oplus}  \sin (i) (5 \cos (2 i)+3) \sin (\omega )}{16 a^4 \left(1-e^2\right)^{5/2}}
\end{equation*}

\begin{align*}
\bar{\mathcal{R}}_{J_4}=&-\frac{3 R_{\oplus} ^4 J_4 \mu_{\oplus} }{128 a^5  \left(1-e^2\right)^{7/2}} \Big[ -35 \sin ^4(i) \left(2 e^2 \cos (2 \omega )-3 e^2-2\right) \\
   & +  20 \sin ^2(i) \left(3 e^2 \cos (2 \omega)-6 e^2-4\right)+8 \left(3 e^2+2\right) \Big]
\end{align*}
The second order averaged solution for $J_2$ ($J_2^2$) can be retrieved from Brouwer's closed form theory \cite{Brou1959}:
\begin{align*}
\bar{\mathcal{R}}_{J_2^2}=&\frac{3 R_{\oplus} ^4 J_2^2  \mu_{\oplus}}{128 a^5 \left(1-e^2\right)^{7/2}} \Big[ \cos ^4(i) \left(30 e^2 \cos (2 \omega )-5 e^2+36 \sqrt{1-e^2}+40\right) \\
&-2 \cos ^2(i) \left(16 e^2 \cos (2 \omega )-9 e^2+12 \sqrt{1-e^2}+4\right) \\
&+2 e^2 \cos (2 \omega )-5 e^2+4 \sqrt{1-e^2} \Big]
\end{align*}

\subsubsection*{Resonant Tesseral harmonics}
The resonant tesseral harmonics for the GEO case are also obtained from Kaula's expansion. Terms associated with $J_{21}$ (i.e. $C_{21}$ and $S_{21}$) are omitted because their strength is considerably smaller. The resonant contribution due to tesseral harmonics yields:
$$
\mathcal{R}_\textrm{tesseral}^{res} = \mathcal{R}^{res}_{J_{22}} + \mathcal{R}^{res}_{J_{31}} + \mathcal{R}^{res}_{J_{32}} + \mathcal{R}^{res}_{J_{33}} + \mathcal{R}^{res}_{J_{41}} +  \mathcal{R}^{res}_{J_{42}} + \mathcal{R}^{res}_{J_{43}} + \mathcal{R}^{res}_{J_{44}}
$$
where for each harmonic only terms satisfying the condition
$$
(l-2p) \dot{\omega} + (l-2p+q)\dot{M} + m (\dot{\Omega} - \dot{\theta}_g) \approx 0
$$
are selected. The resonant condition for satellites at geosynchronous altitude yields:
$$
l-2p+q=m, \quad \textrm{with } m=1,2,3,4 \ldots.
$$
Introducing the \emph{astronomical longitude} as
$$
\lambda = \omega + \Omega + M - \theta_{g}
$$
the resonant contributions read:
\begin{align*}
R_{J_{22}}^{res} =&\frac{R_{\oplus} ^2  \mu_{\oplus} }{a^3}  \bigg( C_{22} F_{220}
   G_{200} \cos (2 \lambda ) + S_{22} F_{220} G_{200} \sin (2 \lambda )\\
   +&C_{22} F_{221} G_{212} \cos (2 (\lambda -\omega )) +S_{22} F_{221} G_{212} \sin (2 (\lambda-\omega )) \\
   +&C_{22} F_{222} G_{224} \cos (2 (\lambda -2 \omega ))+S_{22} F_{222} G_{224} \sin (2 (\lambda -2 \omega )) \bigg)
\end{align*}
\begin{align*}
R_{J_{31}}^{res} =& \frac{R_{\oplus} ^3  \mu_{\oplus} }{a^4}  \bigg(C_{31} F_{310} G_{30-2} \cos (\lambda +2 \omega )+C_{31} F_{312} G_{322} \cos (\lambda -2 \omega ) \\ 
&+C_{31} F_{313} G_{334} \cos (\lambda -4 \omega )+C_{31} F_{311} G_{310} \cos (\lambda ) \\
&+S_{31} F_{310} G_{30-2} \sin (\lambda +2 \omega ) +S_{31} F_{313} G_{334} \sin (\lambda -4 \omega ) \\
   &+S_{31} F_{312} G_{322} \sin (\lambda -2 \omega )+S_{31} F_{311} G_{310} \sin(\lambda )  \bigg)
\end{align*}
\begin{align*}
R_{J_{32}}^{res} =&\frac{R_{\oplus} ^3  \mu_{\oplus} }{a^4}  \bigg( C_{32} F_{320} G_{30-1} \sin (2 \lambda +\omega )+C_{32} F_{322} G_{323} \sin (2 \lambda -3 \omega ) \\
&+C_{32} F_{321} G_{311} \sin (2 \lambda -\omega )-S_{32} F_{320} G_{30-1} \cos (2 \lambda +\omega )\\
&-S_{32} F_{321} G_{311} \cos (2 \lambda -\omega)-S_{32} F_{322} G_{323}\cos (2 \lambda -3 \omega ) \bigg)
\end{align*}
\begin{align*}
R_{J_{33}}^{res} =& \frac{R_{\oplus} ^3  \mu_{\oplus} }{a^4}  \bigg(
C_{33} F_{331} G_{312} \cos (3 \lambda -2 \omega )+C_{33} F_{332} G_{324} \cos (3 \lambda -4 \omega ) \\
&+C_{33} F_{330} G_{300} \cos (3 \lambda )+S_{33} F_{332} G_{324} \sin (3 \lambda -4 \omega )\\
&+S_{33} F_{331} G_{312} \sin (3 \lambda -2 \omega
   )+S_{33} F_{330} G_{300} \sin (3 \lambda )  \bigg) 
\end{align*}
\begin{align*}
R_{J_{41}}^{res} = & \frac{R_{\oplus} ^4  \mu_{\oplus} }{a^5}  \bigg( C_{41} F_{410} G_{40-3} \sin (\lambda +3 \omega )+ C_{41} F_{413} G_{433} \sin (\lambda -3 \omega ) \\
&+C_{41} F_{412} G_{421}
   \sin (\lambda -\omega )+C_{41} F_{411} G_{41-1} \sin (\lambda +\omega ) \\
   &-S_{41} F_{410} G_{40-3} \cos (\lambda +3 \omega)-S_{41} F_{411} G_{41-1} \cos (\lambda +\omega )\\
   &-S_{41} F_{412} G_{421} \cos (\lambda -\omega )-S_{41} F_{413} G_{433}\cos (\lambda -3 \omega ) \bigg)
\end{align*}

\begin{align*}
R_{J_{42}}^{res} =& \frac{R_{\oplus} ^4  \mu_{\oplus} }{a^5}  \bigg(
C_{42} F_{420} G_{40-2} \cos (2 (\lambda +\omega ))+C_{42} F_{422} G_{422} \cos (2 (\lambda -\omega ))\\
&+C_{42} F_{423}
   G_{434} \cos (2 (\lambda -2 \omega ))+C_{42} F_{421} G_{410} \cos (2 \lambda ) \\
   &+S_{42} F_{420} G_{40-2} \sin (2 (\lambda
   +\omega ))+S_{42} F_{423} G_{434} \sin (2 (\lambda -2 \omega )) \\
  &+S_{42} F_{422} G_{422} \sin (2 (\lambda -\omega ))+S_{42} F_{421} G_{410} \sin (2 \lambda ) \bigg)
\end{align*}

\begin{align*}
R_{J_{43}}^{res} =& \frac{R_{\oplus} ^4  \mu_{\oplus} }{a^5}  \bigg(
C_{43} F_{430} G_{40-1} \sin (3 \lambda +\omega )+C_{43} F_{432} G_{423} \sin (3 (\lambda -\omega )) \\
& +C_{43} F_{431} G_{411}\sin (3 \lambda -\omega )-S_{43} F_{430} G_{40-1} \cos (3 \lambda +\omega )\\
   &-S_{43} F_{431} G_{411} \cos (3 \lambda -\omega
   )-S_{43} F_{432} G_{423} \cos (3 (\lambda -\omega )) \bigg)
\end{align*}

\begin{align*}
R_{J_{44}}^{res} =& \frac{R_{\oplus} ^4  \mu_{\oplus} }{a^5}  \bigg(
C_{44} F_{441} G_{412} \cos (4 \lambda -2 \omega )+C_{44} F_{442} G_{424} \cos (4 (\lambda -\omega )) \\
&+C_{44} F_{440}   G_{400} \cos (4 \lambda )+S_{44} F_{441} G_{412} \sin (4 \lambda -2 \omega ) \\
&+S_{44} F_{442} G_{424} \sin (4 (\lambda -\omega
   ))+S_{44} F_{440} G_{400} \sin (4 \lambda ) \bigg)
\end{align*}

\subsection*{Third body perturbations}
The third-body potential implemented in PlanODyn is expanded in powers of the parallactic factor $(a/r_b)$ as in \cite{Kauf1972,Colombo2015}, where $r_b$ is the geocentric distance of the perturber. Terms up to the fourth order ($P_2,P_3,P_4$) in the expansion are retained for both the Sun and the Moon.  
$$
\bar{\mathcal{R}}_\textrm{3body} =\bar{\mathcal{R}}_{P_2\leftmoon} +\bar{\mathcal{R}}_{P_3\leftmoon} +\bar{\mathcal{R}}_{P_4\leftmoon} + \bar{\mathcal{R}}_{P_2\odot} + \bar{\mathcal{R}}_{P_3\odot} + \bar{\mathcal{R}}_{P_4\odot}
$$
The perturbing functions are single-averaged in closed form over the satellite's mean anomaly. This operation yields the following expressions for the disturbing functions \cite{Kauf1972,Colombo2015}:
\begin{equation*}
\bar{\mathcal{R}}_{P_2b} = \frac{a^2 \mu_b}{r_b^3} \left( 3 A^2 e^2+\frac{3 A^2}{4}-\frac{3 B^2 e^2}{4}+\frac{3 B^2}{4}-\frac{3 e^2}{4}-\frac{1}{2} \right)
\end{equation*}

\begin{equation*}
\bar{\mathcal{R}}_{P_3b} =  \frac{a^3 \mu_b}{r_b^4} \left( -\frac{25}{4} A^3 e^3-\frac{75 A^3 e}{16}+\frac{75}{16} A B^2 e^3-\frac{75}{16} A B^2 e+\frac{45 A e^3}{16}+\frac{15 A e}{4} \right)
\end{equation*}

\begin{align*}
\bar{\mathcal{R}}_{P_4b} = & \frac{a^4 \mu_b}{r_b^5} \bigg( \frac{105 A^4 e^4}{8}+\frac{315 A^4 e^2}{16} +\frac{105 A^4}{64}-\frac{315}{16} A^2 B^2 e^4+\frac{525}{32} A^2 B^2 e^2 \\
&+\frac{105 A^2 B^2}{32}-\frac{135 A^2 e^4}{16} -\frac{615 A^2 e^2}{32}-\frac{15 A^2}{8}+\frac{105 B^4 e^4}{64}-\frac{105 B^4 e^2}{32}\\
&+\frac{105 B^4}{64}+\frac{45 B^2 e^4}{32}+\frac{15
   B^2 e^2}{32}-\frac{15 B^2}{8}+\frac{45 e^4}{64}+\frac{15 e^2}{8}+\frac{3}{8} \bigg)
\end{align*}

where $A$ and $B$ are given from \cite{Kauf1972}:
\begin{align*}
A =&-\hat{x}_b \cos (i) \sin (\omega ) \sin (\Omega )+\hat{y}_b \cos (i) \sin (\omega ) \cos (\Omega )+\hat{z}_b \sin (i) \sin (\omega )\\
&+\hat{x}_b \cos
   (\omega ) \cos (\Omega )+\hat{y}_b \cos (\omega ) \sin (\Omega )\\
B =& -\hat{x}_b \cos (i) \cos (\omega ) \sin (\Omega )+\hat{y}_b \cos (i) \cos (\omega ) \cos (\Omega )+\hat{z}_b \sin (i) \cos (\omega ) \\
& -\hat{x}_b \sin  (\omega ) \cos (\Omega )-\hat{y}_b \sin (\omega ) \sin (\Omega )
\end{align*}
and $\boldsymbol{\hat{r}}_b = (\hat{x}_b,\hat{y}_b,\hat{z}_b)$ is the unit vector to the perturbing body.
\subsection*{Solar radiation pressure}
The single-averaged contribution of the solar radiation pressure as implemented in PlanODyn is given from \cite{Kau1962,Krivov1997}:
\begin{align*}
\bar{\mathcal{R}}_{\textrm{SRP}}&=\frac{3}{2} a e P_{SRP} C_R \frac{A}{m} \Big( cos (\varepsilon ) \sin (\lambda_{\odot}) \cos (\omega ) \sin (\Omega )\\
&+\cos (\varepsilon ) \cos (i) \sin(\lambda_{\odot} ) \sin (\omega ) \cos (\Omega ) +\sin (\varepsilon ) \sin (i) \sin (\lambda_{\odot} ) \sin (\omega)\\
&-\cos (i) \cos (\lambda_{\odot} ) \sin (\omega ) \sin (\Omega ) +\cos (\lambda_{\odot} ) \cos (\omega ) \cos (\Omega ) \Big),
\end{align*}
where $\lambda_{\odot}$ is the ecliptic longitude of the Sun, $P_{SRP}$ is the solar radiation pressure per unit area at 1 AU, $C_R$ the satellite's reflectivity coefficient and $A/m$ its area-to-mass ratio.

\subsection*{Earth's general precession}
The long-term contribution of Earth's general precession is described in detail in Appendix B. The perturbing function yields:
\begin{equation*}
\mathcal{R}_{\textrm{prec}} = \sqrt{a \left(1-e^2\right) \mu_{\oplus}} (\mathcal{P}_x \sin (i) \sin (\Omega)-\mathcal{P}_y \sin (i) \cos (\Omega)+\mathcal{P}_z  \cos (i))
\end{equation*}
where $\boldsymbol{\mathcal{P}}=(\mathcal{P}_x,\mathcal{P}_y,\mathcal{P}_z)$ is the angular velocity of a precessing geocentric equatorial frame relative to an inertial (i.e. J2000). 
\subsection*{Equations of motion}
The complete long-term evolution at geosynchronous altitude is described by
$$
\bar{\mathcal{R}}_{\textrm{GEO}} = \bar{\mathcal{R}}_\textrm{zonal} + \mathcal{R}_\textrm{tesseral}^{res} +\bar{\mathcal{R}}_\textrm{3body}+ \bar{\mathcal{R}}_\textrm{SRP} + \mathcal{R}_{\textrm{prec}}.
$$
The equations of motion in orbital elements are then derived via Lagrange's planetary equations \cite{Battin1999}:
\begin{align*}
\frac{da}{dt} & = \frac{2}{na}\frac{\partial\bar{\mathcal{R}}_{\textrm{GEO}}}{\partial M} \\
\frac{de}{dt} & = \frac{1}{n a^2 e}\left((1-e^2)\frac{\partial\bar{\mathcal{R}}_{\textrm{GEO}}}{\partial M}-\sqrt{1-e^2}\frac{\partial \bar{\mathcal{R}}_{\textrm{GEO}}}{\partial \omega}\right) \\
\frac{di}{dt} & = \frac{1}{n a^2 \sin i \sqrt{1-e^2}} \left(\cos i \frac{\partial \bar{\mathcal{R}}_{\textrm{GEO}}}{\partial \omega}-\frac{\partial\bar{\mathcal{R}}_{\textrm{GEO}}}{\partial \Omega}\right) \\
\frac{d\Omega}{dt} & = \frac{1}{n a^2 \sin i \sqrt{1-e^2}} \frac{\bar{\mathcal{R}}_{\textrm{GEO}}}{\partial i} \\
\frac{d\omega}{dt} & = - \frac{1}{n a^2 \sin i \sqrt{1-e^2}} \cos i \frac{\bar{\mathcal{R}}_{\textrm{GEO}}}{\partial i} + \frac{\sqrt{1-e^2}}{n a^2 e} \frac{\bar{\mathcal{R}}_{\textrm{GEO}}}{\partial e} \\
\frac{d M}{dt} & = n - \frac{1-e^2}{na^2e}\frac{\partial\bar{\mathcal{R}}_{\textrm{GEO}}}{\partial e} - \frac{2}{na} \frac{\bar{\mathcal{R}}_{\textrm{GEO}}}{\partial a}.
\end{align*}

\section*{Appendix B: Semi-analytical modelling of Earth's general precession}\label{sec:appndB}

One effect that is usually overlooked in analytical and semi-analytical propagations is the effect of Earth's precession in the orbital evolution. Namely, for the Earth's gravity field representation, the lunisolar contributions and solar radiation pressure, a mean equator and mean equinox of the day frame is usually considered (MOD). All the ephemerides are also computed in this frame. However, this frame is not inertial due to Earth's precession. This effect is negligible for low and medium Earth orbits, but becomes important for GEO and higher-altitude satellites, especially for long-term propagations \cite{Gur2007}. Particularly, it produces a long-term contribution to the inclination and right ascension of the ascending node that we would like to include in our simulations. In the literature, there exist different methods to model this effect \cite{KozKin1973,Efroimsky2005,Gur2007}.

\begin{figure}[!htb]
\centering
\includegraphics[width=0.75\textwidth]{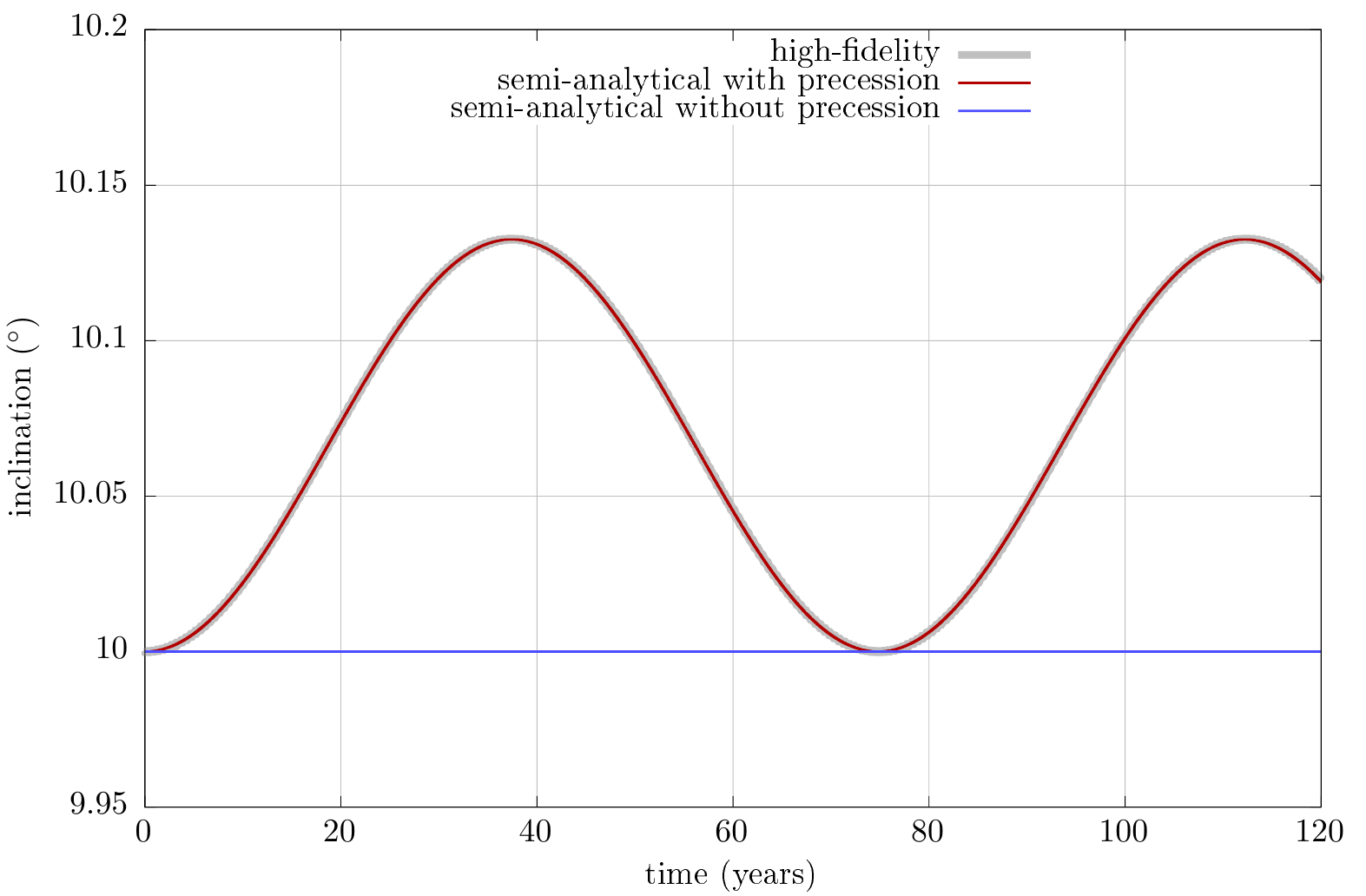}
\caption{Time evolution of the inclination for an orbit at about GEO semi-major axis, showing the contribution of Earth's precession on the evolution. The force model included only the $J_2$ effect and the transformation from MOD to J2000 frames. The three lines corresponds to: a) a high-fidelity propagation (grey line), b) a semi-analytical with only $J_2$ (blue line) and c) a semi-analytical that takes into account the Earth's precession contribution on top of $J_2$ (red line).  }
\label{fig:EPexample}
\end{figure}

Here, an efficient and accurate way to model the Earth's general precession effect is presented. The disturbing function for the Coriolis contribution in the non-inertial MOD frame is given by \cite{Gold1965}:
\begin{equation}
\mathcal{R}_{\textrm{prec}} = \sqrt{a \left(1-e^2\right) \mu_{\oplus}} \hspace{0.1cm} \boldsymbol{\mathcal{P}} \cdot \mathbf{\hat{w}}
\label{eq:Rnonin}  
\end{equation}
where $\mu_{\oplus}$ is the Earth's gravitational parameter, $a,e$ the satellite's semi-major axis and eccentricity, $\mathbf{\hat{w}}$ is the unit vector, normal to the orbital plane, and $\boldsymbol{\mathcal{P}} = (\mathcal{P}_x,\mathcal{P}_y,\mathcal{P}_z)$ is the angular velocity vector of the non-inertial frame relative to an inertial one. In orbital elements the normal vector $\mathbf{\hat{w}}$ becomes:
\begin{equation}
\mathbf{\hat{w}} = \left( \begin{array}{c} \sin{i} \sin{\Omega} \\ - \sin{i} \cos{\Omega} \\ \cos{i} \end{array} \right) 
\end{equation}
where $i$ and $\Omega$ are the satellite's inclination and the right ascension of the ascending node. 
The disturbing function then reads
\begin{equation*}
\mathcal{R}_{\textrm{prec}} = \sqrt{a \left(1-e^2\right) \mu_{\oplus}} (\mathcal{P}_x \sin (i) \sin (\Omega)-\mathcal{P}_y \sin (i) \cos (\Omega)+\mathcal{P}_z  \cos (i))
\end{equation*}
and its contribution to the classical orbital elements, is given by Lagrange planetary equations:
\begin{align}
\label{eq:EPcontrib}
\begin{split}
\left(\frac{di}{dt}\right)_{\textrm{prec}} &= - \mathcal{P}_x \cos{\Omega} - \mathcal{P}_z \sin{\Omega} \\
\left(\frac{d \Omega }{dt}\right)_{\textrm{prec}} &= \mathcal{P}_x \cot{i} \sin{\Omega}  - \mathcal{P}_y \cos{\Omega} \cot{i} - \mathcal{P}_z \\
\left(\frac{d \omega} {dt}\right)_{\textrm{prec}} &= \cos{i} \left( \mathcal{P}_z + \mathcal{P}_y \cos{\Omega} \cot{i} - \mathcal{P}_x \cot{i} \sin{\Omega} \right),
\end{split} 
\end{align}

while the rest of the orbital elements ($a,e,M$) are not directly affected. Keep in mind that the elements appearing in this formulation are non-osculating or \emph{contact elements}. Since short-period terms do not appear in Eqs.~\ref{eq:EPcontrib} the equations of motion can be coupled with the rest of the single-averaged effects. However, one should pay attention when recovering the short-periodic terms of the mixed formulation \cite{Efroimsky2005}. 

To estimate the effect of the Earth's precession, we need now to estimate the angular velocity vector $\boldsymbol{\mathcal{P}}$ of the MOD frame with respect to an inertial frame, which we assume to be the one defined by the mean equator and mean equinox of the epoch J2000. The rotation matrix from the MOD to the J2000 inertial frame is given from 
\begin{equation}
R_{MOD\rightarrow J2000}=  R_z(\zeta) R_y(-\theta) R_z(z)
\end{equation} 
where $R_y,R_z$ denote the clockwise rotation matrices with respect to the y and z axis respectively. The angles $\zeta,\theta,z$ represent the general Earth's precession, i.e. the combined effect of lunisolar attraction on Earth's equatorial bulge and the change of Earth's orbital plane due to planetary attractions, and are computed from Lieske's theory \cite{Lie1977}:
\begin{align*}
\zeta &= 2306.2181 \cdot T + 0.30188 \cdot T^2 + 0.017988 \cdot T^3  \\
\theta &= 2004.3109 \cdot T - 0.42665 \cdot T^2 - 0.041833 \cdot T^3 \\
z &= 2306.2181 \cdot T + 1.09468 \cdot T^2 + 0.018203 \cdot T^3 
\end{align*}
where $T$ is the Julian time past the 01/01/2000 in centuries and the angles are given in arcseconds. Now it suffices to observe that the angular velocity tensor $\Pi$ between the two frames is given by:
\begin{equation}
\Pi =\left(\begin{array}{ccc}
0 & - \mathcal{P}_z & \mathcal{P}_y \\
\mathcal{P}_z & 0 & -\mathcal{P}_x \\
- \mathcal{P}_y & \mathcal{P}_x  & 0
\end{array} \right)  = \dot{R}_{MOD\rightarrow J2000} \cdot R^\intercal_{MOD\rightarrow J2000}
\end{equation}
where the $\dot{R}$ is the time derivative and $R^{\intercal}$ is the transpose of $R$. Thus
\begin{align*}
\mathcal{P}_x &=  \dot{\theta}  \sin{\zeta}  - \dot{z}  \cos{\zeta} \sin{\theta} \\
\mathcal{P}_y &=  \dot{\theta} \cos{z}  + \dot{z}  \sin{\theta} \sin{\zeta} \\
\mathcal{P}_z &= -  \dot{z} \cos{\theta} - \dot{\zeta}
\end{align*}
where the rates are given in arcseconds per century and
\begin{align*}
\dot{\zeta} &= 2306.2181 + 0.60376 \cdot T + 0.053964 \cdot T^2 \\
\dot{\theta} &= 2004.3109 - 0.8533 \cdot T - 0.125499 \cdot  T^2 \\
\dot{z} &= 2306.2181 + 2.18936 \cdot T + 0.054609 \cdot T^2. 
\end{align*}

An illustrative example of this effect is given in Fig.~\ref{fig:EPexample}, where the time evolution of the inclination for a hypothetical orbit at geosynchronous altitude for a time span of 120 years is presented. In order to make the Earth's precession effect clearer, we further account only for the Earth's oblateness ($J_2$) contribution. Under the single-averaged $J_2$ formulation (blue line), the mean inclination is constant. The same initial conditions are propagated also using a high-fidelity modelling (grey line), i.e. using a $J_2$ force model in Cartesian coordinates and NASA's SPICE tool-kit for the Earth's rotation matrices. The evolution now significantly differs, and a non-negligible contribution of Earth's general precession can be observed, producing an oscillation with an amplitude of about 8 seconds of a degree in the inclination evolution. However, this effect can be accurately included in the semi-analytical propagation, by adding the Earth's precession contributions (Eq.~\ref{eq:EPcontrib}) (red line).  

\section*{Compliances with ethical standards}
\textbf{Conflict of Interest Statement} On behalf of all authors, the corresponding author states that there is no conflict of interest.
\bibliographystyle{spmpsci}      
\bibliography{geodyn}   

%
%

\end{document}